\documentclass[a4paper,11pt]{article}
\pdfoutput=1

\usepackage{jheppub} 

\usepackage[T1]{fontenc} 
\usepackage{soul}
\usepackage{subcaption}
\usepackage{dsfont}
\usepackage{soul}
\usepackage{graphicx}
\usepackage{mwe}
\usepackage{comment}
\usepackage[normalem]{ulem}
\makeatletter
\newcommand*{\rom}[1]{\expandafter\@slowromancap\romannumeral #1@}
\makeatother
\newcommand{\haoyu}{\textcolor{blue}}

\newcommand{\footreff}[1]{Footnote~\ref{#1}}

\title{Probing holography in $p$-adic CFT}

\author[a]{Stephen Ebert,}
\author[b]{Hao-Yu~Sun,}
\author[b]{and Meng-Yang~Zhang}

\affiliation[a]{Mani L. Bhaumik Institute for Theoretical Physics,\\ University of California, Los Angeles, CA 90095-1547, USA}
\affiliation[b]{Center for Theoretical Physics and Department of Physics, \\ University of California, Berkeley, CA 94720-7300, USA}

\emailAdd{stephenebert@physics.ucla.edu}
\emailAdd{hkdavidsun@berkeley.edu}
\emailAdd{zhang.mengyang@berkeley.edu}

\abstract{We holographically calculate the partition functions of certain types of isotropic sectors of the CFTs dual to Bruhat-Tits trees and $p$-adic BTZ black holes. Along the way, we propose new spectral decompositions of the Laplacian operator other than the plane-wave basis on these two types of background, with both analytical and numerical evidence. We extract the density of states and hence entropy from the BTZ partition function via the inverse Laplace transform. Then the one-loop Witten diagram is computed in the $p$-adic BTZ black hole background, yielding constraints on the heavy-heavy-light averaged three-point coefficient of its boundary $p$-adic CFT. Finally, for general $p$-adic CFTs (not necessarily holographic), we analyze the representation theory of their global conformal group $PGL\left(2,\mathbb{Q}_p\right)$, and discuss the suitability of different representations as Hilbert spaces of $p$-adic CFT.

\begin{center}
{\it Dedicated to the memory of Steven Scott Gubser.} 
\end{center}
}

\begin{document} 

\maketitle
\flushbottom

\section{Introduction}
\label{sec:intro}
Explorations have emerged in the last three decades between the interplay of algebraic number theory and string theory. Once one defines the $p$-adic norm, a well-known phenomenon appears in string scattering amplitudes from adelic products. We can construct the real Veneziano amplitude $A_4^{(\infty)}(s,t,u)$ for four tachyons scattering in the open bosonic string theory at tree-level from the product over all prime numbers of the $p$-adic Veneziano amplitudes $A^{(p)}(s,t,u)$ \cite{$p$-adicVenezianoAmplitude,BFOW}\footnote{For higher-point scattering amplitudes, see \cite{BogdanScattering, Jepsen:2022pkn} for recent discussions.}
\begin{equation}
\begin{aligned}
\label{eq: 1}
 A_4^{(\infty)}(s,t,u)  = \bigg[ \prod_p A_4^{(p)}(s,t,u) \bigg]^{-1}= \Gamma_\infty (-s-1) \Gamma_\infty(-t -1) \Gamma_\infty(-u-1),
\end{aligned}
\end{equation}
where $\Gamma_\infty(x)$ is the real Euler gamma function, $s,t,u$ are the Mandelstam variables and the $p$-adic Veneziano amplitude is defined as 
\begin{equation}
\begin{aligned}
A_4^{(p)}(s, t, u)  &=\Gamma_p (-s-1) \Gamma_p (-t-1) \Gamma_p(-u-1).
\end{aligned}
\end{equation}
Here $\Gamma_p(x)$ is the $p$-adic Gamma function $\Gamma_p(x) = \zeta_p(x)/\zeta_p(1-x), ~\zeta_p (x) = 1/(1-p^{-x})$ and the $p \rightarrow \infty$ limit reduces to the standard real case $\mathbb{Q}_\infty = \mathbb{R}$. An interpretation of the $p$-adic string is given by \cite{Zabrodin}, where the open string worldsheet is replaced by a Bruhat-Tits tree (defined in Section \ref{sec:BT Tree} there) and its boundary as the $p$-adic numbers. 

Recently inspired by this perspective, Gubser et al. \cite{Gubser1} and Heydeman et al. \cite{Caltech} proposed a non-Archimedean version of a toy model for the Euclidean AdS/CFT correspondence \cite{Maldacena}. In the simplest topology, the usual continuous bulk is replaced by an infinite, symmetric, and homogeneous (i.e., no preferred central vertex) tree of uniform valency $p+1$. This tree, known as the \textit{Bruhat-Tits tree} (or \textit{Bethe lattice}), is expressed as the left coset space 
\begin{equation}
\label{eq: coset}
    T_p\equiv PGL\left(2,\mathbb{Q}_p\right)/PGL\left(2,\mathbb{Z}_p\right),
\end{equation}
where $PGL\left(2,\mathbb{Q}_p\right)$ is the $p$-adic global conformal group\footnote{It is a \textit{totally disconnected locally compact }(TDLC) group, with respect to the $\mathbb{Q}_p$ topology as explained in Section 10.5 in \cite{L-function}, but not compact. Its subgroup $PSL\left(2,\mathbb{Q}_p\right)$ is neither compact nor open. \label{foot:TDLC}}, whose maximal compact open subgroup is $PGL\left(2,\mathbb{Z}_p\right)$. The definition (\ref{eq: coset}) is reminiscent of the hyperbolic 3-space $\mathbb{H}^3\simeq SL(2,\mathbb{C})/SU(2)$ with boundary $\mathbb{P}^1(\mathbb{C})$, describing Euclidean asymptotic AdS$_3$. Additionally, for the \textit{unramified} finite Galois extension $\mathbb{Q}_{p^n}$ of $\mathbb{Q}_{p}$, the tree $T_{p^n}$ has valency $p^n+1$ and boundary $\partial T_{p^n} = \mathbb{P}^1\left(\mathbb{Q}_{p^n}\right)$. Using unramified extensions, we are not limited to just one-dimensional boundaries, but we can think of Euclidean AdS$_{n+1}$ analogous to $T_{p^n}$. 

With this specific discretization of the bulk, one can put physical degrees of freedom on its vertices. The simplest case is to introduce scalars. Furthermore, the tree as well as its dual graph can be identified with tensor networks in order to study bulk reconstruction, quantum error-correction codes \cite{Caltech,MMPS} and holographic RG flow \cite{Ling-Yan}. 

One can study more general fields, such as spins, on the trees. The first realization of spins in $p$-adic AdS/CFT was introduced by Gubser et al. \cite{GubserSpin,SYK} with results on the bulk dual to non-scalar operators and dynamical gauge fields. In particular, they computed the holographic two-point correlator of an operator $\mathcal{O}_\psi$ dual to a spin state $|\psi \rangle$. One of the main conclusions was that the fermionic two-point correlator is of similar form to the scalar two-point correlator up to normalization and a non-trivial sign character resembling the operators' statistics.

There are other exotic and interesting applications in the context of the $p$-adics. An example is to understand higher-order versions of the Klebanov-Tarnopolsky model for both the real and the $p$-adic cases. Recently in \cite{Melonic}, the authors analyzed the situation for $q$ propagators at each interaction vertex to calculate four-point correlators. In addition, \cite{Melonic} provided nice comparisons with matrix field theory regarding the propagators' symmetry group.

There have been more recent uses of $p$-adics including: the Berkovich space to encode the renormalization group flow of the energy spectrum of the theory of a particle-in-a-box \cite{box2}, studying local diffeomorphisms of $p$-adic BTZ black hole and Bruhat-Tits tree backgrounds \cite{Hung123,Hung3regw43} and D-branes \cite{BogdanBranes}.

Given these progresses, the status quo of $p$-adic AdS/CFT seems rather one-sided in the sense that $p$-adic CFT is not well-formulated, because a Hilbert space is absent. Melzer \cite{melzer}, and later Harlow et al. \cite{Symmetree} and Gubser-Parikh \cite{Gubser3}, have shed some light on its OPE structure, but its partition function and local conformal algebra were not thoroughly explored. As mentioned earlier, it is very natural to describe global AdS$_n$ as a Bruhat-Tits tree. One well-known phenomenon studied in 3d gravity is the BTZ black hole. Heydeman et al. \cite{Caltech} formulated a $p$-adic BTZ black hole, which serves as a motivation for this paper in the hope of extracting meaningful information for $p$-adic CFTs. We calculated the bulk partition function and showed it has many key features as in \cite{MaloneyWitten}, such as Bekenstein-Hawking area law in 3d gravity. We hope this partition function could initiate future works to match the boundary CFT data. 

A meaningful direction to gain more insight on the holographic $p$-adic CFT's structure is to study the constraints on the averaged three-point coefficients for $p$-adic BTZ black holes as done in regular BTZ black holes \cite{KrausMaloney}. We found the averaged three-point coefficient for a $p$-adic BTZ black hole in the limit of large horizon $l$ to obey an exponentially decaying behavior $e^{-\Delta l}$ similar to that for regular BTZ black holes \cite{KrausMaloney}, where $\Delta$ is a boundary CFT data. One would hope to recover this result purely from the Lie algebra representation of the holographic $p$-adic CFT. However, we make a strong argument against the existence of a local algebra, and therefore we turn to group representations, where a classification theorem comes in handy. We analyze each case, and propose a way of checking which representation of $p$-adic CFT fits the genus-1 bulk calculation.

The rest of this paper is organized as follows. In Section \ref{sec:2}, we review mathematical and physical concepts relevant to $p$-adic AdS/CFT. In Section \ref{sec:3}, we solve isotropic Laplace problems on Bruhat-Tits trees and $p$-adic BTZ black hole geometries via linear recurrence, and therefore obtain the partition functions of the rotation-singlet sectors, whose various implications are discussed. In Section \ref{sec:4}, we calculate the one-loop Witten diagram describing the 1-to-2 scattering between two types of bulk scalars dual to light primary fields on the boundary in the background of a $p$-adic BTZ black hole, and the result imposes a constraint on potentially precise formulations on $p$-adic CFTs. In Section \ref{sec:5}, we review the representation theory on $PGL\left(2,\mathbb{Q}_p\right)$. Furthermore, we present an analysis on possible group representations as Hilbert spaces for $p$-adic CFTs. Finally, we conclude with a discussion of the results and future directions in Section \ref{sec:6}. The full spectrum, including all anisotropic eigenvalues, of the Laplacian on Bruhat-Tits trees is presented in the Appendix \ref{sec:spectrumBTTree}, without computing the corresponding full partition function.

\section{Summary of \texorpdfstring{$p$}{}-adic basics}
\label{sec:2}
%Now before presenting our main results, we familiarize ourselves with necessary ingredients.
\subsection{\texorpdfstring{$p$}{}-adic numbers}
As mentioned in the introduction, in constructing the $p$-adic AdS/CFT correspondence, the non-Archimedean field $\mathbb{Q}_p$ plays an important role. We briefly review Archimedean and non-Archimedean fields before discussing $\mathbb{Q}_p$. Let $\mathbb{F}$ be any field with a norm $|\cdot|_\mathbb{F}$ which obeys the standard axioms\footnote{Rigorously speaking, in algebraic geometry and algebraic number theory, these axioms define the term ``\textit{valuation}'' or ``\textit{absolute value}'', differing from the ``norm'' in functional analysis, whose \textit{absolute homogeneity} replaces the second axiom here. However, we still abuse the term ``norm'' throughout this paper.} for any $x,y \in \mathbb{F}$ \cite{Milne}:

1. $|x|_\mathbb{F} \geq 0$ \text{and is saturated when} $x \equiv 0$;

2. $|x \cdot y|_\mathbb{F} = |x|_\mathbb{F} \cdot |y|_\mathbb{F}$;

3. $|x+y|_\mathbb{F} \leq |x|_\mathbb{F} + |y|_\mathbb{F}$ (triangle inequality).

When $\mathbb{F}$ is \textit{Archimedean}, its norm obeys $\operatorname{sup} \left\{|n|_\mathbb{F} : n \in \mathbb{Z}\right\} = \infty$; whereas when $\mathbb{F}$ is \textit{non-Archimedean}, its norm obeys $\operatorname{sup}\left\{|n|_\mathbb{F} : n \in \mathbb{Z}\right\} = 1$. The major difference between Archimedean and non-Archimedean fields is that only the latter has ultrametricity \cite{L-function}: 
\begin{equation}
\label{eq: ultra}
    |x + y|_\mathbb{F} \leq \operatorname{sup} \left(|x|_\mathbb{F}, |y|_\mathbb{F} \right),
\end{equation}
implying that all triangles over an non-Archimedean field are isosceles.

Characteristic of $\mathbb{F}$ is defined as the least $n$ such that when one adds up $n$ copies of $1\in\mathbb{F}$, one obtains zero. Naturally, $\mathbb{Q}$, $\mathbb{R}$, and $\mathbb{C}$ are fields of characteristic zero, while the set of residue classes modulo a prime $p$ is a field of characteristic $p$ \cite{Koblitz}. We are concerned with $\mathbb{Q}_p$, a \textit{characteristic zero non-Archimedean field}. To obtain degree-$n$ unramified extensions $\mathbb{Q}_{p^n}$, we adjoin $\mathbb{Q}_p$ by a primitive $(p^n -1)^{\text{th}}$ root of unity \cite{Koblitz}.

For any prime number $p$, $\mathbb{Q}_p$ is the completion of $\mathbb{Q}$ with respect to the $p$-adic norm $|\cdot|_p$ \cite{L-function}. To define $|\cdot|_p$, we note that any $x \in \mathbb{Q}_p \backslash \{0\}$ has a unique $p$-adic expansion
\begin{equation}
 \label{expansion}
     x = \underbrace{... a_{3} a_{2} a_{1} a_{0}}_{\text { in } \mathbb{Z}_{p}}. \underbrace{a_{-1} a_{-2} ... a_{v_{p}}}_{\text { fractional part of } x} \equiv \sum_{n = v_p}^\infty a_n p^n,
 \end{equation}
 where $a_n \in \{0, 1, \cdots, p-1 \}$, and $v_p$ is the smallest integer index such that $a_{v_p}\neq 0$ \cite{BFOW}. The $p$-adic norm of $x$ is then defined as
 \begin{equation}
 \label{eq: norm}
     |x|_p = p^{-v_p}.
 \end{equation}
Notice that although $0\in\mathbb{Q}_p$ has no $p$-adic expansion, we naturally define $|0|_p = 0$. 
 
One can ask do other completions of $\mathbb{Q}$ exist? The answer is given by Ostrowski's theorem \cite{L-function}: \emph{the only non-trivial norms on $\mathbb{Q}$ are those equivalent to the $|\cdot|_p$ or the ordinary norm $|\cdot|_\infty$.} In other words, $\mathbb{Q}_p$ and $\mathbb{R}$ are the only completions of $\mathbb{Q}$. For unramified extensions of Ostrowski's theorem for $\mathbb{Q}_{p^n}$, see \cite{Milne,UCLAOstro}.

Here we list the notations for subsets of $\mathbb{Q}_p$ used in later sections. We denote the multiplicative group of the $p$-adic field by $\mathbb{Q}_p^\times \equiv \mathbb{Q}_p \backslash \{0\}$, the ring of integers of $\mathbb{Q}_p$ by $\mathbb{Z}_p \equiv \{x\in \mathbb{Q}_p : |x|_p \leq 1\}$, and the set of units in $\mathbb{Q}_p$ by $\mathbb{U}_p \subset \mathbb{Z}_p$ such that $\forall
x \in \mathbb{U}_p,|x|_p =1$.

\subsection{Bruhat-Tits tree}
\label{sec:BT Tree}
The Bruhat-Tits tree is an infinite tree structure built on equivalence classes of the $\mathbb{Q}_p^2$-lattice $\mathcal{L}$ 
which are spanned by two linearly independent vectors $u,v \in \mathbb{Q}_p^2$:
\begin{equation}
    \label{lattice}
    \mathcal{L} \equiv \left\{au+bv\in \mathbb{Q}_p^2|a,b\in \mathbb{Z}_p\right\}.
\end{equation}
The equivalence relation between the two $\mathbb{Q}_p^2$-lattices $\mathcal{L}$ and $\mathcal{L}'$ is defined as: $\mathcal{L}\thicksim \mathcal{L}'$ if $\mathcal{L} = c \mathcal{L}'$ for some $c\in \mathbb{Q}_p^\times$.

Based on these definitions, a Bruhat-Tits tree is then constructed by assigning each equivalence class of the $\mathbb{Q}_p^2$-lattice to one vertex on the tree. It is straightforward to see that by applying the $PGL\left(2,\mathbb{Q}_p\right)$ group actions on a lattice equivalence class in the following fashion 
\begin{equation}
    M:l=(u,v)\rightarrow (Mu,Mv),\quad M\in GL(2,\mathbb{Q}_p),
\end{equation}
we obtain another new equivalence class. Any subgroup which is conjugate to $PGL\left(2,\mathbb{Z}_p\right)$ will leave a lattice equivalence class invariant, so the Bruhat-Tits tree $T_p$ is identified with the coset $PGL\left(2,\mathbb{Q}_p\right)/PGL\left(2,\mathbb{Z}_p\right)$. 

On the tree we also need to clarify the meaning of an edge between two vertices. Therefore, a relation between two lattice equivalence classes $\mathcal{L}$ and $\mathcal{L}'$ is introduced as described in \cite{Zabrodin} and reviewed in the Appendix of \cite{padic}: they are called \textit{incident} if $p\mathcal{L}\subset \mathcal{L}'\subset \mathcal{L}$, and we connect them by an edge.

Using this incident relation to define edges on the Bruhat-Tits tree has two advantages. First, this relation is reflexive, so the Bruhat-Tits tree becomes unoriented, with exactly one edge between two adjacent vertices. Second, the action of $PGL\left(2,\mathbb{Q}_p\right)$ in the tree preserves the incident relation between any two lattice classes, leaving the number of edges between any two vertices invariant. If we use the edge number as a natural metric in the tree, then we see that $PGL\left(2,\mathbb{Q}_p\right)$ is its \textit{isometry group}. This fact is significant because, in usual AdS/CFT, the bulk isometry group is to be identified with the boundary conformal group. Indeed, the suitable conformal group for the tree boundary $\mathbb{P}^1(\mathbb{Q}_p)$ is the same $PGL\left(2,\mathbb{Q}_p\right)$, acting in a fractional linear fashion. Therefore, we consider the Bruhat-Tits tree as a natural candidate for the $p$-adic AdS bulk's asymptotics.\footnote{Iterative refinements on vertices of a Bruhat-Tits tree in the context of holography is proposed in Section 5.3 of \cite{Gubser1}, and is later extended in \cite{pAdS}.}

\begin{figure}[h]
    \centering
    \includegraphics[width = 8cm]{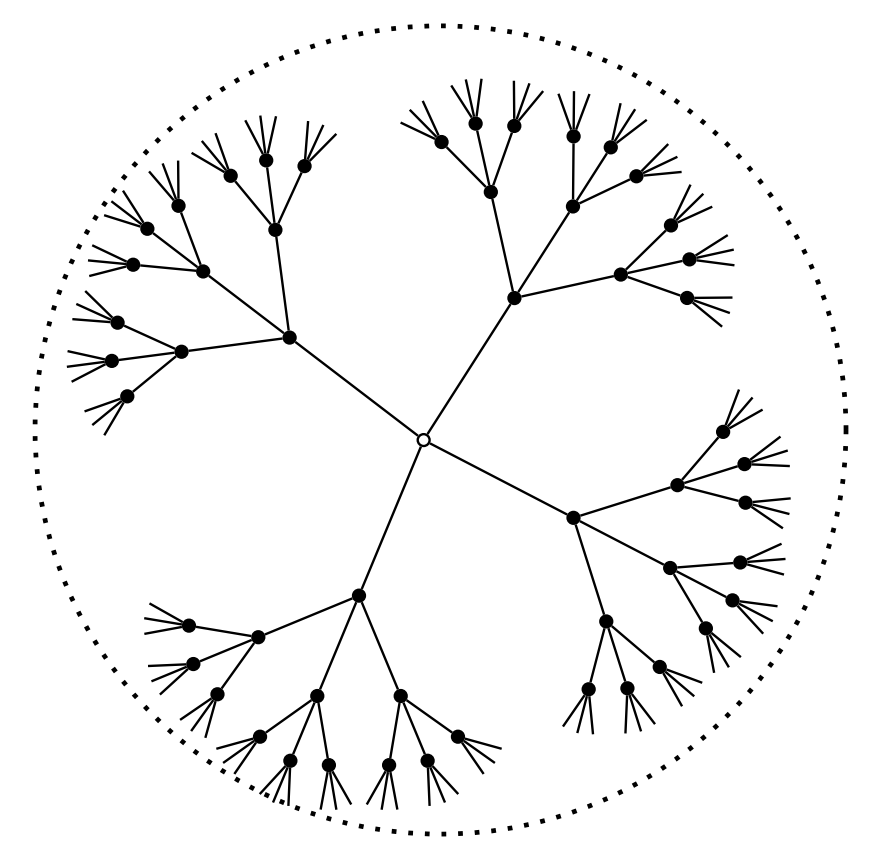}
    \caption{The Bruhat-Tits tree for the 3-adic numbers.}
    \label{fig:BTtree}
\end{figure}

Apart from the formal definition, a Bruhat-Tits tree is also visualized as Figure \ref{fig:BTtree}
in the representation as follows. From \cite{padic}, we know that incidentally to any lattice class $(u,v)$, there are always $p+1$ other lattice classes: $(pu,v)$ and $(u+nv,pv)$ where $n\in \mathbb{F}_p$ takes $p$ possible values, indicating that the Bruhat-Tits tree is homogeneous with valency $p+1$. 

Given the valency, there is a good way to translate the tree into $p$-adic numbers. Because any $p$-adic number has a unique expansion (\ref{expansion}), it is determined by a unique sequence of $(a_n)$, $a_n\in\mathbb{F}_p$. We assign coordinates $(z,z_0)$ on the Bruhat-Tits tree, where $z_0$ is the prime number $p$'s exponent, regarded as a level in the tree, and $z$ is a $p$-adic number up to $\mathcal{O}(z_0)$ precision. Therefore, each path on the Bruhat-Tits tree from $(z_0\rightarrow\infty)$ to the boundary $\mathbb{P}^1\left(\mathbb{Q}_p\right)$ located at $z_0 \rightarrow 0$ uniquely represents a $p$-adic number. This is graphically presented in terms of a ``trunk'' and ``branches'' in \cite{Gubser1}.

\subsection{An invitation to \texorpdfstring{$p$}{}-adic CFTs}

The majority of CFTs of our interests are ``one-dimensional'' ones; however, we will see that all higher-dimensional $p$-adic CFTs are very similar to ordinary 2d CFTs. We review Melzer's axioms \cite{melzer} on $p$-adic CFTs. They must have operator product expansion algebras (OPA) just like ordinary CFTs. The main difference between ordinary and $p$-adic CFTs is that local derivatives do not exist in the latter due to $\mathbb{Q}_p$ being totally disconnected.\footnote{By totally disconnected for the $p$-adic numbers, we mean that two open sets are totally disjoint. Whereas the Archimedean field $\mathbb{R}$ is a connected metric space.} More explicitly, this is seen by applying Leibniz's rule to $\mathbb{C}$-valued characteristic (or indicator) functions over $\mathbb{Q}_p$, all of which are locally constant \cite{melzer}. Finally, to make the OPA complete, all fields are primary (\ref{eq: primary}): 
\begin{equation}
\label{eq:primary}
    \phi_a^\prime\left(x^\prime\right)\left(dx^\prime\right)^\Delta = \phi_a(x) (dx)^\Delta,
\end{equation}
and the following OPE must exist
\begin{equation}
    \phi_{m}(x) \phi_{n}(y)=\sum_{a} C_{m n}^{a}(x, y) \phi_{a}(y)
\end{equation}
with $C^a_{m n}(x,y) \in \mathbb{R}$.

Here $\Delta$ is the conformal dimension, $dx$ is the Haar measure defined on $\mathbb{Q}_p$, and the transformation $x \rightarrow x^\prime\in \mathbb{P}^1\left(\mathbb{Q}_p\right)$ is a fractional linear one:
\begin{equation}
    x \rightarrow x^{\prime}=\frac{a x+b}{c x+d}, \quad \left(\begin{array}{ll}{a} & {b} \\ {c} & {d}\end{array}\right) \in
    GL\left(2,\mathbb{Q}_p\right),
\end{equation}
so the Haar measure and scalar field transform respectively as:
\begin{equation}
    d x \rightarrow d x^{\prime}=\left|\frac{a d-b c}{(c x+d)^{2}}\right|_p d x,
\end{equation} 
\begin{equation}
\label{eq: primary}
    \phi_a(x) \rightarrow \phi_a^\prime \bigg( \frac{ax + b}{cx + d}\bigg) =\left|\frac{a d-b c}{(c x+d)^{2}}\right|_p^{-\Delta} \phi_a(x).
\end{equation}
Since the bulk is a Bruhat-Tits tree and the boundary consists of $p$-adic numbers, evaluating correlators is more convenient than in the ordinary case. For instance, the general two- and three-point functions for local operators $\mathcal{O}_1, \mathcal{O}_2, \mathcal{O}_3, \dots$with different conformal dimensions $\Delta_1, \Delta_2, \Delta_3, \dots$respectively are of similar form to real CFTs' \cite{Gubser1}:
\begin{equation}
    \langle \mathcal{O}_1(z_1) \mathcal{O}_2(z_2) \rangle = \frac{C_{\mathcal{O}_1\mathcal{O}_2}}{|z_{12}|^{2\Delta_1}_p}, \quad \quad  \langle \mathcal{O}_1(z_1) \mathcal{O}_2(z_2) \mathcal{O}_3(z_3) \rangle = \frac{C_{\mathcal{O}_1\mathcal{O}_2\mathcal{O}_3}}{|z_{12}|^{\Delta_{12}}_p |z_{23}|^{\Delta_{23}}_p |z_{31}|^{\Delta_{31}}_p},
\end{equation}
up to contact terms. Here, the dependence $z_{ij} \equiv z_i - z_j$, $\Delta_{12} \equiv \Delta_1 + \Delta_2 - \Delta_3$, and the dependence $z_i$ is completely fixed by the invariance under fractional linear transformations. Ultrametricity constrains three- and four-point functions to be exact in cross-ratios in the $p$-adic norm, unlike the usual ones \cite{melzer, Gubser1}. The OPE coefficients form an associative algebra and primary operators can have \textit{arbitrary} dimensions, but the identity operator must have dimension 0.

Another property worth mentioning about $p$-adic CFTs is that they are automatically \textit{unitary} unlike their Archimedean counterparts. However, as opposed to representations of $sl(2,\mathbb{C})$ in the usual 2d CFTs, the $p$-adic global conformal group $PGL\left(2,\mathbb{Q}_p\right)$ lacks a Lie algebra, leading to the absence of a central charge or a good notion of state-operator correspondence.\footnote{Examples of ordinary 2d CFTs with $c=0$ include special classes of logarithmic CFTs, see, e.g., \cite{Gurarie,Ludwig}.} Despite lacking both local conformal algebra and descendants, we discuss in Section \ref{sec:5} on allowed group representations of a $p$-adic CFT.

\subsection{\texorpdfstring{$p$}{}-adic AdS/CFT and BTZ black hole}
\label{sec:BTZ}
In order to construct a $p$-adic version of the BTZ black hole, we first review the ordinary BTZ black hole, a classic black hole solution to the 3d Einstein equation \cite{BTZ}. A \textit{non-rotating} Euclidean BTZ black hole is described by the following complete Riemannian metric \cite{KOS}:
\begin{equation}
    \label{BTZmetric}
    ds^2 = \left(r^2-r_+^2\right)dt^2+\frac{1}{r^2-r_+^2}dr^2+r^2d\phi^2,
\end{equation}
where $r_+$ is the outer horizon radius, related to the ADM energy and central charge of the boundary 2d CFT by \cite{KrausMaloney}
\begin{equation}
    r_{+}=\sqrt{\frac{12 E}{c}-1}.
\end{equation}
Similarly, a $p$-adic BTZ black hole can also be formulated by solving classical equations of motion. In \cite{Edge}, Gubser et al. proposed to use edge length dynamics to formulate ``gravity'' (beyond linearized regime) on Bruhat-Tits trees, and even though large diffeomorphisms were seemingly not included there, this ``gravity'' does result in BTZ black holes with non-uniform lengths, incorporating topological changes by the 1-cycle. Their idea has been generalized to weighted graphs \cite{bai2020sum,huang2020bounds}.

However, to avoid technicalities above, we choose to review the $p$-adic BTZ black hole constructed instead by \textit{Schottky uniformization} as proposed in \cite{Caltech}, in which the black hole is a quotient of the Bruhat-Tits tree (analogue of the zero-temperature AdS$_3$), similar to the construction of a regular Euclidean BTZ black hole \cite{BTZasQuotient}.

In Euclidean AdS$_3$/CFT$_2$ at zero temperature, the bulk is identified with the hyperbolic space $\mathbb{H}^3$ and the boundary is the \textit{sphere at infinity} $S^2_{\infty}$, on which its conformal group is $PSL(2,\mathbb{C})$, same as the isometry group of $\mathbb{H}^3$. \textit{Schottky uniformization} provides us with a way to construct elliptic curves of higher genus on the conformal boundary. In this complex case, a genus-1 closed curve corresponds to $T^2$ torus and the solid torus bulk is topologically equivalent to the BTZ black hole. Generally for a genus-$n$ curve, \textit{Schottky uniformization} starts by picking a discrete $PSL(2,\mathbb{C})$ subgroup called \textit{Schottky group} $\Gamma$ with $n$ generators $\{\gamma_1,\cdots,\gamma_n\}$. Each $\gamma_i$ has fixed points in $S^2_{\infty}$, and the genus-$n$ curve is constructed as $S^2_{\infty}/\Gamma$ after removing those fixed points. The authors in \cite{Caltech, MMPS} extended this procedure to construct the $p$-adic BTZ black hole, which we will review and follow.

For a genus-1 boundary, $\Gamma\equiv q^{\mathbb{Z}}$ is generated by $q \in \mathbb{C}^{\times}$. Fixed points $0, \infty$ of the action by $q$ need to be removed from $\mathbb{P}^1(\mathbb{C})$ before taking the quotient. We define the \textit{domain of discontinuity} $A = \mathbb{P}^1(\mathbb{C})\setminus \{0,\infty\}$ and hence the quotient $C\equiv A/q^{\mathbb{Z}}$. Meanwhile, we also take the quotient of the bulk $\mathbb{H}^3$, and the total quotient space is $\mathbb{H}^3/q^{\mathbb{Z}}\cup C$, which is visualized as a solid torus. We should mention that the generator $\gamma$ can be written in terms of parameter $q = e^{2\pi i\tau }$, where $\tau\in \mathbb{C}$ is the torus' moduli.

In the BTZ black hole (\ref{BTZmetric}), $r_+$ is a solution-classifying parameter to be realized in \textit{Schottky uniformization}. Note that the \textit{Schottky group} $q^{\mathbb{Z}}$'s generator $\gamma$ can be written as \cite{Caltech,MaloneyWitten}:
\begin{equation}
    \label{eq:qmatrix}
    \left(\begin{array}{cc}{q^{\frac{1}{2}}} & {0} \\ {0} & {q^{-\frac{1}{2}}}\end{array}\right) \in PSL(2,\mathbb{C}).
\end{equation}
The \textit{Schottky parameter} $q$ is written in terms of horizon radius $q = e^{2\pi r_+}$ \cite{Caltech,MMPS}, so $r_+ = \frac{1}{2\pi} \log q$, proportional to the Bekenstein-Hawking entropy.

A torus $T^2$ is the same as a complex lattice $\mathbb{Z}+\tau\mathbb{Z}$, $\tau \in \mathbb{C}$, while in the $p$-adic case, this viewpoint is not true due to $p^{\infty} \rightarrow 0$ forcing many lattice equivalence classes to be 0. However, we could still select one \textit{Schottky group} $\Gamma$, a discrete subgroup of $PGL\left(2,\mathbb{Q}_p\right)$ to form genus-$n$ curves from $\mathbb{P}^1\left(\mathbb{Q}_p\right)$. The genus-one curve is the Tate uniformized elliptic curve $E_q = \mathbb{Q}_p^\times / q^\mathbb{Z}$ and genus-$n$ curve is the Mumford curve. We demonstrate the genus-one example by picking $\Gamma$ generated by $q \in \mathbb{Q}_p^{\times}$, so that
\begin{equation}
    \Gamma=\left\langle\begin{pmatrix}
    q & 0\\ 0 & 1
    \end{pmatrix}\right\rangle.
\end{equation}

\begin{figure}[h]
    \centering
    \includegraphics[width = 8 cm]{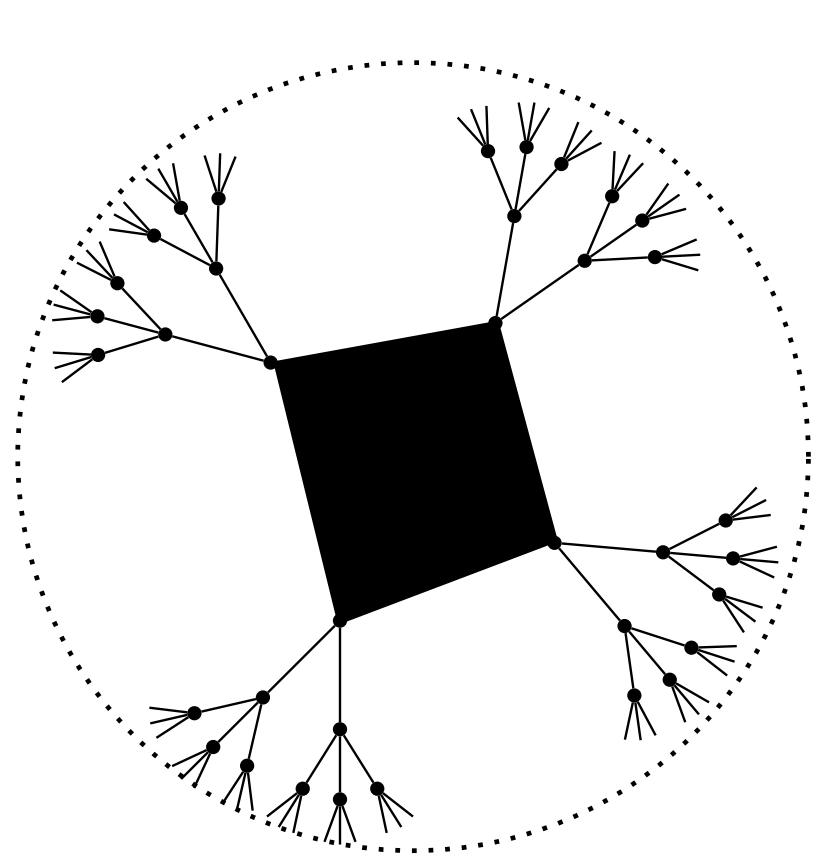}
    \caption{($l = 4$, $p=3$) BTZ black hole is at the center. The dotted lines represent the Bruhat-Tits tree structure repeating itself in a fractal fashion.}
    \label{fig:p-adicBTZ}
\end{figure}

Again we remove its fixed points, which are still $\{0,\infty\}$, from $\mathbb{P}^1\left(\mathbb{Q}_p\right)$, then the total space including bulk and boundary is $B = T_p \cup \left(\mathbb{P}^1\left(\mathbb{Q}_p\right)\setminus \{0,\infty\}\right)$, where $T_p$ is the Bruhat-Tits tree from Section \ref{sec:BT Tree}. The quotient $B/q^{\mathbb{Z}}$ is visualized as a graph with one regular polygon at the center. On each vertex of the polygon, a ``Bruhat-Tits'' inhomogeneous subtree is attached as seen in Figure \ref{fig:p-adicBTZ}. 

This graph could also be considered as a $p$-adic BTZ black hole, whose horizon area is the number of edges $l$ of the central polygon, with $l$ related to the Schottky parameter $q$ via $l = \log_p |q|_p$.\footnote{The $\log_p$ denotes the ordinary logarithm with base $p$, not the $p$-adic logarithm.} This also adds a restriction: $|q|_p >1$. In Sections \ref{sec:3} and \ref{sec:4}, we will use the above graph as the $p$-adic BTZ black hole and perform calculations on it.

\section{Path integrals for the isotropic sector}
\label{sec:3}
In this section, we try to calculate the partition function of the boundary $p$-adic CFT directly from the bulk by resorting to the Gubser-Klebanov-Polyakov-Witten (GKPW) dictionary. Recall for a boundary CFT local operator $\mathcal{O}$ \cite{GKP, WittenHolography} 
\begin{equation}
    Z_{\text{grav}}[\phi_{\partial}^i(x);\partial M]=\Bigg\langle\exp\left(-\sum_i\int_{\partial M} d^dx\phi^i_{\partial}(x)\mathcal{O}^i(x)\right) \Bigg \rangle_{\text{CFT on }\partial M},
\end{equation}
with the boundary condition on bulk scalar field $\phi^i(z,x)=z^{d-\Delta}\phi_{\partial}^i(x)+\left(\text{subleading}\right)$ as $z\rightarrow0$, where $z$ is the radial coordinate.

When we set field values $\phi_{\partial}^i$ on the conformal boundary to be zero, it is expected to calculate the CFT partition function, see, e.g., Eq.(72) in \cite{Gubser1}.

For simplicity, we restrict to the non-extended case, i.e., $q=p^1$, so the bulk path integral on a Bruhat-Tits tree $T_p$ is 
\begin{equation}
    Z_{\text{tree}}=\int\mathcal{D}\phi_a e^{-S_{\text{tree}}[\phi_a]},
\end{equation}
where the action $S_{\text{tree}}[\phi_a]$ is for massive scalar fields with sources on the tree, and the subscript ``$a$'' labels vertices. Naturally, this action is \cite{Gubser1}
\begin{equation}
\label{eq:action}
    S_{\text{tree}}[\phi_a] = \sum _ {\langle ab \rangle} \frac{1}{2} \left(\phi_a - \phi_b\right)^2 + \sum_a \bigg( \frac{1}{2} m_p^2 \phi_a^2 - J_a \phi_a \bigg)
\end{equation}
with $a$ and $b$ labeling the tree's vertices and $\sum_{\langle ab \rangle}$ refers to summing over adjacent vertices on the tree, and $J_a$ is a source.

As expected, the linearized equations of motion for a scalar field $\phi_a$ are
\begin{equation}
\label{eq:eom}
    \left(\square + m_p^2\right) \phi_a = J_a,
\end{equation}
but with a modification to the regular Laplacian. The modification is that the Laplacian here is the lattice/graph Laplacian\footnote{Connection Laplacian \cite{Gubser1} and Hodge Laplacian \cite{Edge,Caltech} are proved to be equivalent on Bruhat-Tits tree.} and is defined as a positive definite operator
\begin{equation}
\label{eq: discrete}
    \square \phi_a \equiv \sum_{\substack{\langle ab \rangle}} (\phi_a - \phi_b).
\end{equation}
With this Laplacian at our disposal, the desired partition function is easily calculable via
\begin{equation}
    Z_{\phi} = \frac{1}{\sqrt{\text{det}'\left(\square+m_p^2 \right)}},
\end{equation} 
where the superscript $'$ means omitting zero modes, which is absent as we will see later.

Another way to obtain the partition function is through the use of a tensor network formulation for $p$-adic AdS/CFT by \cite{Ling-Yan}. The authors put a tensor network on the Bruhat-Tits tree, similar to \cite{Caltech} but different from the dual graph in \cite{MMPS}. Then by making analogies with ordinary \textit{diagonal} CFTs\footnote{\textit{``Diagonal''} means that torus partition functions are diagonal invariants, such as Liouville theory and $(A,A)$-series minimal models, e.g., Ising model. Non-diagonal CFTs are the majority, and include logarithmic CFTs, $\widehat{su}(2)$ WZW models in $D$ and $E$ series, and $(A,D)$-, $(A,D)$-, $(A,E)$- and $(E,A)$-seires Virasoro minimal models, where $(A_4,D_4)$, i.e., the 3-state Potts model being the simplest one.}, their proposed ``torus'' partition function is\footnote{To be precise, it is a genus-1 Tate curve on the boundary of the Bruhat-Tits tree.}:
\begin{equation}
    \sum_a|q|^{\Delta_a}.
\end{equation}
Here $a$ labels all primary fields, and $\Delta_a$'s correspond to arbitrary scaling dimensions according to Melzer's axioms, and are compatible with the associative operator product algebra. This expression seems to be encompassing, but not explicit. %Conspicuously, multiplicities here are all one, which is not the case for ordinary non-diagonal 2d CFTs.

In the rest of this paper with our path integral approach, we restrict to the ``isotropic'' sector\footnote{This means that $\phi$ only has a nontrivial radial profile.} of the full partition function \eqref{eq: discrete}, and treat the anisotropic sector on another occasion \cite{spectrum}. The motivations for focusing on this sector include:
\begin{itemize}
    \item First of all, the Bruhat-Tits tree enjoys translational and rotational invariance under the global action of $PGL(2,\mathbb{Q}_p)$, in spite of the center which is chosen artificially.
    \item As we can see from Appendix \ref{sec:spectrumBTTree}, the multiplicity of anisotropic modes is overwhelmingly larger than that of isotropic ones, so it is much less cumbersome to deal with isotropic modes.
    \item The above simplification can be justified by drawing analogy between the $p$-adic CFT here and ordinary $2$d CFTs. In the latter case, it is often rewarding to study just the scalar or spinless sector \cite{Afkhami-Jeddi:2019zci,Hartman:2019pcd}. In $p$-adic CFT, although there is no right- or left-movers (holomorphic or anti-holomorphic modes), there is a precise counterpart to the spinless states defined by $L_0=\bar{L}_0$ in $2$d CFTs, which is the singlet part under the global rotation grounp $PGL(2,\mathbb{Q}_p)$. Furthermore, as shown in examples such as the Narain CFT, the so-called ``primary counting partition function'' has a convenient spectral decomposition property \cite{Benjamin:2021ygh}, and each sector of the full partition function contains valuable information about the full system.
    \item In Section \ref{sec:5} we will make preliminary connections between the forthcoming computations and the representation theory of $PGL(2,\mathbb{Q}_p)$, so as a beginning step, it is more natural to consider the sector which preserves the isometry of the spacetime. Anisotropic sectors would correspond to inserting a nontrivial $PGL(2,\mathbb{Q}_p)$ element into the trace \eqref{eq:sign}.
\end{itemize}

A caveat is that our calculations are only for bulk scalar fields and not for the real gravitational contributions to the presumably full bulk path integral.\footnote{Attempts at formulating gravity on Bruhat-Tits trees include \cite{Edge}, but our techniques do not apply to calculating gravitational partition functions there.} In the following three subsections, we first turn off the mass $m_p^2$, and then turn it back on near the end of this section.

\subsection{Laplace problem on Bruhat-Tits trees}
\label{sec:Laptree}
As promised, in this subsection and the next, we study massless scalars, which are dual to boundary marginal operators in the usual AdS/CFT context \cite{WittenHolography,Caltech}. For now we turn off the source $J$ in \eqref{eq:eom}; we will deal with $J\neq0$ in Section \ref{sec:4.2}.

We first define a few concepts on the Bruhat-Tits tree to be used in later sections. On this homogeneous tree, one can arbitrarily pick the central point and assign any vertex with ``depth $n$,'' the number of edges going outwards from the center to that vertex, and the center has depth 0.

When we talk about scalar fields on the Bruhat-Tits tree, we refer to a real-valued scalar function globally defined on each vertex of the tree. The spectrum has been considered in the literature to some extent, for example in \cite{Lubotzky}, and here we solve the problem in more different settings.

Now in the spectrum we find all {\it isotropic} modes, i.e., those which lack angular profiles, as follows: one starts from the conformal boundary placed at a fictitious finite radial cut-off, which will later be taken to infinity, with the Dirichlet boundary condition $\phi|_{\partial T_p}\equiv\phi_N=0$, then $p$ of them connect to one inner point with value $\phi_{N-1}$. This point connects to a point further inwards with field value $\phi_{N-2}$. Following the definition of Laplacian (\ref{eq: discrete}) and denoting the eigenvalue of the function $\phi_i,i=1,\dots,N$ as $\lambda$, there is a local recursion relation around the valency-$(p+1)$ vertex for the sourceless case $J=0$:
\begin{equation}
\label{eq:isotropic}
    p(\phi_{N-1}-0)+(\phi_{N-1}-\phi_{N-2})=\lambda\phi_{N-1},
\end{equation}
implying $\phi_{N-2}=(p+1-\lambda)\phi_{N-1}$. Now at the depth $n=N-1$, for another point connecting to the inner point with value $\phi_{N-2}$, we suppose it has another value $\tilde{\phi}_{N-1}\neq\phi_{N-1}$. This value must satisfy the same relation (\ref{eq:isotropic}) with a fixed $\phi_{N-2}$. Thus, we have $\tilde{\phi}_{N-1}=\phi_{N-1}$. However, this kind of argument fails to arrive at the isotropy at smaller depths, and this is where we choose to restrict to the isotropic sector, where field values at the same depth $n$ are equal, denoted as $\phi_n$. The full spectrum including all other anisotropic modes are presented in Appendix \ref{sec:spectrumBTTree}.

The recursion relation starting from $n=2$ for isotropic $\phi_n$ now reads
\begin{equation}
\label{eq:recursion}
    p(\phi_{n-1}-\phi_n)+(\phi_{n-1}-\phi_{n-2})=\lambda \phi_{n-1},\quad 2\leq n\leq N-1
\end{equation}
whose characteristic equation has two roots:
\begin{equation}
\label{eq:roots}
    \alpha_\pm=\frac{1+p-\lambda\pm\sqrt{(\lambda-p-1)^2-4p}}{2p}.
\end{equation}

Field value at depth $n$ equals the general solution to the linear recurrence \eqref{eq:recursion}
\begin{equation}
\label{eq:phitree}
    \phi_n=c_+\alpha_+^n+c_-\alpha_-^n,
\end{equation}
and we solve for coefficients $c_{\pm}$ with two initial conditions at depths 1 and 2:
\begin{equation}
\label{eq:initial}
    \phi_1=\left(1-\frac{\lambda}{p+1}\right)\phi_0,\quad\phi_2=\frac{p+1-\lambda}{p}\phi_1-\frac{\phi_0}{p}=\left(1-\frac{2\lambda}{p}+\frac{\lambda^2}{p+p^2}\right)\phi_0,
\end{equation}
where $\phi_0$ at the center is not fixed.
The coefficients are 
\begin{equation}
    c_{\pm}=\left[\frac{1}{2}\pm\frac{p^2-1-\lambda p+\lambda}{2(p+1)\sqrt{(p+1-\lambda)^2-4p}}\right]\phi_0.
\end{equation}

Now we treat \eqref{eq:phitree} as an degree-$n$ polynomial equation in $\lambda$. Numerically we see that, somewhat surprisingly, all roots of the equations for any $n$ and $p$ (primes and non-primes alike) are real. And in particular, when $n$ is odd, there is one universal root $\lambda=p+1$. Also, the constant term in the polynomial $\phi_n(p,\lambda,\phi_0)$ is always $\phi_0$, while the coefficient of the highest-degree term is always $(-1)^N\phi_0/\left(p^N+p^{N-1}\right)$. Then by applying the Vieta's formula to $\phi_N=0$, the product of all roots of the degree-$N$ polynomial $\phi_N(\lambda)$ is 
\begin{equation}
\label{eq:coincide}
p^N+p^{N-1}
\end{equation}
which is in fact insensitive to the exact boundary value of $\phi_N$.

The full spectrum of the graph Laplacian on the Bruhat-Tits tree requires further explanation as we will only consider a subset of the full spectrum when the eigenvectors are isotropic (i.e., the field value on the vertices with the same distance to the center share the same field value) and omit the anistropic ones for simplicity in this section. We refer the reader to these details in Appendix \ref{sec:spectrumBTTree}. We are motivated by the isotropic case due to having translational and rotational invariance on the tree.

Since $-\log\text{det}\left(\Box\right)$ is divergent in radius $\sim N$, in principle we are supposed to regularize it by local counterterms. We notice that the number of boundary points is also $p^N+p^{N-1}$, which dominates the number of points in the bulk for large $N$:
\begin{equation}
\label{eq:volume}
    \frac{(p+1)p^N-2}{p-1}\xrightarrow{N\rightarrow\infty}\frac{p}{p-1}\left(p^N+p^{N-1}\right).
\end{equation}
Giving this observation, let us first recall that in the usual AdS$_3$/CFT$_2$, there are several places where various divergences appear. First, in the one-loop determinant of $\Box+m^2$ for a massive scalar on $\mathbb{H}^3$ \cite{Xi}, 
\begin{equation}
\frac{1}{2}\text{Vol}\left(\mathbb{H}^3\right)\int\frac{dt}{t}\frac{e^{-(m^2+1)t}}{(4\pi t)^{3/2}},
\end{equation}
there are $1/t$ UV divergence and Vol$\left(\mathbb{H}^3\right)$ IR divergence, both removable by local counterterms. In another context, for the on-shell Einstein-Hilbert action with constant metric:
\begin{equation}
    \frac{1}{16\pi G}\int d^3x\sqrt{g}\left(R-2\Lambda\right)=\frac{V}{4\pi Gl^2},
\end{equation}
where the cosmological constant $\Lambda=-1/l^2$ with $l$ being the AdS$_3$ radius, and $V$ is the spacetime volume, one can introduce a height cutoff $\epsilon$ in the upper-half space model. Then the regularized volume becomes \cite{Krasnov}:
\begin{equation}
    V_{\epsilon}(r)=\pi l^3\left(\frac{r^2}{2\epsilon^2}-\frac{1}{2}-\ln\frac{r}{\epsilon}\right),
\end{equation}
where the first boundary-area divergence can be removed by adding a boundary term local in boundary metric, and the second logarithmic divergence can be removed by a local counterterm as well.

In our case, the situation is different from the usual cases, since our boundary area appears in $e^S$ instead of the action $S$. The naive speculation is that the volume (i.e., number of vertices) on a Bruhat-Tits tree grow exponentially instead of power-law. By mimicking the removal of boundary-area divergence in ordinary AdS$_3$ above, we propose the partition function:
\begin{equation}
\label{eq:treepart}
    \boxed{Z_{\text{tree}}=\left(\frac{p}{p-1}\right)^{1/2}}.
\end{equation}

\begin{figure}[h!]
    \centering
    \begin{subfigure}[t]{.5\textwidth}
    \centering
    \includegraphics[height=2in]{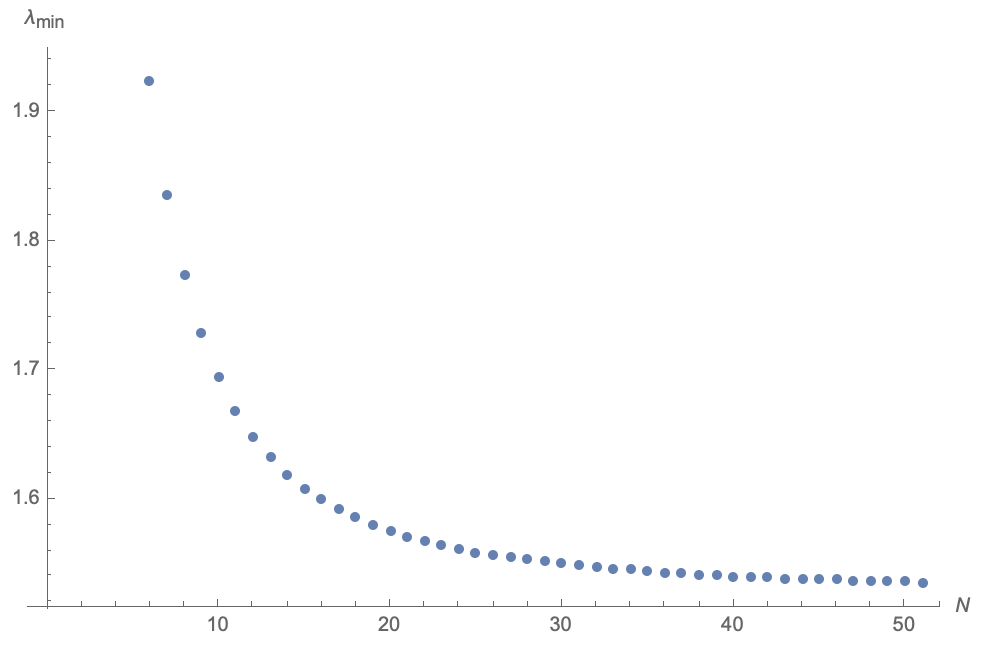}
    \caption{Convergence of the smallest eigenvalues.}
    \end{subfigure}%
    ~
    \begin{subfigure}[t]{.5\textwidth}
    \centering
    \includegraphics[height=2in]{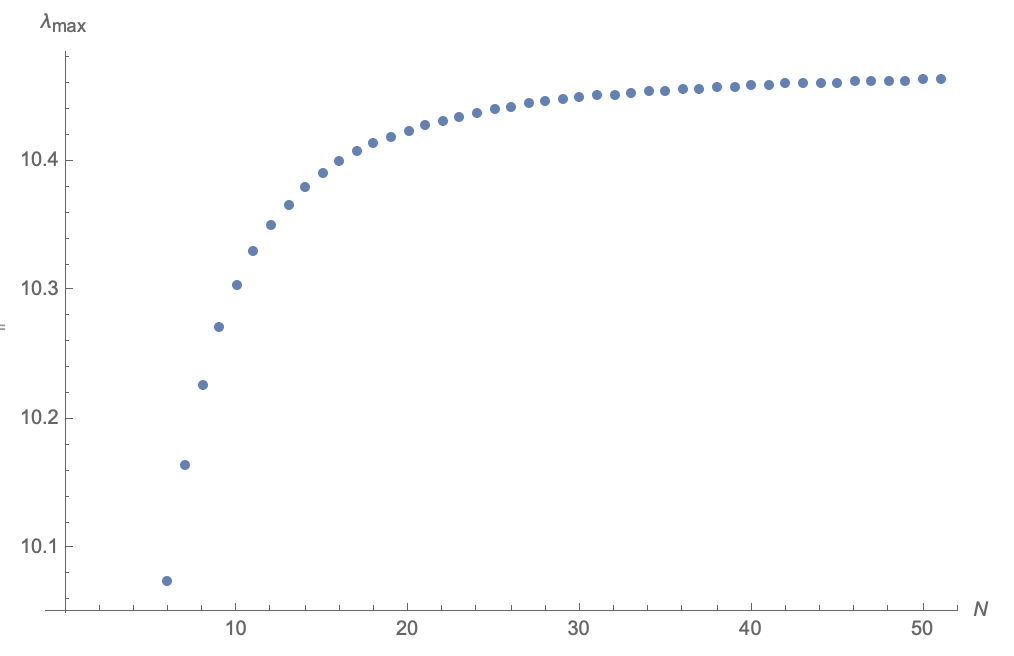}  
    \caption{Convergence of the largest eigenvalues.}
    \end{subfigure}
    \caption{Numerical bounds on the smallest and the largest eigenvalues via \textit{Mathematica}'s \texttt{NSolve}, as the fictitious boundary cutoff $N$ increases up to 51. They agree with results from Newton's method.}
    \label{fig:nsolve}
\end{figure}

We then investigate the behavior of the smallest and the largest eigenvalues of the Laplacian $\square$ as $N\rightarrow\infty$ at a fixed $p$. We used Newton's method to find the upper bound on $\lambda_1$ and the lower bound on $\lambda_N$, and they seem to converge numerically; although intermediate eigenvalues do not converge, which is natural since the amount of them increases as $N$ increases. For example, see Figure \ref{fig:nsolve} when $p=5$ and $N=3,\dots,51$ for their convergence. By Newton's method, we obtain the lower bound $\sim$1.52786 after 8036 iterations, and the upper bound $\sim$10.4721 after 474 iterations.

Now we seek to find the eigenfunctions on Bruhat-Tits trees. Unlike discrete Laplacians on a multidimensional regular rectangular grid with Dirichlet boundary conditions, the universal solutions to the second-order linear recurrence cannot be expressed in terms of a linear combination of Chebyshev polynomials of the first and second kinds due to the nontrivial topology of Bruhat-Tits trees. The first expression in (\ref{eq:initial}) contains a constant term $\phi_0$, so there is no inner product over a finite real interval $[-a,a]$ which makes $\phi_1$ and $\phi_2$ orthogonal to each other. Another way to see this impossibility is that there is a linear term in $\lambda$ for the second expression in (\ref{eq:initial}).

\begin{figure}[h!]
\centering
    \begin{subfigure}[t]{.5\textwidth}
    \centering
    \includegraphics[height=2in]{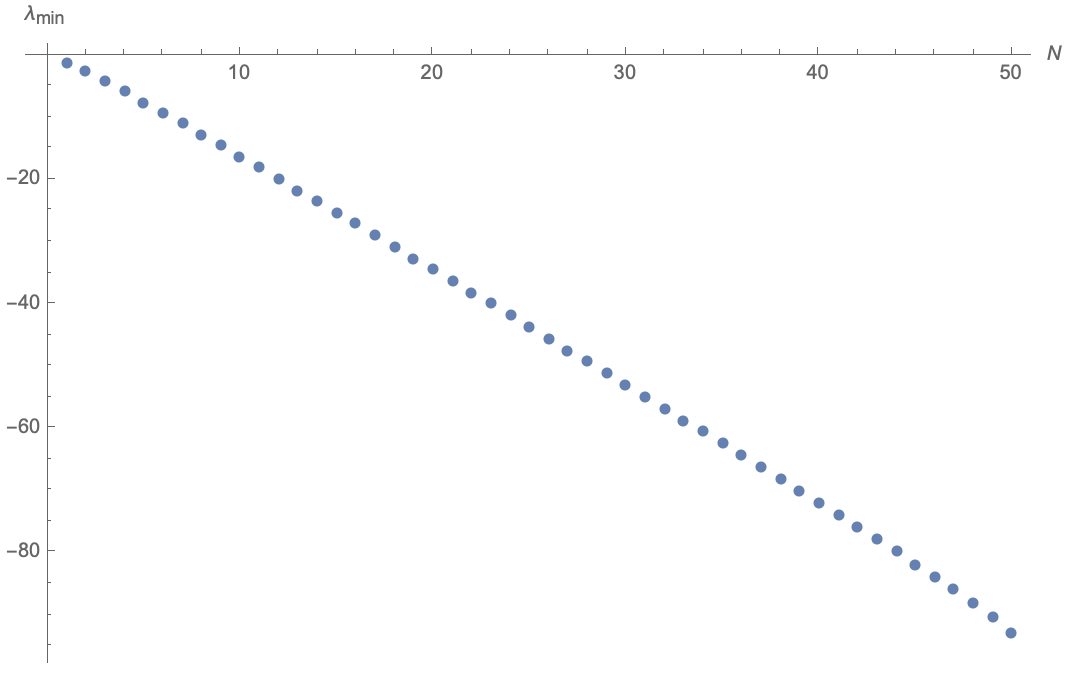}
    \caption{Asymptotics at the smallest eigenvalue.}
    \end{subfigure}%
\centering
    \begin{subfigure}[t]{.5\textwidth}
    \centering
    \includegraphics[height=2in]{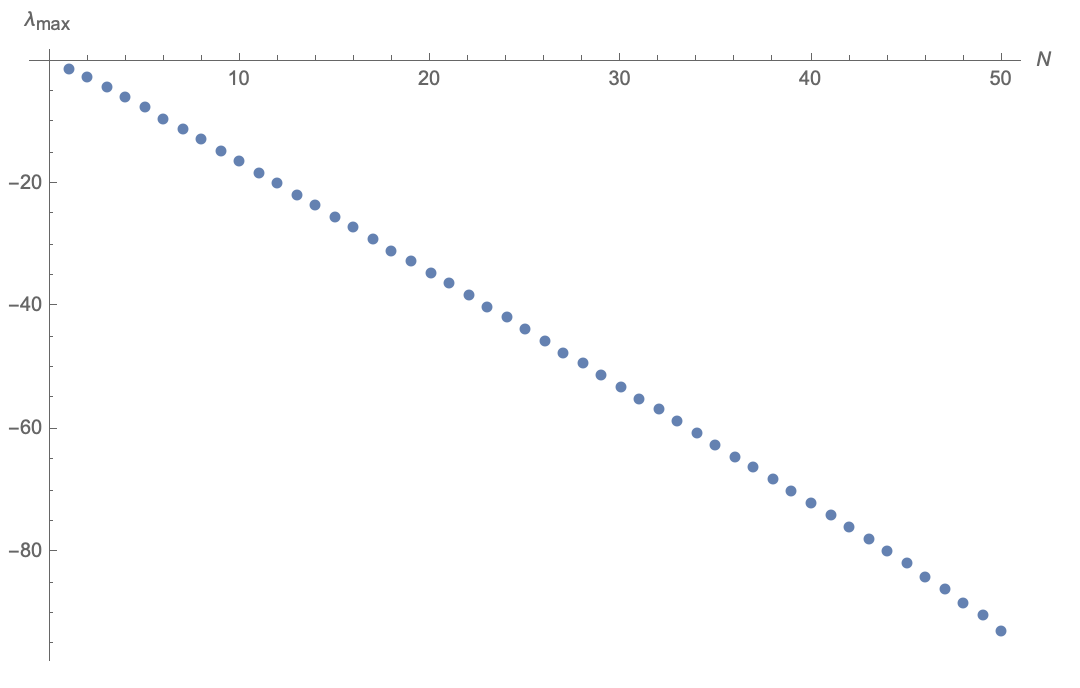}
    \caption{Asymptotics at the largest eigenvalue.}
    \end{subfigure}%\vspace{10pt}
 
\centering
    \begin{subfigure}[t]{.5\textwidth}
    \centering
    \includegraphics[height=2in]{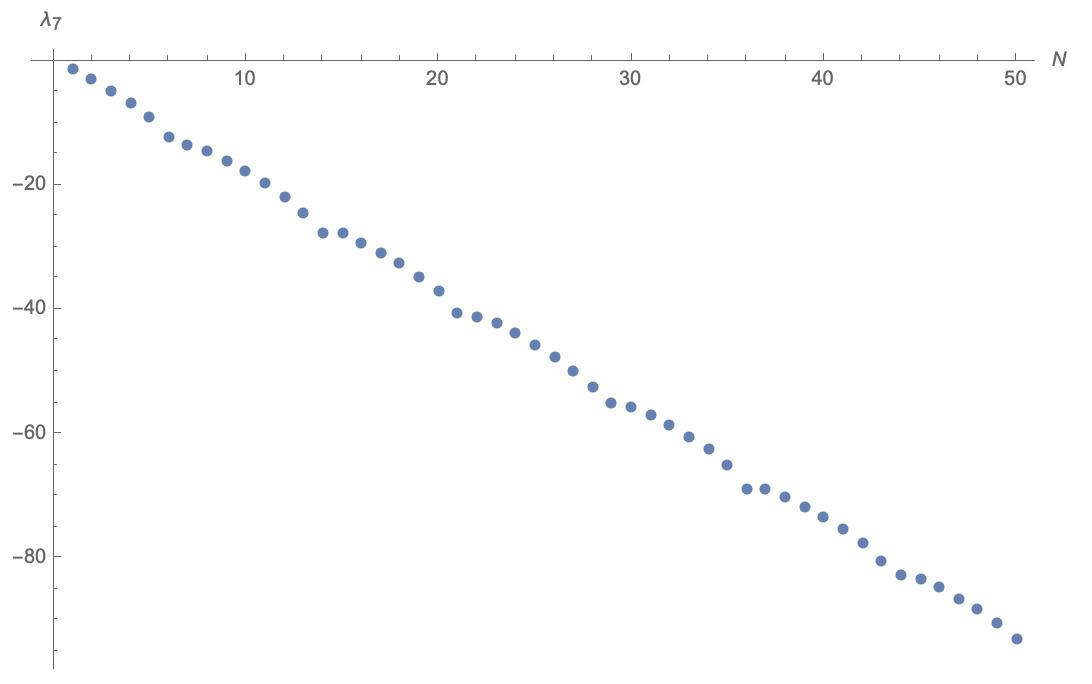}
    \caption{Asymptotics  at the 7$^{\text{th}}$ largest eigenvalue.}
    \end{subfigure}%
\centering
    \begin{subfigure}[t]{.5\textwidth}
    \centering
    \includegraphics[height=2in]{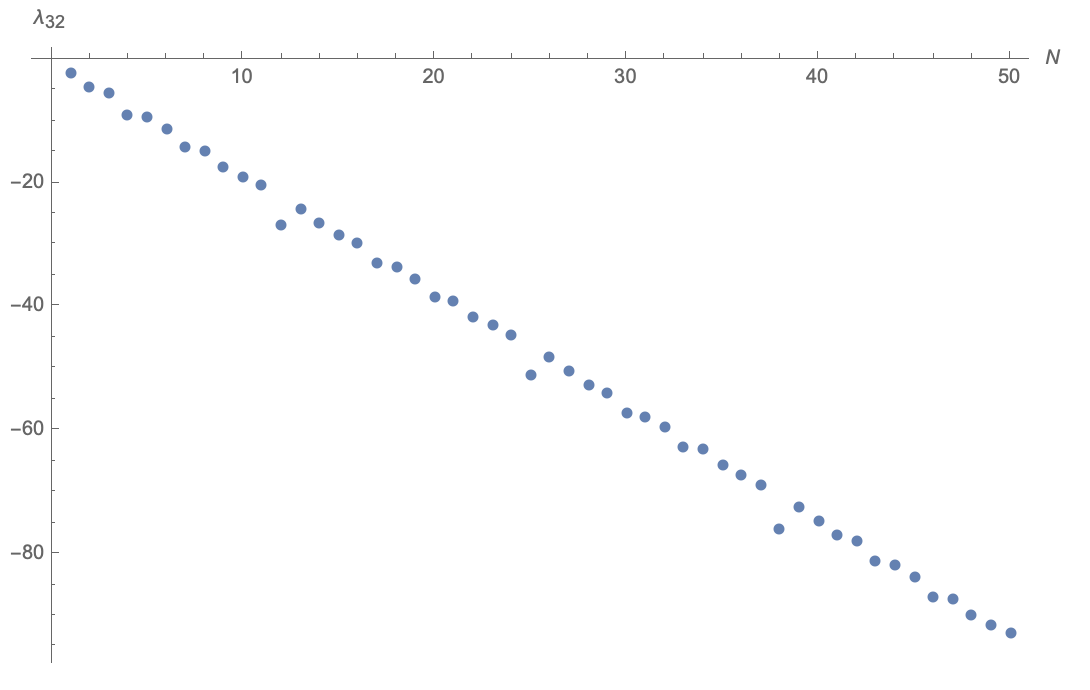}
    \caption{Asymptotics at the 32$^{\text{th}}$ largest eigenvalue.}
    \end{subfigure}%
\caption{Asymptotics of $\text{Re}\left[\log\left(\phi_n/\phi_0\right)\right]$ evaluated at different eigenvalues as the cutoff $N$ increases, with $p=41$.}
    \label{fig:log}
\end{figure}

Numerically, we observe that the decay of the field value is almost exponential, but faster than the asymptotically decay $\sim z^{-1/2}$ of Bessel functions of the first and second kinds $J_{\alpha}(z)$ and $Y_{\alpha}(z)$. In Figure \ref{fig:log}, we plot the real part\footnote{The field value $\phi_n$ can be negative at many different depths $n$.} of $\log\left(\phi_n/\phi_0\right), n=1,\dots,51, N=51$. The large but finite negative value is an artifact that we can only compute for finite $N$; ideally we should get $\log 0$. Notice that although their semi-log plots look almost the same, at least to the naked eye, if one plots their face values, they look quite different and consistent with the approximate (asymptotic) orthogonality. 

\begin{figure}[h]
    \centering
    \begin{subfigure}[t]{.5\textwidth}
    \centering
    \includegraphics[height=1.91in]{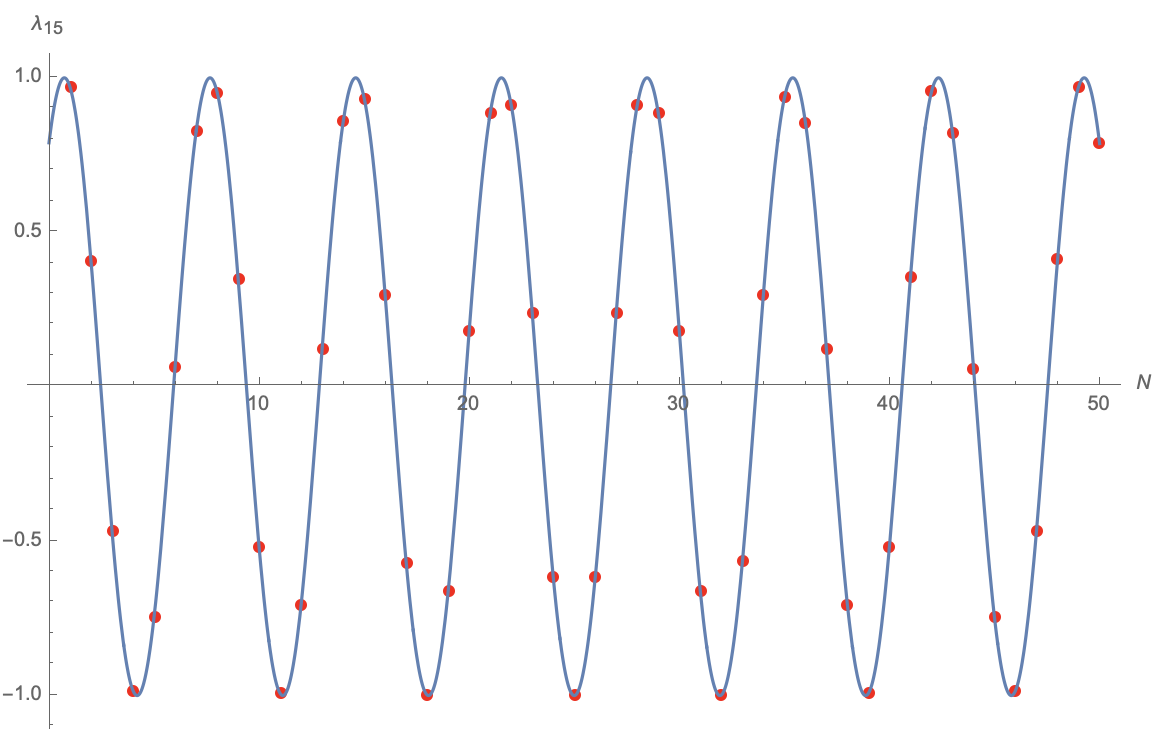}
    \caption{Oscillation of $\phi_n/\phi_0$ at the $15^{\text{th}}$ largest eigenvalue for $p=239$.}
    \end{subfigure}%
    ~
    \begin{subfigure}[t]{.5\textwidth}
    \centering
    \includegraphics[height=1.91in]{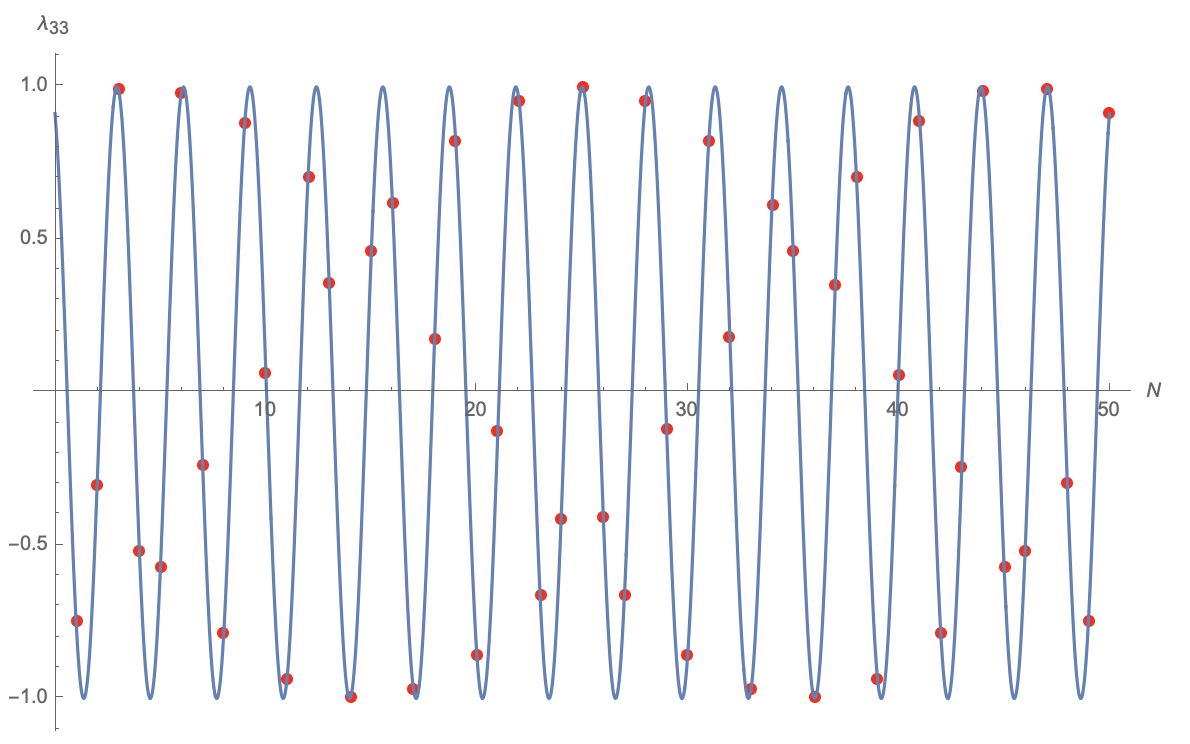}
    \caption{Oscillation of $\phi_n/\phi_0$ at the $33^{\text{th}}$ largest eigenvalue for $p=239$.}
    \end{subfigure}
    \caption{Oscillations of eigenvalues over the cutoff $N$, where red dots are data points from \textit{Mathematica}'s \texttt{NSolve}, and blue sinusoidal curves with phase shifts are fittings with frequencies $n\frac{i-1}{N-1}\pi$ for the $\phi_n/\phi_0$ at the $i^{\text{th}}$ largest eigenvalue, $n=1,\dots,N-1$, $i=1,\dots,N$.}
    \label{fig:oscillation}
\end{figure}

On the other hand, within the exponentially decaying envelope, $\phi_n$ oscillates discretely around zero as $n$ increases. This oscillatory behavior is shown in Figure \ref{fig:oscillation} after the exponential envelope is removed.

\begin{figure}[h]
    \centering
    \includegraphics[width = 8cm]{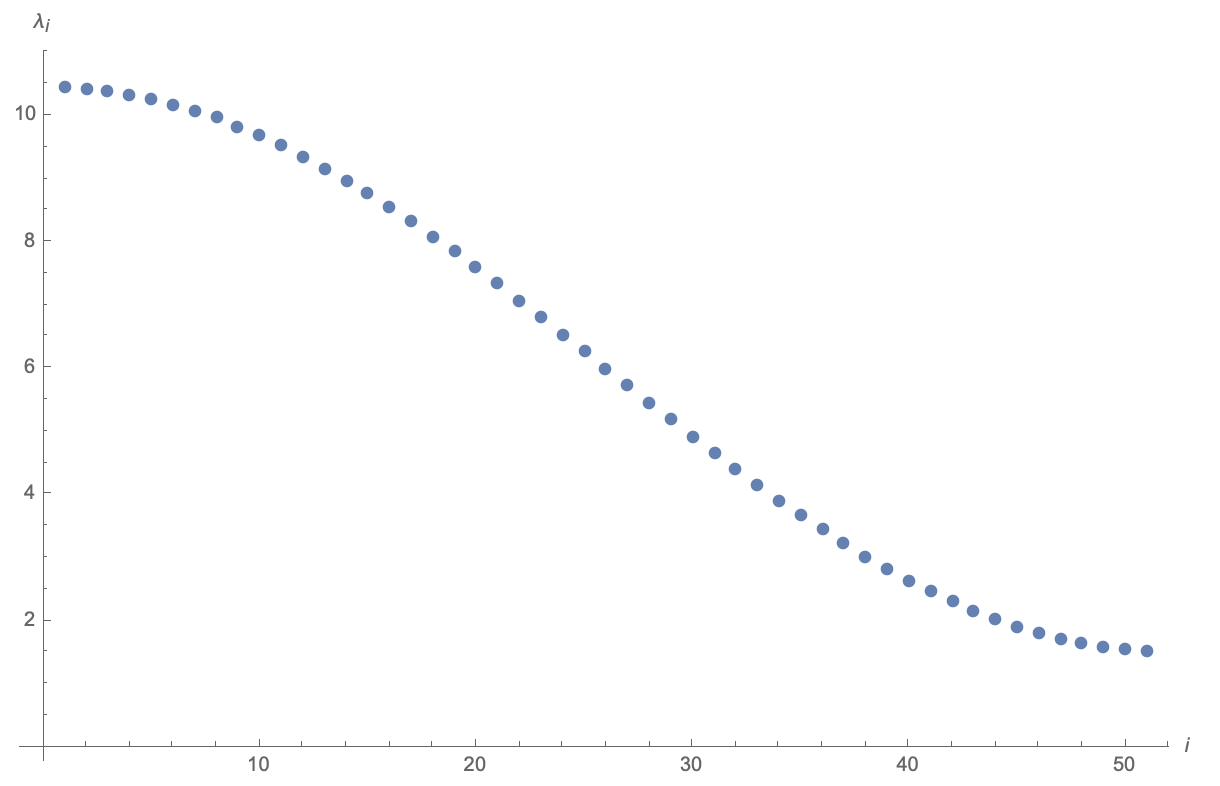}
    \caption{The spectrum $\{\lambda_i\}$ of Laplacian $\Box$ when the cutoff is $N=51$, ordered from the largest to the smallest, agreeing with (\ref{eq:eigenvalue}) with $k=1$. The horizontal axis is $1\leq i\leq N$, not cutoff $N$ or depth $n$.}
    \label{fig:profile}
\end{figure}

Numerically, for a radial cutoff at depth $N$, we propose the following ansatz:
\begin{equation}
\label{eq:ansatz}
    \boxed{\phi_{n,i}=p^{-n/2}\cos{\left(kn\frac{i-1}{N-1}\pi+\psi\right)}\phi_{0,i}},
\end{equation}
where $1\leq i\leq N$ labels $N$ eigenvalues, $n$ is the depth, and $k$ and $\psi$ are to be determined. After plugging this ansatz for $\phi_{n, i}$ into the recurrence relation (\ref{eq:recursion}), we obtain:
\begin{equation}
\begin{split}
    0=&p^{1/2}\sin\left(kn\frac{i-1}{N-1}\pi+\psi\right)+(\lambda_i-p-1)\sin\left(k\frac{(i-1)(n-1)}{N-1}\pi+\psi\right)\\&+p^{1/2}\sin\left(k\frac{(i-1)(n-2)}{N-1}\pi+\psi\right)\\
    =&\sin\left(k\frac{(i-1)(n-1)}{N}\pi+\psi\right)\left[2p^{1/2}\cos\left(k\frac{i-1}{N-1}\pi\right)+(\lambda_i-p-1)\right].
\end{split}
\end{equation}
The eigenvalues are asymptotically
\begin{equation}
\label{eq:eigenvalue}
    \boxed{\lambda_i=p+1-2p^{1/2}\cos\left(k\frac{i-1}{N-1}\pi\right)}.
\end{equation}
The integer $k$ in the frequency in (\ref{eq:ansatz}) can freely vary \emph{ab initio}, but by simply plotting the spectrum $\{\lambda_i\}$ agaisnt $i$ at a fixed $N$, we can see that the profile is monotonically decreasing as in Figure \ref{fig:profile}. Hence $k$ is fixed to be 1. The validity of this frequency is numerically tested up to $p=2477$ (larger $p$'s do not increase computational complexity significantly). However, the phase shift $\psi$ in (\ref{eq:ansatz}) has to be determined numerically and is conveniently unimportant for us.

The eigenvalues (\ref{eq:eigenvalue}) are exact only if they correspond to $\phi_{n,i}$ in (\ref{eq:ansatz}) at large depth $n$ (i.e., far away from the initial condition (\ref{eq:initial}) at the center) and $N\rightarrow\infty$. For $p=5$, we see that the largest and the smallest eigenvalues are asymptotically $6\pm2\sqrt{5}$. These are consistent with results from Newton's method as well as Vieta's formula in the sense that the summation of the eigenvalues (\ref{eq:eigenvalue}) is exactly $(p+1)N$. Additionally, all the eigenvalues are confined within an interval $\left[-2\sqrt{p},2\sqrt{p}\right]$.\footnote{Similarly looking bounds on eigenvalues in the context of principal series representation of $GL\left(2,\mathbb{Q}_p\right)$ without boundary conditions on a Bruhat-Tits tree were obtained in \cite{Lubotzky} (Theorem 5.4.2).}

Overall, this is a different spectral decomposition of Laplacian on the Bruhat-Tits tree from the plane-wave basis \cite{Zabrodin, Lubotzky, Caltech}, in that eigenfunctions here may oscillate around zero. We call it the ``evanescent wave'' basis. Also a key feature of discrete Laplacian here on trees is that solutions to the Laplace equation averaged over the circular boundary $\mathbb{P}^1\left(\mathbb{Q}_p\right)$ is not equal to the value at the center, as opposed to the continuous Laplacian.

Finally, it is a trivial exercise to change the valence to $p^n+1$ in the recurrence (\ref{eq:recursion}) and repeat everything above if one wants to study the scalar on $T_{p^n}$ that models AdS$_{n+1}$.

\subsection{Laplace problem on BTZ graphs}
\label{subsec:BTZ}
We now turn to studying the Laplace problem for BTZ black holes. Conceptually, to calculate the determinant of Laplacian $\Box$, we are not able to use its heat kernel as did in \cite{Xi} for continuous AdS$_3$, because the BTZ graph is essentially a constant-time slice \cite{Caltech,MMPS}, and there is no good notion of ``time.''

Compared to the Bruhat-Tits tree, in terms of the defining recursion relations for $\Box$, the only modifications on the linear recurrence for the scalar field $\phi$ on a BTZ graph are the initial conditions on $\phi_1$ in terms of $\phi_0$, as explained below.

In particular, the key difference between a $p$-adic BTZ black hole and the Bruhat-Tits tree is that the field values on the event horizon (depth 0) could be different. Given the horizon's area $l$, the field values are denoted as $\phi_{0,0},\phi_{0,1},\dots,\phi_{0,s},\dots,\phi_{0,l-1}$, where a specific $s$ labels a horizon vertex as well as the entire subtree rooted at that vertex, as shown in Figure \ref{fig:p-adicBTZ}.

\begin{figure}[h]
    \centering
    \includegraphics[width = 6cm]{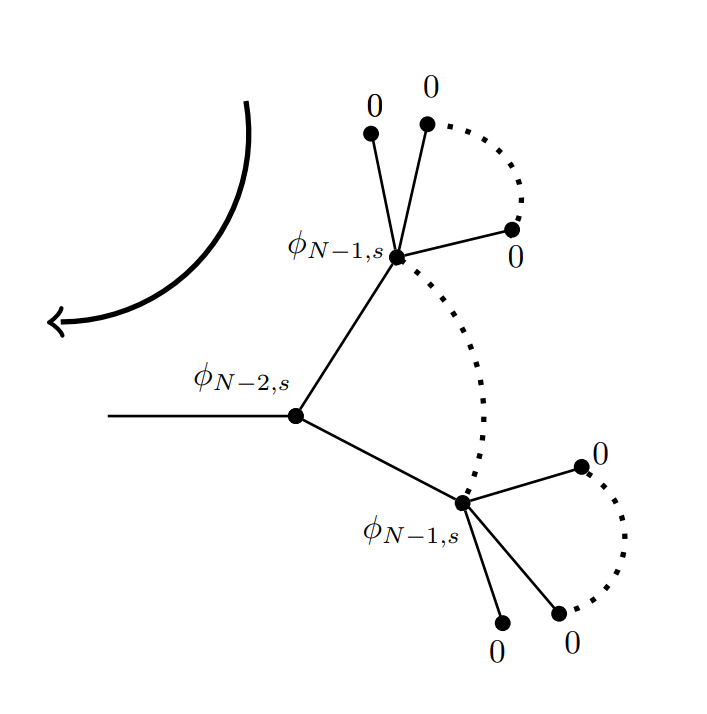}
    \caption{Going from the boundary toward the center, with the initial condition (\ref{eq:bdryinitial}).}
    \label{fig:bdry}
\end{figure}

Now, as shown in Figure \ref{fig:bdry}, we travel inwards from the boundary (at depth $N$) where all the fields vanish, and denote the field value on the layer next to the boundary as $\phi_{N-1}$. Without loss of generality, we start from the subtree rooted at vertex $s$ on the horizon. In the following, we will focus on the {\it partially isotropic} sector of the spectrum, namely all field values at the same depth within the same subtree are equal [in the same manner as discussed right below \eqref{eq:isotropic}], but different subtrees can have nontrivial relative overall scalings on their field values, and nothing more.\footnote{Formally we are breaking the full rotation group $\mathbb{Z}_p\rtimes \mathbb{Z}_l$ of the BTZ graph down to its normal subgroup $\mathbb{Z}_p$ (visually $\mathbb{Z}_p\rtimes \mathbb{Z}_l$ is the symmetry group of an $l$-gon whose each vertex hosts a $p$-gon distinguishable from the one at another vertex, a ``$p$-''hedreal group). The notion of a semidirect product of $\mathbb{Z}_l$ acting on $\mathbb{Z}_p$ requires that there is a group homomorphism $f:\mathbb{Z}_p\rtimes \mathbb{Z}_l\rightarrow \mathbb{Z}_l$ which is identity on $\mathbb{Z}_l$, and its kernel is $\mathbb{Z}_p$. This can be seen by holding the $l$-gon still and arbitrarily rotating all the $l$ numbers of $p$-gons, and the latter action constitutes the kernel.} Intuitively, this sector of interest is somewhere in between the full spectrum and the fully isotropic sector (all subtrees have identical field values).

The initial condition on the boundary of the subtree $s$ is
\begin{equation}
\label{eq:bdryinitial}
    \phi_{N-2,s}=(p+1-\lambda_t)\phi_{N-1,s},\quad t=0,\dots,l-1,
\end{equation}
where $\phi_{N,s}$ was a free parameter already set to be $0$, and the subscript $t$ in eigenvalue $\lambda_t$ of the graph Laplacian \eqref{eq: discrete} will be explained later below \eqref{eq:modes}.\footnote{Although $t$ shares the same range as $s$, it has a \textit{different} physical meaning, and by definition it is independent of $s$ , which is obvious because $\lambda_t$ is a global quantity.\label{foot:t}}
Since we are dealing with fully isotropic modes on each {\it individual} subtree, as argued as explained right below \eqref{eq:isotropic}, the linear recursion relation toward the central horizon is exactly the same as (\ref{eq:recursion}):
\begin{equation}
\label{eq:reverse}
    \phi_{n-2,s}+(\lambda_t-p-1)\phi_{n-1,s}+p\phi_{n,s}=0,\quad 2\leq n\leq N-1,
\end{equation}
in the ``reverse'' order, and the field values are 
\begin{equation}
\label{eq:value}
    \phi_{n,s}=c_{+,t}\left(\phi_{N-1,s}\right)\cdot\alpha_{+,t}^{N-1-n}+c_{-,t}\left(\phi_{N-1,s}\right)\cdot\alpha_{-,t}^{N-1-n},
\end{equation}
where both coefficients $\left[c_{+,t}\left(\phi_{N-1,s}\right),c_{-,t}\left(\phi_{N-1,s}\right)\right]$ and solutions $\left(\alpha_{+,t},\alpha_{-,t}\right)$ to the characteristic equation of \eqref{eq:reverse} are pairs of Galois conjugates as before.\footnote{Although they will not enter the rest of our analysis, we have
\begin{equation}
    \alpha_{\pm,t}=\frac{(1+p-\lambda_t)\pm\sqrt{(1+p-\lambda_t)^2-4p}}{2}.
\end{equation}
}

Now we denote the ratio between field values on the first layer (depth 1) and those on the horizon as $k\equiv \phi_{1,s}/\phi_{0,s}$. Although $\phi_{1,s}$ or $\phi_{0,s}$ can vary between subtrees rooted at different horizon vertices $s$, the linear recurrence \eqref{eq:reverse} implies that $k$ must be isotropic around the loop, i.e., without a subscript $s$, because it is determined solely by the recursion relation for $n=2$. Incidentally, the overall scaling between field values on subtrees rooted at $s_1$ and $s_2$ is $\phi_{0,s_1}/\phi_{0,s_2}=\phi_{1,s_1}/\phi_{1,s_2}$.

\begin{figure}[h]
    \centering
    \includegraphics[width = 8cm]{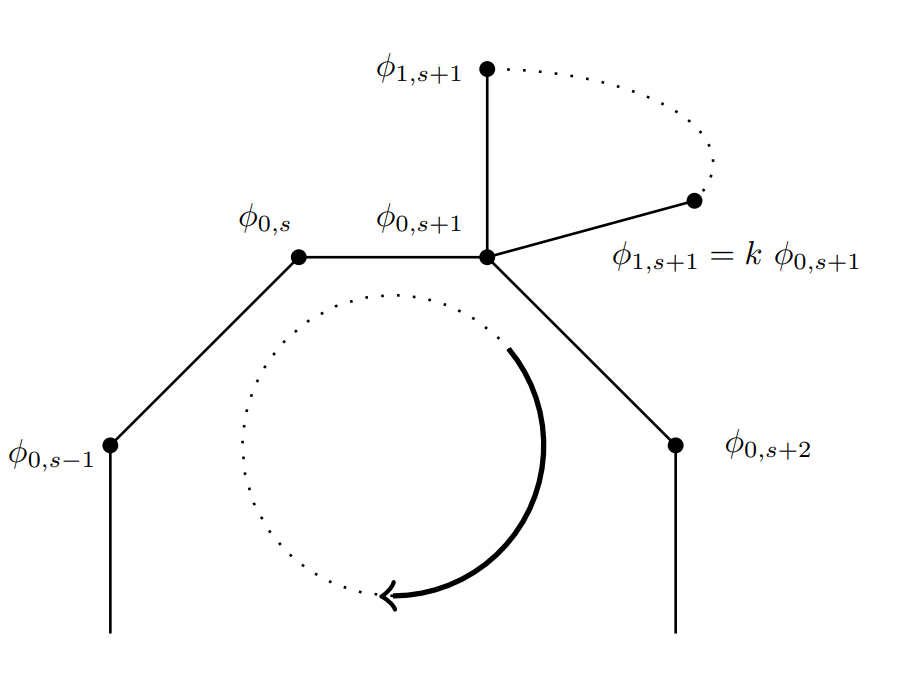}
    \caption{Going around the horizon with recursive relation (\ref{eq:recursion2}).}
    \label{fig:horizon}
\end{figure}

However, $k$ still depends on $\alpha_{\pm,t}$ and therefore $\lambda_t$, so we denote it by $k_t(\lambda_t)$. We examine the recursion relation around the event horizon:
\begin{equation}
\label{eq:recursion2}
    \phi_{0,s+2}-\left[(p-1)(1-k_t(\lambda_t))-\lambda_t+2\right]\phi_{0,s+1}+\phi_{0,s}=0,\quad s=0,\dots,l-1,
\end{equation}
with the periodic boundary condition\footnote{We might consider anti-periodic boundary conditions for fermions as in \cite{GubserSpin}, and intuitively all $l$ later on will be replaced by $2l$.} $\phi_{0,0}=\phi_{0,l}$, as shown in Figure \ref{fig:horizon}.

On the other hand, the necessary and sufficient condition for the existence of periodicity in a second-order linear recurrence like (\ref{eq:recursion2}) is that the two solutions $r_+,r_-$ to its quadratic characteristic equation are roots of unity (not necessarily primitive). Suppose $r_+=e^{2\pi i\frac{w}{q}}$ is the $q^{\text{th}}$ root of unity and $r_-=e^{2\pi i\frac{w'}{q'}}$ is the $q'^{\text{th}}$ root of unity, then their period is $\text{lcm}(q,q')$. In our case, the period is $l$, the horizon length.

The solutions to the characteristic equation of (\ref{eq:recursion2}) are
\begin{equation}
\label{eq:r1r2}
    r_{\pm,t}=\frac{1}{2}\left\{\left[(p-1)\left(1-k_t(\lambda_t)\right)-\lambda_t+2\right]\pm\sqrt{\left[(p-1)(1-k_t(\lambda_t))-\lambda_t+2\right]^2-4}\right\},
\end{equation}
then it is clear from Vieta's formula that
\begin{equation}
    (p-1)(1-k_t(\lambda_t))-\lambda_t+2=+2\cos\left(\frac{2\pi t}{l}\right)
    \end{equation}
    and
    \begin{equation}
    \sqrt{4-\left[(p-1)(1-k_t(\lambda_t))-\lambda_t+2\right]^2}=+2\sin\left(\frac{2\pi t}{l}\right).
\end{equation}
If we denote the discriminant in (\ref{eq:r1r2}) as $\delta$, then we note that it is impossible to have
\begin{equation}
    +\frac{\sqrt{\delta}}{2i}=\sin\frac{2\pi}{q}\,\,\,\text{and}\,\,\, -\frac{\sqrt{\delta}}{2i}=\sin\frac{2\pi}{q'},\quad 0<q\neq q'\leq l,\quad l=\text{lcm}(q,q')>2,
\end{equation}
i.e., different denominators in the exponents of roots of unity $r_+$ and $r_-$, because 
\begin{equation}
    \sin\frac{2\pi}{q}+\sin\frac{2\pi}{q'}=2\sin\left(\frac{q+q'}{qq'}\pi\right)\cos\left(\frac{q'-q}{qq'}\pi\right)=0
\end{equation}
indicates that $(q+q')/qq'=0,1$ or $(q'-q)/qq'=1/2,3/2$. The first equation implies that $q=q'=2$ and the second equation implies that $q=1,q'=2$. Hence, $r_+$ and $r_-$ are both $l^{\text{th}}$ roots of unity, and are complex conjugates to each other. 

Then we have
\begin{equation}
\label{eq:modes}
    k_t(\lambda_t)=1-\frac{1}{p-1}\left(2\cos\left(\frac{2\pi t}{l}\right)+\lambda_t-2\right),\quad t=0,\dots,l-1,
\end{equation}
with double degeneracies $k_t\left(\lambda_t\right)=k_{l-t}(\lambda_{l-t})$, and $t$ now labels global ``oscillation'' modes (for all subtrees), answering \footreff{foot:t}. To avoid overcounting, we observe that pairs -- $\left[k_t(\lambda_t),\lambda_t\right]$ and $\left[k_{l-t}(\lambda_{l-t}),\lambda_{l-t}\right]$ -- correspond to the same mode along the horizon, because $t\Longleftrightarrow l-t$ is equivalent to swapping solutions $r_{+,t}$ and $r_{-,t}$ to (\ref{eq:r1r2}), so that after solving the initial conditions $\phi_{0,0}=A+B$ and $\phi_{0,1}=Ar_{+,t}+Br_{-,t}$, all $\phi_{0,s}$'s are invariant under this swapping. Then the maximum value of $t$ should be $\lfloor l/2 \rfloor$.

Let us take a deeper look into this $k_t(\lambda_t)$, by stepping outward away from the horizon. Starting from depth 1, we adopt the same recursion as used in the isotropic case on the Bruhat-Tits tree. Therefore, the recursion relation here stays the same as (\ref{eq:recursion}) for any depth $n>2$, implying that solutions $\alpha_{\pm}$ to the characteristic equation are the same as (\ref{eq:roots}). When $n=2$, the field value $\phi_{n-2}$ in \eqref{eq:recursion} is replaced by $\phi_{0,s},s=0,\dots,l-1$, and $\phi_{n-1}$ in \eqref{eq:recursion} becomes $\phi_{1,s}=k_t(\lambda_t)\phi_{0,s}$. Then, the initial condition here gives:
\begin{equation}
\begin{split}
    \tilde{c}_{\pm,t}\left(\phi_{0,s}\right)=\left(\frac{1}{2}\pm\frac{(p+1)(p+1-\lambda_t)-4p\cos\left(\frac{2\pi t}{L}\right)}{2(p-1)\sqrt{(1+p-\lambda_t)^2-4p}}\right)\phi_{0,s}.
\end{split}
\end{equation}
Numerically, we observe that the coefficient of the highest degree in $\lambda_t$ for the polynomial $\phi_{N,s}=\left(\tilde{c}_{+,t}\left(\phi_{0,s}\right)\cdot \alpha_{+,t}^N+\tilde{c}_{-,t}\left(\phi_{0,s}\right)\cdot\alpha_{-,t}^N\right)\phi_{0,s}$ is $(-1)^{N}\phi_{0,s}/\left(p^N-p^{N-1}\right)$, where $\alpha_{+,t}$ and $\alpha_{-,t}$ are the same as in (\ref{eq:value}). Thus, the constant term is
\begin{equation}
\label{eq:constant}
    \frac{1}{p^N-p^{N-1}}\left(p^N+p^{N-1}+2\sum_{i=0}^{N-2}p^i-2\cos\left(\frac{2\pi t}{l}\right)\sum_{i=0}^{N-1}p^i\right)\phi_{0,s}.
\end{equation}
The product of all roots is \textit{independent} of index $s$:
\begin{equation}
\label{eq:rootproduct}
    p^N+p^{N-1}+2\frac{p^{N-1}-1}{p-1}-2\frac{p^N-1}{p-1}\cos\left(\frac{2\pi t}{l}\right).
\end{equation}
Note that (\ref{eq:rootproduct}) is the product of eigenvalues for one specific $t$. In order to account for all modes when computing $\det\Box$, we must multiply contributions from all $t=0,\dots,\lfloor l/2 \rfloor$, and for convenience we shift $t$ by 1 in the product.  

To multiply $\lceil l/2\rceil$ terms of \eqref{eq:rootproduct} together, we recall that roots of $T_n(x)$, the Chebyshev polynomial of the first kind of degree $n$, are
\begin{equation}
    x_k=\cos\left(\frac{2k-1}{2n}\pi\right),\quad k=1,\dots,n,
\end{equation}
called \textit{Chebyshev nodes} in interval $[-1,1]$, and hence (see, e.g., \cite{Zwillinger})
\begin{equation}
    T_n(x)=2^{n-1}\prod^n_{k=1}\left[x-\cos\left(\frac{(2k-1)\pi}{2n}\right)\right].
\end{equation}
Then it is not hard to see, using the reflection symmetry $T_n(-x)=(-1)^nT_n(x)$, for coprime $\alpha$ and $\beta$, we have
\begin{equation}
    \boxed{\prod^{\beta}_{k=1}\left[2x\pm2\cos\left(\frac{2\pi k\alpha}{\beta}+\theta\right)\right]=2\left[T_{\beta}(x)+(\pm1)^{\beta}(-1)^{\alpha\beta+\alpha}\cos(\beta\theta)\right]},
\end{equation}
which leads us to the desired product:
\begin{equation}
\label{eq:Cheby}
\begin{split}
    &\prod_{t=1}^{\lceil l/2\rceil}\left\{\frac{p^N-1}{p-1}\left[\left(2\frac{p^{N-1}-1}{p-1}+p^{N-1}+p^N\right)\bigg/\frac{p^N-1}{p-1}-2\cos\left(\frac{2\pi t}{l}\right)\right]\right\}\\
    =& 
    \begin{cases}
    \left(\frac{p^N-1}{p-1}\right)^{\frac{l}{2}}\left[2T_l\left(\frac{p^{N-1}(p^2+1)-2}{2\left(p^N-1\right)}\right)-2\right]^{\frac{1}{2}}\quad\quad\quad\quad\quad\quad\quad\quad\quad\quad\quad\quad\quad\quad\quad\,\,\,\,\, l\text{ even},\\
    \left(\frac{p^N-1}{p-1}\right)^{\frac{l}{2}}\left[2T_l\left(\frac{p^{N-1}(p^2+1)-2}{2\left(p^N-1\right)}\right)-2\right]^{\frac{1}{2}}\left[\frac{p^{N+1}+p^{N-1}-2+2(p^N-1)\cos\left(\frac{\pi}{l}\right)}{p-1}\right]^{\frac{1}{2}}\quad l\text{ odd}.
    \end{cases}
\end{split}
\end{equation}

For large $N$, we have:
\begin{equation}
\label{eq:Chebyshev}
\begin{split}
\begin{cases}
    \sqrt{2}\left(\frac{p^N}{p-1}\right)^{\frac{l}{2}}\left[T_l\left(\frac{p^2+1}{2p}\right)-1\right]^{\frac{1}{2}}\quad\quad\quad\quad\quad\quad\quad\quad\quad\quad\quad\quad\quad\,\,\,\, l\text{ even,}\\
    \sqrt{2}\left(\frac{p^N}{p-1}\right)^{\frac{l}{2}}\left[T_l\left(\frac{p^2+1}{2p}\right)-1\right]^{\frac{1}{2}}\left[\frac{p^{N-1}(p^2+1+2p\cos\left(\pi/l\right))}{p-1}\right]^{\frac{1}{2}}\quad\,\,\,\,\,\,\, l\text{ odd}.
    \end{cases}
    \end{split}
\end{equation}
Since $N$ is really an infinite quantity, we need to fully forget all subleading terms in (\ref{eq:Chebyshev}). Consequently, there are no descendants and agree with Melzer's axioms on non-Archimedean CFTs \cite{melzer}, and Chebyshev polynomials do not serve as counterparts of the usual degeneracy-counting function $1/\eta(-1/\tau)$ in 2d CFTs.

Furthermore, if $l$ and $p$ are not small, we use the explicit expression
\begin{equation}
T_l(x)=\cosh \left(l \text{arccosh}x\right),\quad x\geq1,
\end{equation}
then we obtain
\begin{equation}
\label{eq:BTZdet}
    \boxed{
    \begin{cases}
    \left(\frac{p^{N+1}}{p-1}\right)^{\frac{l}{2}}\quad\quad\quad\quad\quad l\text{ even,}\\
    \left(\frac{p^{N+1}}{p-1}\right)^{\frac{l+1}{2}}\left(\frac{p+1}{p}\right)\quad l\text{ odd.}
    \end{cases}}
\end{equation}
Now we can already see that $\det\left(\Box\right)$ is divergent exponentially as $p^{lN}$ when $N\rightarrow\infty$, which is very different from the number of boundary points $l(p-2)(p-1)^{N-1}$, or the total number of points in the BTZ graph $lp^N$. So we cannot directly obtain a finite answer using the similar argument which leads to (\ref{eq:treepart}), and the unregularized partition function is:\footnote{Since our divergence originates from a divergent number of eigenvalues as $N\rightarrow\infty$, one might try zeta function regularization. However, since eigenvalues here are complicated factors of Chebyshev polynomials, we do not see an easy way out; we hope to revisit this issue in the future.}
\begin{equation}
\label{eq:BTZpart}
    \boxed{
    Z_{\text{BTZ}}=\begin{cases}
    \left(\frac{p-1}{p^{N+1}}\right)^{\frac{l}{4}}\quad\quad\quad\quad\quad l\text{ even,}\\
    \left(\frac{p-1}{p^{N+1}}\right)^{\frac{l+1}{4}}\left(\frac{p}{p+1}\right)\quad l\text{ odd.}
    \end{cases}}
\end{equation}

Apart from the divergence, (\ref{eq:BTZpart}) is very similar to the partition function of a  BTZ black hole in the usual Euclidean AdS$_3$ at leading order, as reviewed in Appendix \ref{app:modular}. 

In summary, we have to undergo three recurrences to solve the Laplace problem for the partially isotropic sector on a $p$-adic BTZ black hole:

\textit{1. From the asymptotic boundary to the horizon\footnote{Skipping \textit{Step 1} results in a messy situation, as explained in Appendix \ref{app:messy}.}, using recurrence (\ref{eq:reverse});\footnote{The sole purpose of recurrence \eqref{eq:reverse} is to show the isotropy of $\phi_{n,s}$ within the subtree $s$.}}

\textit{2. Go around the horizon once, using recurrence (\ref{eq:recursion2});}

\textit{3. From the horizon to the asymptotic boundary, using recurrence (\ref{eq:recursion}).}

Since the recurrence relation (\ref{eq:reverse}) for depth $n>2$ is the same as the one in the Bruhat-Tits tree (\ref{eq:recursion}), the asymptotic behavior of the eigenfunction and the eigenvalues stay the same as in (\ref{eq:ansatz}) and (\ref{eq:eigenvalue}), respectively. We are still in the ``evanescent wave'' basis as in Section \ref{sec:Laptree}.
Unlike the full spectrum on the Bruhat-Tits tree, the full spectrum on the BTZ graph will not be presented in Appendix \ref{sec:spectrumBTTree} due to its involved nature, but in principle the same techniques there are still applicable, and we will report on it in the near future \cite{spectrum}.

Now we perform the non-Wick-rotated inverse Laplace transform on the partition function (\ref{eq:BTZpart}) to obtain the density of states. To this end, we need to do two radical things: 
\begin{itemize}
\item First, we strip off the divergent factor in (\ref{eq:BTZpart}) by hand, since otherwise the density of states to be obtained would be very negative numbers;
\item Second, we regard $l$ as ``$1/\beta\sim i/\tau>0$'' for a non-rotating BTZ. Although in our $p$-adic setup, there is no mathematically rigorous $\tau\in\mathbb{C}$, in order to do the integral transform, we need to turn on an auxiliary imaginary part of the inverse temperature momentarily, so that $\tilde{\beta}=\beta+i\beta',\beta'\in\mathbb{R}$.
\end{itemize}
Then going from the canonical ensemble to the microcanonical ensemble, we have
\begin{equation}
\label{eq:dos1}
    \rho(E)=\mathcal{L}^{-1}\left\{Z_{\text{BTZ}}\left(\tilde{\beta}\right)\right\}(E)=
    \begin{cases}
    \frac{1}{2\pi i}\int_{\beta-i\infty}^{\beta+i\infty}d\tilde{\beta} e^{\tilde{\beta} E}\left(p-1\right)^{1/4\tilde{\beta}}\quad\quad\quad\quad\quad\quad l\text{ even},\\
    \frac{1}{2\pi i}\int_{\beta-i\infty}^{\beta+i\infty}d\tilde{\beta} e^{\tilde{\beta} E}\left(p-1\right)^{1/4\tilde{\beta}+1/4}\left(\frac{p}{p+1}\right) \quad l\text{ odd}.
    \end{cases}
\end{equation}
However, the second expression cannot be evaluated explicitly, so we focus on the high-temperature limit as $\beta\rightarrow0$ so that $1/4\tilde{\beta}+1/4\approx1/4\tilde{\beta}$, and from now on we do not treat even and odd $l$ separately, because they only differ by a factor $\frac{p}{p+1}$. Then we get
\begin{equation}
\label{eq:dos2}
    \rho(E)=\frac{\ln(p-1)}{8}     
    {}_0F_1\left(;2;\frac{E\ln(p-1)}{4}\right)+\delta(E),
\end{equation}
for all primes $p$, where ${}_0F_1$ is the \textit{confluent hypergeometric limit function}, and is related to the modified Bessel function of the first kind as
\begin{equation}
    I_{\alpha}(x)=\frac{(x/2)^{\alpha}}{\Gamma(\alpha+1)}{}_0F_1\left(;\alpha+1;\frac{x^2}{4}\right).
\end{equation}
In (\ref{eq:dos2}), we have $\propto I_1\left(\sqrt{E\ln(p-1)}\right)$, and it goes to zero as $E\rightarrow0$. Its asymptotic behavior of ${}_0F_1$ as $x\rightarrow\infty$ is
\begin{equation}
\label{eq:asymptotics}
    {}_0F_1\left(;\alpha;x\right)\approx x^{-(\alpha-1)/2}\Gamma(\alpha)\frac{e^{2\sqrt{x}}}{\sqrt{2\pi\sqrt{x}}}\left(1-\frac{4(\alpha-1)^2-1}{16\sqrt{x}}+\dots\right)
\end{equation}
so in semi-classical limit, for positive energy, we discard Dirac delta and its derivative in (\ref{eq:dos1}). When $p>3$, we have
\begin{equation}
\label{eq:approxDoS}
\boxed{\rho(E)\approx
\frac{\ln^{1/4}(p-1)}{\sqrt{2\pi}}e^{\sqrt{E\ln(p-1)}}E^{-3/4}\left(1-\frac{3}{8\sqrt{E\ln(p-1)}}+\mathcal{O}(E^{-1})+\dots\right)}.
\end{equation}

Finally and straightforwardly, the Bekenstein-Hawking-like entropy is
\begin{equation}
\label{eq:entropy}
    S\approx\sqrt{E\ln(p-1)}-\frac{3}{4}\ln E+\frac{1}{4}\ln\left(\ln(p-1)\right)-\frac{1}{2}\ln(2\pi)-\dots,
\end{equation}
where the second term is the famous logarithmic correction terms previously discovered in \cite{Carlip,Log}. This result is also consistent with the ``species problem'' \cite{Sorkin} because we are calculating scalar fields all the time. One can also derive the Cardy-like formula \cite{Cardy1,Cardy2,Cardy3} via saddle point approximation on (\ref{eq:dos1}).

The usual Benkenstein-Hawking entropy of black holes from Cardy-like formula has $4\pi\sqrt{Ek}$ as the leading term \cite{MaloneyWitten}, where $k$ is proportional to the Brown-Henneaux central charge $3l/2G_N$ \cite{BH}. By comparing this with (\ref{eq:entropy}), we see that our $\ln(p-1)$ is like $k$. However, this raises a puzzle because increasing the valency of the tree should increase the curvature, corresponding to decreasing $k$ in the continuous AdS$_3$.\footnote{Since the Bruhat-Tits tree has no holonomy, defining a Riemann tensor is arduous. Yau et al. \cite{Yau} were able to define a Ricci curvature $\kappa_{xy}$ on graphs without a Riemann tensor, but in terms of the edge lengths $a_{xy}$, from which Gubser et al. \cite{Edge} found that on-shell the tree has a constant negative Ricci curvature
$\kappa_{xy} = -2 \frac{p-1}{p+1}$
and the edge length fluctuations are massless modes.} We will discuss this near the end.

Another stand-alone case of (\ref{eq:dos2}) is $p=3$, since $\ln2<0$, and the asymptotic expansion (\ref{eq:asymptotics}) is only true when $|\arg x|<\pi/2$. Now ${}_0F_1$ is related to the Bessel function of the first kind as
\begin{equation}
    J_{\alpha}(x)=\frac{(x/2)^{\alpha}}{\Gamma(\alpha+1)}{}_0F_1\left(;\alpha+1;-\frac{x^2}{4}\right),
\end{equation}
and $J_{\alpha}(x)$ has the following asymptotics for real $x\rightarrow\infty$:
\begin{equation}
    J_{\alpha}(x)\approx\sqrt{\frac{2}{\pi x}}\cos\left(x-\frac{\alpha\pi}{2}-\frac{\pi}{4}\right),
\end{equation}
so the semiclassical limit of density of states is
\begin{equation}
    \rho(E)|_{p=2}\approx 2\sqrt{2}\frac{(-\ln2)^{3/4}}{\sqrt{\pi}}E^{-3/4}\cos\left(\sqrt{-E\ln2}-\frac{3\pi}{4}\right),
\end{equation}
which is a pathological result due to the oscillatory nature. It seems that a 3-adic BTZ black hole is unstable.

The continuous integral transform (\ref{eq:dos1}) is justified because in the high temperature regime $l\rightarrow\infty$, the separation between two adjacent discrete inverse temperatures is $\sim 1/l^2$. On the other hand, if we do not perform coarse-graining, we need to perform the discrete inverse Laplace transform. Superficially, the discrete inverse Laplace transform has the same expression as the one used in going from canonical partition function $Z_N(\beta)$ for $N$ particles to grand partition function $\mathcal{Z}(\beta,\mu)$:
\begin{equation}
    \mathcal{Z}(\beta,\mu)=\sum_{N=0}^{\infty}\left(e^{\mu \beta}\right)^NZ_N(\beta),
\end{equation}
but here the temperature is held fixed, and particle number is the analogue of $p$-adic discrete temperature.\footnote{This transform is also called a unilateral $Z$-transformation, with the less common but equivalent definition where powers are positive, same as probability generating functions.} Unfortunately in our case, the Z-transform does not yield a closed form so we stick to the continuous approximation (\ref{eq:dos1}).

Let us examine more details on the density of states. At low energy $E_0$, we integrate the density of states (\ref{eq:dos2}) over the interval $[E_0,E_0+\epsilon]$ with a small but finite $\epsilon$
\begin{equation}
\label{eq:degeneracy}
    \int_{E_0}^{E_0+\epsilon}dE\rho(E)=\left.\frac{\ln(p-1)}{8}{}_0F_1\left(;2;\frac{E\ln(p-1)}{4}\right)\right\vert^{E_0+\epsilon}_{E_0},
\end{equation}
although there is no particle interpretation in ordinary 2d CFTs (roughly because their correlators have no simple poles), and we expect so in $p$-adic CFT, in the bulk we can view the tree as a lattice, and number of vertices equals the number of degrees of freedom (or ``particles''), which is $lp^N$.
The low-energy limit of (\ref{eq:degeneracy}) is
\begin{equation}
    \frac{1}{8}{}_0F_1\left(;2;\frac{\ln(p-1)E_0}{4}\right)\ln(p-1)\epsilon+\frac{1}{128}{}_0F_1\left(;3;\frac{\ln(p-1)E_0}{4}\right)\ln^2(p-1)\epsilon^2+\mathcal{O}(\epsilon^3).
\end{equation}
Small-argument behavior of ${}_0F_1$ is just 1, so we have:
\begin{equation}
    \frac{1}{8}\ln(p-1)\epsilon+\frac{1}{128}\ln^2(p-1)\epsilon^2+\mathcal{O}(\epsilon^3)<\frac{1}{16}\ln(p-1)\sum_{i=1}^{\infty}(i+1)\epsilon^i=\frac{\epsilon(2-\epsilon)}{16(\epsilon-1)^2}\ln(p-1),
\end{equation}
which is a constant polynomial in total number of ``particles,'' hence satisfying the sparsity condition on in \cite{Vazirani1, Vazirani2} on the number of low-energy eigenstates in a gapless 1D system with a local Hamiltonian\footnote{We thank Ning Bao for pointing out these references.}, hence in principle one is able to approximate the Hilbert subspace near the ground state in the supposedly dual $p$-adic CFT. This may be worth investigating in the future.

\subsection{Turning on the scalar mass}
Here we again turn off the source $J$ in \eqref{eq:eom}, and now we have a Helmholtz-like wave equation
\begin{equation}
    \left(\square + m_p^2\right) \phi_a =0.
\end{equation}

The on-shell mass squared of a bulk scalar in (\ref{eq:action}) is real\footnote{Here $n$ is the degree of unramified extension $\mathbb{Q}_{p^n}$ of $\mathbb{Q}_{p}$, so that the Burhat-Tits tree is now $T_{p^n}$.} \cite{Gubser1,Caltech}:
\begin{equation}
\label{eq:mass}
    m^2_p = - \frac{1}{\zeta_p(\Delta - n) \zeta_p(-\Delta)} = - (p+1) + 2p^{n/2} ~ \text{cosh} \bigg [ \bigg( \Delta - \frac{n}{2}\bigg) \text{ln} ~ p \bigg],
\end{equation}
and invariant under $\Delta \rightarrow n -\Delta$, where the $p$-adic or ``finite'' local zeta function $\zeta_p(s)$ is defined as:
\begin{equation}
    \zeta_p(s)\equiv\frac{1}{1-p^{-s}},
\end{equation}
which obtains its name because the real Riemann zeta function $\zeta_\infty(s)$ can be constructed from Euler's adelic product:
\begin{equation}
  \zeta_\infty (s) \equiv \sum_{n=1}^{\infty}\frac{1}{n^s}=\prod_{p} \zeta_p(s) = \prod_p \frac{1}{1 - p ^{-s}}.
\end{equation}

Then the Breitenlohner-Freedman (BF) bound is $m_{\text{BF},p}^2=-1/\zeta_p(-n/2)^2$, with $\Delta=n/2$. For $m_p^2$ above this bound, two possible $p$ can satisfy \eqref{eq:mass}: 
\begin{equation}
    p^\pm=\frac{1}{2}\left[(1+p^n+1/m_p^2)\pm\sqrt{(1+p^n+1/m_p^2)^2-4p^n}\right].
\end{equation}
We adopt the same convention on the solutions to (\ref{eq:mass}) as in \cite{Gubser1}, i.e., $\Delta=\Delta_+>n/2$. Then for massless scalars, $\Delta=n$, so we are restricted to $\Delta=0,1$ when $n=1$.

%\begin{figure}[h]
 %   \centering
 %   \begin{subfigure}[t]{.5\textwidth}
 %   \centering
 %   \includegraphics[height=1.9in]{mass1p5.png}
 %   \caption{$p=5$, $n=1$.}
 %   \end{subfigure}%
 %   ~
 %   \begin{subfigure}[t]{.5\textwidth}
 %   \centering
 %   \includegraphics[height=1.9in]{mass2p5.png}  
 %   \caption{$p=5$, $n=2$.}
 %   \end{subfigure}
%    \caption{Scalar mass $m_p^2$ as a function of conformal dimension $\Delta$. $m_p^2>0$ %when $\Delta>n$.}
%    \label{fig:mass}
%\end{figure}

Now we hope to calculate partiton function when $\phi$ is massive, which amounts to calculating the determinant of $\Box+m_p^2\mathds{1}$. We relate the field polynomial $\phi_N^{\text{tree}}(\lambda)$ ($\phi_{n,s}^{\text{BTZ}}(\lambda_t)$ for BTZ black holes) resulting from the boundary condition $\phi|_{\partial T}\equiv \phi_N=0$ with the ``monic'' (up to $(-1)^N$) characteristic polynomial $P_N(\lambda)=\prod_{i=1}^N\left(\lambda_i-\lambda\right)=\det\left(\Box-\lambda \mathds{1}\right)$ of the Laplacian $\Box$. What we have calculated in the previous two subsections are essentially $P_N(0)$, the constant term of $P_N(\lambda)$, and now we perturbatively investigate $P_N\left(-m^2\right)$, i.e., the determinant $\det\left(\Box+m_p^2\mathds{1}\right)=\prod_{i=1}^N\left(\lambda_i+m_p^2\right)$.

It is important that $\lambda_i$'s are always greater than the BF bound $m_{\text{BF},p^n}=-1/\zeta_p(-n/2)^2$, which is $m_{\text{BF},p}=-\left(\sqrt{p}-1\right)^2$ for $n=1$, whose absolute value is strictly smaller than all eigenvalues for both Bruhat-Tits trees and BTZ black holes in (\ref{eq:eigenvalue}). Hence, we will not encounter issues of alternating signs upon calculating the determinant of $\Box+m_p^2\mathds{1}$.

In principle, one could possibly use minimal polynomials for Gaussian integers to study powers of Galois conjugates. However, we will proceed in a more combinatorial approach.

\subsubsection{On Bruhat-Tits trees}
Since the polynomial $\phi_N(\lambda)$ in $\lambda$ always has the constant term 1, we need to rescale it to be monic up to $(-1)^{N}$:
\begin{equation}
    \boxed{P_N^{\text{tree}}(\lambda)\equiv\phi_N^{\text{tree}}(\lambda)/\phi^{\text{tree}}_0\prod_{i=1}^N\lambda_i=\left(p^N+p^{N-1}\right)\phi^{\text{tree}}_N(\lambda)/\phi^{\text{tree}}_0},
\end{equation}
where $P_N^{\text{tree}}(\lambda)$ is defined in (\ref{eq:phitree}), so that $P_N^{\text{tree}}(0)=p^N+p^{N-1}$.

By denoting $x\equiv p-\lambda+1$, we can rewrite $P_N^{\text{tree}}(\lambda)$ as
\begin{equation}
\begin{aligned}
    \frac{1}{2(2p)^N}\sum_{k=0}^n\binom{N}{k}x^k&(x^2-4p)^{\frac{N-k-1}{2}}\\
    &\times\left\{\left[(x^2-4p)^{\frac{1}{2}}\left(1+(-1)^{N-k}\right)\right]+\frac{p-1}{p+1}x\left[\left((1+(-1)^{N-k-1}\right)\right]\right\}.
    \end{aligned}
\end{equation}
Repeatedly applying the binomial theorem in a nested fashion gives us the following results:
\begin{itemize}
\item The linear term of $P_{N}^{\text{tree}}(\lambda)$ is, since $p\neq1$:
\begin{equation}
    \left(-Np^{N-1}-2\sum_{i=1}^{N-1}ip^{i-1}\right)\lambda=\frac{(N+2p-Np^2)p^N-2p}{p(p-1)^2}\lambda,
\end{equation}
which goes to $-N\frac{p+1}{p-1}p^{N-1}\lambda$ when $N$ is large;

\item The quadratic term is
\begin{equation}
\begin{split}
    &\sum_{i=0}^{N-2}p^i\left[\frac{(i+1)(i+2)}{2}+(i+1)(i+2)(N-i-2)\right]\lambda^2\\
    =&\frac{1}{2p(p-1)^4}\left[(N^2-N)p^{N+3}-(N^2+5N-6)p^{N+2}-(N^2-5N-6)p^{N+1}\right.\\
    &\quad\quad\quad\quad\quad\quad\quad\left.+(N^2+N)p^N-(4N+6)p^2+(4N-6)p\right]\lambda^2,
    \end{split}
\end{equation}
which goes to $N^2\frac{p+1}{2(p-1)^2}p^{N-1}\lambda^2$ when $N$ is large.
\end{itemize}
So for small $|m_p^2|<1$, we have the unregularized partition function $Z_{\text{tree}}(m\rightarrow 0)$:
\begin{equation}
\begin{split}
    \det\left(m_p^2\mathds{1}+\Box\right)=P_N^{\text{tree}}\left(-m^2\right)=&\left(p^N+p^{N-1}\right)\left(1+\frac{N}{p-1}m_p^2+\frac{1}{2}\left(\frac{N}{p-1}m_p^2\right)^2+\dots\right)\\
    =&\boxed{\left(p^N+p^{N-1}\right)e^{\frac{Nm_p^2}{p-1}}},
\end{split}
\end{equation}
where the regularization factor $p^N+p^{N-1}\propto$ (\ref{eq:volume}) is now manifest.

For completeness, we look into the large-mass limit, where only high-degree terms in $P_N^{\text{tree}}(\lambda)$ matter.
\begin{itemize}
\item The $\lambda^{N-1}$ term is $-(-1)^NN(p+1)\lambda^{N-1}$. So in order to ignore the $\lambda^{N-2}$ term, we need $m^2$ to be larger than $N$;

\item The $\lambda^{N-2}$ term is
\begin{equation}
    \frac{1}{2}(-1)^N\left[N(N-1)p^2+2(N-1)^2p+N(N-1)-2\right]\lambda^{N-2},
\end{equation}
which goes to $\frac{1}{2}(-1)^NN^2(p+1)^2\lambda^{N-2}$ when $N$ is large;

\item The coefficient of $\lambda^{N-3}$, a degree 3 polynomial in $p$ involves first-order linear recurrence with variable coefficient for $p^i$ coefficients $f_{N}$, such as
\begin{equation}
    f_N=f_{N-1}+N(N-1)/2,
\end{equation}
but in the end we have
\begin{equation}
\begin{split}
    &-(-1)^N\left\{\frac{N(N+1)(N-4)}{6}+2+\frac{N(N-2)(N-3)}{2}p\right. \\
    &\quad\quad\quad\quad\left. +\left[\frac{N(N^2-5N+8)}{2}-2\right]p^2+\frac{N(N-1)(N-2)}{6}p^3\right\},
\end{split}
\end{equation} 
which goes to $-\frac{1}{6}(-1)^NN^3(p+1)^3\lambda^{N-3}$  when $N$ is large.
\end{itemize}
Then collectively we have the unregularized partition function:
\begin{equation}
\begin{split}
&\quad \,\, Z_{\text{tree}}(m\rightarrow\infty)\\
&=\left(p^N+p^{N-1}\right)m_p^{2N}\left(1+\frac{N(p+1)}{m_p^2}+\frac{1}{2}\left(\frac{N(p+1)}{m_p^2}\right)^2+\frac{1}{6}\left(\frac{N(p+1)}{m_p^2}\right)^3+\dots\right)\\
    &=\boxed{\left(p^N+p^{N-1}\right)m_p^{2N}e^{\frac{N(p+1)}{m_p^2}}},
    \end{split}
\end{equation}

Now we discuss the conditions on $\Delta$ when $|m_p^2|$ is small. In order to have $0<-m_p^2\ll1$, we write $\Delta=1+\epsilon$ where $\epsilon\ll1$. So we have
\begin{equation}
    \left(1-p^{n-\Delta}\right)\left(p^{\Delta}-1\right)\ll1,
\end{equation}
where $n$ denotes the unramified extension $\mathbb{Q}_{p^n}$, then we get
\begin{equation}
    \epsilon\ll\frac{\ln\left[\frac{p^{1-n}}{2}\left(2+p^n-\sqrt{p^{2n}+4}\right)\right]}{\ln p}.
\end{equation}
and similarly, for $\Delta=1-\epsilon$, we need $-1\ll-m^2<0$, and we get
\begin{equation}
    \epsilon\ll\frac{\ln\left[\frac{p}{2}\right(1-p^{-n}\sqrt{p^n(p^n-4)}\left)\right]}{\ln p}
\end{equation}
From this expression we also see that when $n=1$, the smallest prime $p$ is 5, consistent with the result from density of states in Section \ref{subsec:BTZ}.

\subsubsection{On BTZ black holes}
The characteristic polynomial for Laplacian on BTZ black hole is different from $P_N^{\text{tree}}(\lambda)$. It is rescaled from the field polynomial\footnote{Here the subscript is ``$s$'' not ``$t$'', because this polynomial depends on the initial field value $\phi_{0,s}$ on horizon, as written above (\ref{eq:constant}).} $\phi^{\text{BTZ}}_{N,s}(\lambda_t)$ at the cutoff depth $N$ to
\begin{equation}
\boxed{P_N^{\text{BTZ}}(\lambda_t)\equiv\prod_{t=1}^{\lceil l\rceil}\tilde{P}_{N,t}^{\text{BTZ}}(\lambda_t)=\left(p^N-p^{N-1}\right)^l\prod_{t=1}^{\lceil l\rceil}\phi^{\text{BTZ}}_{N,s}\left(\lambda_t\right)/\phi_{0,s}^{\text{BTZ}}},
\end{equation}
so that $P_N^{\text{BTZ}}(\lambda_t)$ and $\tilde{P}_{N,t}^{\text{BTZ}}(\lambda_t)$ are monic up to $(-1)^{Nl}$ and $(-1)^{N}$, respectively, and $P_N^{\text{BTZ}}(0)$ agrees with (\ref{eq:Cheby}).

Let us first consider when the mass $|m_p^2|$ is small. The linear term in $\lambda_t$ in $\tilde{P}_{N,t}^{\text{BTZ}}(\lambda_t)$ for one specific $t$ is
\begin{equation}
\begin{split}
    &\left[-Np^{N-1}+\left(\cos\left(\frac{2\pi t}{l}\right)-1\right)\sum_{i=1}^{N-1}2i(N-i)p^{N-i-1}\right]\lambda_t\\
    =&-\left(Np^{N-1}+4\sin^2\left(\frac{\pi t}{l}\right)\frac{N(p^N+1)(p-1)-(p^N-1)(p+1)}{(p-1)^3}\right)\lambda_t,
\end{split}
\end{equation}
which goes to
\begin{equation}
    -\left(\frac{Np^N}{(p-1)^2}+4Np^{N-1}\sin^2\left(\frac{\pi t}{l}\right)\right)\lambda_t
\end{equation}
when $N$ is large.

For small $m_p^2$, we only calculate $\tilde{P}^{\text{BTZ}}_{N,t}\left(-m_p^2\right)$ up to the linear term in $\lambda_t$, written in shorthand:
\begin{equation}
\begin{split}
    &A\cos\left(\frac{2\pi t}{l}\right)+B
    \end{split}
\end{equation}
where
\begin{equation}
    A\equiv-\frac{2 m_p^2 \left(N (p-1) \left(p^N+1\right)-(p+1) p^N+p+1\right)}{(p-1)^3}-\frac{2 \left(p^N-1\right)}{p-1},
\end{equation}
\begin{equation}
\begin{aligned}
    \hspace{-43pt}B\equiv m_p^2 N p^{N-1}&+\frac{2 m^2 \left(N (p-1) \left(p^N+1\right)-(p+1) p^N+p+1\right)}{(p-1)^3}\\
    &+p^{N-1}+\frac{2 \left(p^{N-1}-1\right)}{p-1}+p^N,
    \end{aligned}
\end{equation}
then $\lceil l\rceil$ terms multiply together to be
\begin{equation}
P^{\text{BTZ}}_{N}\left(-m_p^2\right)=
\begin{cases}
    \sqrt{2}(-A/2)^{\frac{l}{2}}\left[T_l\left(-B/A\right)-1\right]^{\frac{1}{2}} \quad\quad\quad\quad\quad\quad\quad\quad\quad\quad l\text{ even},\\
    \sqrt{2}(-A/2)^{\frac{l}{2}}\left[T_l\left(-B/A\right)-1\right]^{\frac{1}{2}}\left(A\cos\left(\pi/l\right)+B\right)^{\frac{1}{2}} \,\,\,\quad l\text{ odd},
    \end{cases}
\end{equation}
where $-B/A$ at $\mathcal{O} (m_p^2)$ is
\begin{equation}
\begin{aligned}
    \frac{p^{N+2}+p^N-2 p}{2 p \left(p^N-1\right)}+\frac{p^{N-1} \left(p^{N+1}+p^N-2 N p+2 N-p-1\right)}{2 \left(p^N-1\right)^2}&m^2+\mathcal{O}\left(m_p^4\right)\\\xrightarrow{N\rightarrow\infty}
    &\hspace{-2pt}\frac{p^2+1}{2p}+\frac{p+1}{2p}m_p^2.
    \end{aligned}
\end{equation}
Because $dT_l(x)/dx=lU_{l-1}(x)$, where $U_l(x)$ is the Chebyshev polynomial of the second kind, when both $l$ and $p$ are not small, we get the unregularized BTZ partition function:
\begin{equation}
\begin{split}
    Z_{\text{BTZ}}(m_p \rightarrow 0) &= P^{\text{BTZ}}_{N}\left(-m_p^2\right) \\&\boxed{\approx 
    \begin{cases}
    \left(1+\frac{lm_p^2}{2p}\right)^{\frac{1}{2}}\left(\frac{p^{N+1}}{p-1}\right)^{\frac{l}{2}}\left(1+\frac{Nm_p^2}{(p-1)^2}\right)^{\frac{l}{2}} \quad\quad\quad\quad\quad\quad\quad\, l\text{ even},\\
    \left(1+\frac{lm_p^2}{2p}\right)^{\frac{1}{2}}\left(\frac{p^{N+1}}{p-1}\right)^{\frac{l}{2}}\left(1+\frac{Nm_p^2}{(p-1)^2}\right)^{\frac{l}{2}}\left(A\cos\left(\frac{\pi}{l}\right)+B\right) \, l\text{ odd},
  \end{cases}}
  \end{split}
\end{equation}
which recovers (\ref{eq:BTZdet}) when $m_p^2=0$.

For large mass $|m_p^2|$, we calculate the $\lambda_t^{N-1}$ term in $\tilde{P}_{N,t}^{\text{BTZ}}(\lambda_t)$ to be
\begin{equation}
    (-1)^N\left(2\cos\left(\frac{2\pi t}{l}\right)-N(p+1)\right)\lambda_t^{N-1},
\end{equation}
and the $\lambda_t^{N-2}$ term is
\begin{equation}
    (-1)^N\left(\frac{N(N-1)}{2}(p^2+1)+(N-1)^2p+1-2(N-1)\cos\left(\frac{2\pi t}{l}\right)\right)\lambda_t^{N-2},
\end{equation}
so we have terms with the three highest degrees added up to
\begin{equation}
\begin{split}
    \tilde{P}^{\text{BTZ}}_{N,t}\left(-m_p^2\right)&=m_p^{2N}+m_p^{2N-2}\left(N(p+1)-2\cos\left(\frac{2\pi t}{l}\right)\right)\\+&m_p^{2N-4}\left(\frac{N(N-1)}{2}(p^2+1)+(N-1)^2p+1-2(N-1)\cos\left(\frac{2\pi t}{l}\right)\right)+\dots,
    \end{split}
\end{equation}
and when $N$ is large it is
\begin{equation}
    C\cos\left(\frac{2\pi t}{l}\right)+D,
\end{equation}
where 
\begin{equation}
    C\equiv -2m_p^{2N}\left(\frac{1}{m_p^2}+\frac{N}{m_p^4}\right),\,
    D\equiv m_p^{2N}\left(1+\frac{N(p+1)(2m_p^2+N+Np)}{2m_p^4}\right),
\end{equation}
then $\lceil l\rceil$ terms multiply together to
\begin{equation}
    P^{\text{BTZ}}_{N}\left(-m^2\right)=
\begin{cases}
    \sqrt{2}(-C/2)^{\frac{l}{2}}\left[T_l\left(-D/C\right)-1\right]^{\frac{1}{2}} \quad\quad\quad\quad\quad\quad\quad\quad\quad\quad l\text{ even},\\
    \sqrt{2}(-C/2)^{\frac{l}{2}}\left[T_l\left(-D/C\right)-1\right]^{\frac{1}{2}}\left(C\cos\left(\pi/l\right)+D\right)^{\frac{1}{2}} \,\,\,\quad l\text{ odd},
    \end{cases}
\end{equation}
where $-D/C$ at $\mathcal{O}(m_p^2)$ is
\begin{equation}
    \frac{N}{4}(p+1)^2+\frac{1}{4}\left(1-p^2\right)m_p^2+\mathcal{O}\left(m_p^4\right)+\dots
\end{equation}
so explicitly the unregularized BTZ partition function for very large $m^2$ is
\begin{equation}
    \boxed{Z_{\text{BTZ}}(m_p \rightarrow \infty)\approx 
    \begin{cases}
    m_p^{lN-l}\left(1+\frac{N}{m_p^2}\right)^{\frac{l}{2}}\left(\frac{N(p+1)^2}{2}+\frac{(1-p^2)m_p^2}{2}\right)^{\frac{1}{2}} \quad\quad\quad\quad\quad\quad\quad\,\,\, l\text{ even},\\
    m_p^{lN-l}\left(1+\frac{N}{m_p^2}\right)^{\frac{l}{2}}\left(\frac{N(p+1)^2}{2}+\frac{(1-p^2)m_p^2}{2}\right)^{\frac{1}{2}}\left(C\cos\left(\frac{\pi}{l}\right)+D\right)\,\, l\text{ odd},
  \end{cases}}
\end{equation}

\section{One-loop Witten diagrams}
\label{sec:4}
In the work by Kraus and Maloney \cite{KrausMaloney}, they proposed a duality between higher-energy states on the conformal boundary and semi-classical gravity in AdS$_3$ for the BTZ black hole. They showed that a bulk Witten diagram with two types of perturbative (i.e., not massive conical defects) scalar fields in the bulk is equivalent to the average value of the three-point coefficient $\overline{\langle E|\mathcal{O}| E\rangle}$, where $|E\rangle$ is the high-energy state dual to the BTZ black hole, and $\mathcal{O}$ is the operator dual to one type of the light scalars. Here, the average of the three-point coefficient is taken over all states with energy $E$ \begin{equation}
    \overline{\langle E|\mathcal{O}| E\rangle} \equiv \frac{\langle E|\mathcal{O}| E\rangle}{\rho(E)},
\end{equation} 
where $\rho (E)$ is the density of states given explicitly by the asymptotic Cardy formula \cite{Cardy1, Cardy2, Cardy3}. In Section \ref{sec:2}, we reviewed a way to construct a $p$-adic version of the BTZ black hole as the quotient space of the Bruhat-Tits tree by the $p$-adic Schottky group $q^{\mathbb{Z}}$. In this section, we propose to use Kraus-Maloney's technique in $p$-adic BTZ configuration and calculate the analogous Witten diagram.\footnote{Another name for Witten diagrams in $p$-adic AdS are called ``subway diagrams'' \cite{Gubser1}.} This calculation provides a dual interpretation for the boundary $p$-adic CFT averaged three-point coefficient, which in principle could be independently derived from a pure CFT calculation. 

\subsection{Review on BTZ black hole calculation by Kraus-Maloney}
In this section, we provide a brief overview of Kraus and Maloney's results \cite{KrausMaloney} on the bulk and boundary sides, as well as list their assumptions. 

\subsubsection{Cardy formula for three-point coefficients in 2d CFTs}

The high and low energy spectra of a CFT are related by modular invariance, i.e., $\mathcal{Z}(\beta)=\mathcal{Z}\left((2 \pi)^2/\beta\right)$. Analogously, modular invariance can be used to refer to high- and low-dimensional operators as ``heavy'' and ``light,'' respectively. This can be used to obtain results on the asymptotic spectral density weighted by OPE coefficients. Kraus and Maloney used modular invariance in the torus one-point function to estimate light-heavy-heavy three-point coefficients $\langle E| \mathcal{O} |E\rangle$ for a BTZ black hole. They proved that the averaged three-point coefficients from the bulk in the large-horizon limit and from the boundary in the high-temperature limit agree. 

The three-point coefficients are easily found by taking the inverse Laplace transform and using the saddle point approximation in the high-temperature limit for a primary operator $\mathcal{O}$
\begin{equation}
    \langle \mathcal{O} \rangle = \operatorname{Tr}_{\mathcal{H}_{S^1}} \mathcal{O} ~ e^{-\beta H} = \sum_i \langle i | \mathcal{O} | i \rangle ~ e^{-\beta E_i},
\end{equation}
where we trace over CFT states on the thermal circle and these coefficients are constrained by modular invariance.

The asymptotic behavior of the light-heavy-heavy coefficient is exponentially suppressed. The suppression depends on the central charge $c$ and conformal dimensions of operators $\mathcal{O}$ and $\chi$, which are light primary operators dual to AdS$_3$ bulk scalars $\phi_\mathcal{O}$ and $\phi_\chi$, with energy $E_\mathcal{O}, E_{\chi} \ll \frac{c}{12}$. To compute the averaged three-point function coefficient, the last ingredient we need is the density of states which is given by the Cardy formula in the large $E$ limit \cite{Cardy1, Cardy2, Cardy3}. In this limit, the final result of the averaged three-point function coefficient is
\begin{equation}
\label{eq:CFT3pt}
    \overline{\langle E|O| E\rangle}  \approx C_{\mathcal{O} \chi \chi} r^{\Delta_\mathcal{O}}_+ e^{-2 \pi \Delta_\chi r_+},
\end{equation}
which matches precisely in the bulk calculation done in Section \ref{section: KM-AdS}.

\subsubsection{Witten diagram calculation in AdS\texorpdfstring{$_3$}{}}
\label{section: KM-AdS}
\begin{figure}[h]
    \centering
    \includegraphics[width = 7cm]{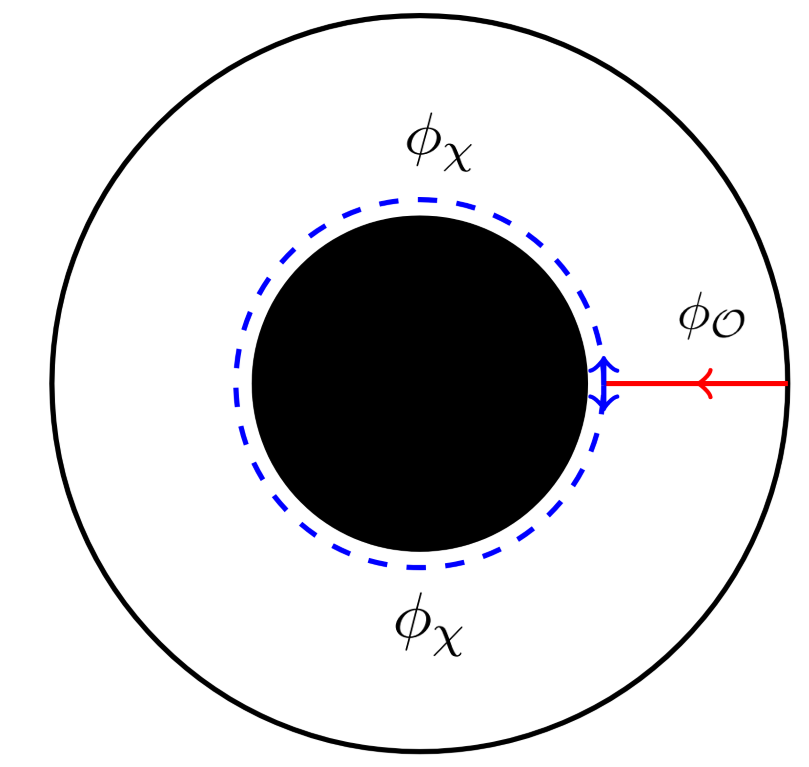}
    \caption{As illustrated in the Witten diagram for the regular BTZ black hole, a light scalar field $\phi_\mathcal{O}$ is emanated from the boundary to the horizon and splits into a pair of light fields $\phi_\chi$ that wrap around the horizon.}
    \label{fig:KMW}
\end{figure}

The bulk theory has an interaction term $\phi_\mathcal{O} \phi_\chi^2$ with coupling $C_{\mathcal{O} \chi \chi}$. The cubic vertex integrated over the entire BTZ AdS spacetime in Figure \ref{fig:KMW} is
\begin{equation}
\label{eq: KM_BTZ}
    \overline{\langle E|O| E\rangle}=C_{\mathcal{O} \chi \chi} \int d r d t_{E} d \phi ~ r ~ G_{b b}\left(r ; \Delta_{\chi}\right) G_{b \partial}\left(r, t_{E}, \phi ; \Delta_{\mathcal{O}}\right).
\end{equation}

We want to match the integral (\ref{eq: KM_BTZ}) in the large $r_+$ limit to the CFT result (\ref{eq:CFT3pt}) for the asymptotic three-point coefficient. The BTZ black hole is obtained from global AdS$_3$ via periodic identifications (i.e., AdS$_3 / \mathbb{Z}$ under $\phi \sim \phi + 2\pi$), which allows us to perform the method of images to obtain the BTZ black hole propagator from global AdS$_3$. The BTZ black hole propagator is
\begin{equation}
    G_{b b}\left(r, r^{\prime}\right)=-\frac{1}{2 \pi} \sum_{n = -\infty}^{\infty} \frac{e^{-\Delta \sigma_{n}\left(r, r^{\prime}\right)}}{1-e^{-2 \sigma_{n}\left(r, r^{\prime}\right)}},
\end{equation}
where $\sigma_n(r,r^\prime)$ is the geodesic distance between $r$ and the $n^\text{th}$ image of $r^\prime$. There is an apparent UV-divergent tadpole for the $n = 0$ term; however, this can be easily canceled by a local counterterm and other terms $n\neq 0$ are finite. As we will see in Section \ref{sec:4.2}, this type of UV divergence is absent in the case for $p$-adic BTZ black holes because of the form of the Green function, but a tadpole term remains present. Additionally, Kraus and Maloney considered the scalar fields to be massive: $E_\mathcal{O} \approx m_\mathcal{O} \gg 1, E_\chi \approx m_\chi \gg 1$ such that $m_\mathcal{O}, m_\chi \ll c$.

In the large $r_+$ limit, the averaged three-point coefficient is 
\begin{equation}
\label{eq: KrausMaloneywitten}
    \overline{\langle E|O| E\rangle} \approx C_{\mathcal{O} \chi \chi} r^{\Delta_\mathcal{O}}_+ e^{-2 \pi \Delta_\chi r_+}.
\end{equation}

\subsection{\texorpdfstring{$p$}{}-adic Witten diagram calculation}
\label{sec:4.2}
Previously, we reviewed that the $p$-adic BTZ black hole is constructed as a quotient space of the Bruhat-Tits tree and is visualized as a central polygon with a sub-Bruhat-Tits rooted tree attached to each vertex of the polygon. The central polygon is the horizon of the $p$-adic BTZ black hole with area $l = -ord_{p} (q) = \log_p|q|_p$ and $q$ is the generator of Schottky group $q^{\mathbb{Z}}$. Considering the construction of the $p$-adic BTZ black hole, we chose a new set of coordinates $(n,h)$ to parametrize the bulk points. The labels of vertices on the horizon, to which bulk points are attached (directly or indirectly), are represented by $n = 0,1, \cdots,l-1$. Whereas $h = 0,1, \cdots, \infty$ represents the number of edges between the attached central vertex and that bulk point.

Under this parametrization, in order to calculate the similar Witten diagram mentioned in \cite{KrausMaloney}, we replace the original integration over AdS space with a summation over all bulk points $(n,h)$ on the quotient space of the Bruhat-Tits tree
\begin{equation}
    \label{wittendiagram}
     \overline{\langle E|\mathcal{O}| E\rangle} \approx C_{\mathcal{O} \chi \chi} \sum_{(n,h)} d(n,h) G_{bb} (n,h;\Delta_{\chi}) G_{b \partial} (n,h;x, \Delta_\mathcal{O}),
\end{equation}
where $x\in \mathbb{Q}_p$ is the boundary coordinate of the operators $\mathcal{O}$, $\Delta_\chi$ and $\Delta_{\mathcal{O}}$ are scaling dimensions of operators $\chi$ and $\mathcal{O}$. $d(n,h)$ counts the number of vertices sharing the same coordinate $(n,h)$.

There are two different cases that we need to calculate separately. The first case is both the bulk and boundary points are attached to the same central vertex. The second case is both the bulk and boundary points are attached to different vertices. We denote the central vertex attached by the boundary point as vertex 0, such that these two cases are $n =0$ and $n\neq 0$.
\subsubsection{Propagators revisited in BTZ background}
\label{sec:revisit}
In Section \ref{sec:2}, we introduced the $p$-adic BTZ black hole as the quotient space $T_p / q^{\mathbb{Z}}$, which is different from the original Bruhat-Tits tree $T_p$. One obvious distinction is that the quotient space loses some global symmetries.\footnote{Global symmetries under action by the isometry group, e.g., $PGL\left(2,\mathbb{Q}_p\right)$ in the context of Bruhat-Tits trees. When we quotient $\mathbb{P}^1\left(\mathbb{Q}_p\right)$ by the Schottky group $q^{\mathbb{Z}}$, the isometry group is then broken to a subgroup of $PGL\left(2,\mathbb{Q}_p\right)$.} Remember that the normal Bruhat-Tits tree has a perfect homogeneity, and in principle, we could choose any local vertex to be a central point. However, the $p$-adic BTZ background certainly has some predetermined central vertices, which has been shown in Figure \ref{fig:p-adicBTZ} as vertices of the central polygon.

Given the global symmetry breaking, we should question whether the theory defined on the $p$-adic BTZ black hole would deviate from the normal Bruhat-Tits tree theory defined by the action (\ref{eq:action}), and more importantly, whether the propagators (i.e., Green functions as the main characters of Witten diagram calculation shown above) would also change. Fortunately, by observations, we find that even though the global symmetry is broken by a topological change, the local features of the graph are still preserved. In other words, the valency of each vertex is still $p+1$, same as on the Bruhat-Tits tree. Meanwhile, since the $p$-adic BTZ black hole is also an undirected graph with an infinite number of vertices, we should expect the action (\ref{eq:action}) to still be valid in the BTZ black hole background. However, when we compute the propagators, the equations of motion has sources inserted on some vertices. The symmetry loss of the BTZ black hole will also cause the symmetry loss to the solutions of these equations of motions. For instance, on the Bruhat-Tits tree, no matter where we insert the source, due to homogeneity of the tree, the solution will be homogeneous. However, in the BTZ black hole case, the depth of vertices, where we insert the source, from the horizon will indeed affect the solutions and subsequently the solutions will be different from those on a normal Bruhat-Tits tree.

One approach to compute the propagators in the background of an ordinary Euclidean BTZ black hole is the method of images \cite{Xi, KrausMaloney}, which will be demonstrated in the next subsection. Instead, we can also straightforwardly start from the solution to the equation of motion with a source insertion. This provides us a sanity check for the use of method of images. In general, due to the loss of symmetries, solving the equation of motion with sources inserted in arbitrary vertices on the $p$-adic BTZ is arduous, but we can still use the residual symmetries to evaluate a simple case.

Suppose we use the same action (\ref{eq:action}) for the $p$-adic BTZ background. Meanwhile, we restrict our calculations to the case where only one current source $J$ is coupled to the vertex $0$ on the horizon, without other source couplings. The equation of motion is then:
\begin{equation}
 \label{sourcedeom}
 \left(\square + m_p^2\right) \phi_{i} = \begin{cases}J & i = C_0\\
 0 & \text{otherwise}\end{cases},
\end{equation} 
yielding the propagator:
\begin{equation}
    \label{propagator}
    G_{bb}\left(C_0,a\right)=\frac{\phi_a}{J},
\end{equation}
where $\phi_a$ is the field value to an arbitrary vertex $a$ and $C_0$ represents the vertex $0$ on the horizon. 

We should mention that the solution depends on the specified boundary condition. In order to find the same class of solutions as those on the Bruhat-Tits tree, we specify the boundary condition:
\begin{equation}
\label{boundarycondition}
    \lim_{i\rightarrow\partial T_p} \phi_i = 0.
\end{equation}
For simplicity, we set the mass $m_p$ of the scalar field $\phi_i$ to be $0$. 

In Section \ref{sec:3}, we demonstrated a way to solve Laplace's equation by using linear recursion in the scalar fields. Here, we follow a similar technique. We denote the vertices on the horizon as $C_n$ where $n = 0,\cdots, l-1$. Consider one specific vertex $C_i$, the subtree rooted at $C_n$ is solved by using a recursion relation:
\begin{equation}
    (p+1)\phi_{h,n} = p\phi_{h+1,n}+\phi_{h-1,n},
\end{equation}
where the vertices on the subtree are parametrized by $h$, the depth of a vertex with respect to $C_i$. From Section \ref{sec:3}, we know the solution to this recursion relation is
\begin{equation}
    \phi_{h,n} = a+bp^{-h},
\end{equation}
where $a, b$ are two free variables that are fixed by the boundary conditions. We first enforce the boundary condition (\ref{boundarycondition}) to set $a = 0$, so $\phi_{h,n} = \phi_{C_n}p^{-h}$.

We also need to determine all the field values $\phi_{C_n}$ on the horizon. This requires us to use the recursive equations on the horizon for $n\neq0$:
\begin{equation}
    \label{eq: horizoneq}
    (p+1)\phi_{C_{n}}=\phi_{C_{n-1}}+\phi_{C_{n+1}}+\frac{p-1}{p}\phi_{C_{n}}
\end{equation}
The equation on vertex $0$ is modified by the source: 
\begin{equation}
\label{eq:source}
(p+1)\phi_{C_{0}}=\phi_{C_{l-1}}+\phi_{C_{1}}+\frac{p-1}{p}\phi_{C_{0}}+J.
\end{equation}

These linear equations can be solved either numerically or analytically. We demonstrate a simple example where $l = 3$ and obtain the following solutions to (\ref{eq: horizoneq}):
\begin{equation}
\label{eq:solution}
    \begin{aligned}
    \phi_{C_0} &= \frac{1}{p-\frac{1}{p}} \left(1+\frac{2}{p^3-1}\right)J\\
    \phi_{C_1} &= \phi_{C_2} = \frac{1}{p-\frac{1}{p}} \frac{p^2+p}{p^3-1}J.
    \end{aligned}
\end{equation}
In (\ref{eq:mass}), we gave a correspondence between the mass of a bulk scalar field and the scaling dimension of a boundary operator. For a massless scalar, the corresponding scaling dimension is $\Delta = 1$. Then we rewrite the propagators (\ref{eq:solution}) in a convenient way 
\begin{equation}
\label{eq: eomsolution}
\begin{aligned}
G_{bb}(C_0,C_0) &= \frac{\zeta_p(2\Delta)}{p^{\Delta}}\left(1+\frac{2}{p^{\Delta l}-1}\right)\\
G_{bb}(C_0,C_n) &= \frac{\zeta_p(2\Delta)}{p^\Delta}\frac{p^{n}+p^{l-n}}{p^{\Delta l}-1}.
\end{aligned}
\end{equation}
In the subsequent subsections, we will see directly that these results are consistent with the results given by method of images in \cite{Caltech} for both bulk-to-bulk and bulk-to-boundary propagators.

\subsubsection{\texorpdfstring{$n = 0$}{} case}
\begin{figure}[h]
    \centering
    \includegraphics[width = 8cm]{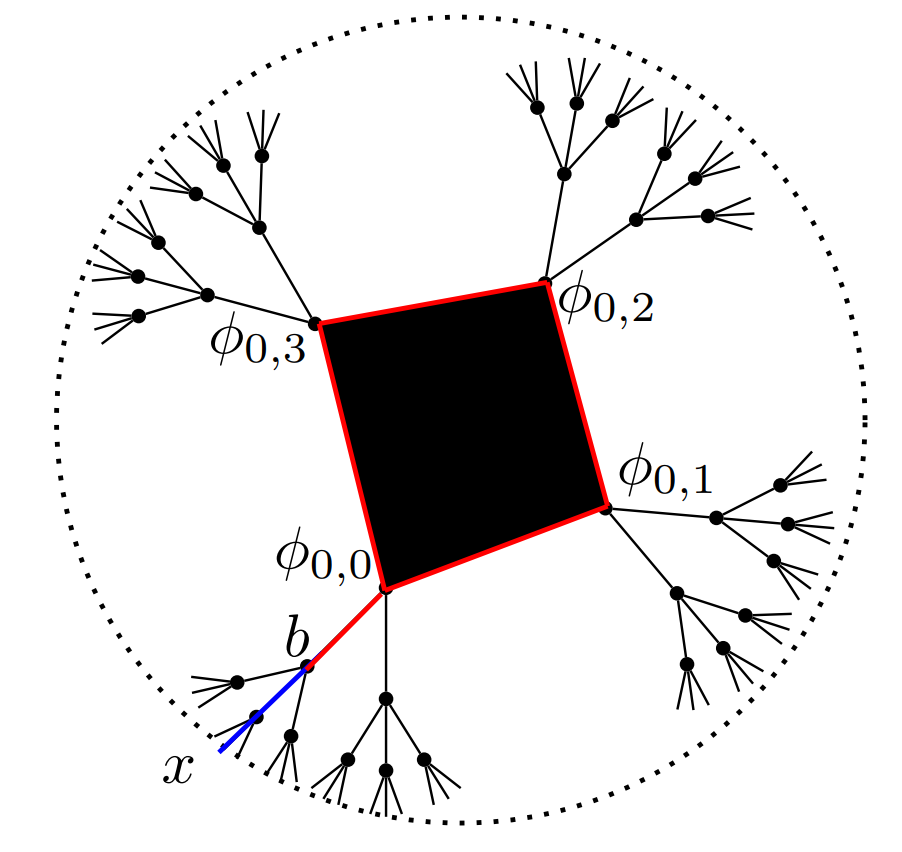}
    \caption{Witten diagram in the $p$-adic BTZ black hole ($p=3,l=4,n=0$). Red line: the bulk-to-bulk propagator. Blue line: the bulk-to-boundary propagator.}
    \label{fig:n0}
\end{figure}

For the $n =0$, the boundary point $x$ and the bulk point $b$ are in the same subtree rooted at, without loss of generality, the central vertex 0. The Witten diagram in Figure \ref{fig:n0} is what is needed to calculate the averaged three-point coefficient. 

To calculate this Witten diagram, we must determine two main factors: the bulk-to-bulk and bulk-to-boundary propagators. Since both fields $\chi$ and $\mathcal{O}$ are normal perturbative scalar fields, we directly derive the bulk-to-bulk propagator on the Bruhat-Tits tree by finding the tree Laplacian's Green function, which has a simple form\footnote{Here we omit the normalization factor $\frac{\zeta_{p}(2\Delta)}{p^\Delta}$ in \cite{Gubser1}.} \cite{Gubser1,Caltech}
\begin{equation}
    \label{green}
    G_{bb} (z,z_0;w,w_0)=p^{-\Delta_{\chi} d(z,z_0;w,w_0)},
\end{equation}
where the function $d(\cdot,\cdot)$ gives the geodesic distance. In the previous subsection, we provided a way to compute the Green function in $p$-adic BTZ background by solving the sourced equation of motion (\ref{eq:source}). In general, that approach is doable but complicated. Fortunately, the $p$-adic BTZ background is realized as the quotient space of the normal Bruhat-Tits tree, so we use the method of images to solve the equations given the solutions in the parent space. Following \cite{KrausMaloney}, we use the method of images to derive the bulk-to-bulk propagator from vertex $b$ to itself. Using the $(n,h)$ parametrization as mentioned before, we obtain
\begin{equation}
  \label{bulkbulk}
     G_{bb}(n,h) = p^{-\Delta_\chi d(b,b)}+2\sum_{i=1}^{\infty}p^{-2\Delta_\chi h}p^{-i\Delta_\chi l} = 1+\frac{2p^{-2\Delta_\chi h}}{p^{\Delta_\chi l}-1},
\end{equation}
where the summation is over all images of $b$ under the action of the Schottky group, and the index $i$ is regarded as the winding number around the horizon. Comparing this result with the solution (\ref{eq: eomsolution}) by setting $h = 0$, we see that the two results agree up to a normalization factor $\zeta_{p}(2\Delta)/p^\Delta$ we omitted in \eqref{green}. Notice that a constant $1$ appears in the bulk-to-bulk propagator. This is the tadpole term which usually causes divergence in the normal continuum AdS spacetime. Although it does not cause a divergence in our case, it is still unphysical. As shown in Figure \ref{fig:KMW}, the tadpole term does not probe the black hole geometry, because the $\phi_\chi$ loop in the tadpole case shrinks to a point instead of wrapping around the black hole horizon. Moreover, the same tadpole contribution also appears in the empty AdS calculation \cite{KrausMaloney}, which is about the semiclassical vacuum one-point function $\langle 0|\mathcal{O}|0\rangle$. In order to renormalize this one-point function, we need to cancel the tadpole term.\footnote{This does not contradict the fact that $\langle E|\mathcal{O}|E\rangle\neq0$ for $E\gg c/12$ due to Hawing radiation in \cite{KrausMaloney}.} Fortunately, this can be done by adding the ordinary local counterterm $\sum_i c_i \phi_i$ into the action, where $i$ is the label of bulk vertices. This counterterm mimics the $\int d^4x\, \phi(x)$ in the continuum case. The renormalized bulk-to-bulk propagator is:
\begin{equation}
    \label{bulkbulkrenormalized}
    G_{bb}^{\text{renorm}}(n,h) = \frac{2p^{-2\Delta_{\chi}h}}{p^{\Delta_{\chi}l}-1}
\end{equation}

The bulk-to-boundary propagator is derived from the bulk-to-bulk propagator by moving one point to the boundary.\footnote{This limiting process is safe here, but it would be naively wrong when one were to calculate two-point correlators, as explained in Section 4 of \cite{Gubser1}.} Notice that if we were to directly take this limit in (\ref{green}), it would vanish due to $d(z,z_0;w,w_0)\rightarrow \infty$. Therefore, we need to perform a regularization prescription as provided in Section 3 of \cite{Gubser1}. The bulk-to-boundary propagator on the Bruhat-Tits tree is derived via \cite{Gubser1}:
\begin{equation}
\label{Prescription}
G_{b\partial}(z,z_0;x) = \lim_{\delta_x \rightarrow 0} |\delta_x|_p^{-\Delta}G_{bb}(z,z_0;w,w_0).
\end{equation}
Given a bulk point $(w,w_0)$, we denote any boundary point which is reached by an oriented path $(z,z_0) \rightarrow (w,w_0)$ as $y$. The supremum of $|y-x|_p$ is denoted by $\delta_x$. When we move $(w,w_0)$ to the boundary point $x$, the limit is taken as $\delta_x \rightarrow 0$. Clearly, some prescription factor $|\delta_x|_p^{-\Delta} \rightarrow \infty$ is required so that the bulk-to-boundary propagator does not vanish. 

In \cite{Caltech, Zabrodin}, another regularization procedure is provided. Instead of taking the asymptotic limit of the bulk-to-bulk propagator, they regularized the geodesic distance. The main feature there is that Zabrodin defined $d_{reg}(C,x) = 0$ \cite{Zabrodin}, where $C$ is a vertex on the horizon and $x$ is the boundary point in the subtree rooted at $C$. By inspection, we realize that these two regularization methods are equivalent and both are consistent with the recursive derivation in Section \ref{sec:revisit}. We then say that these regularizations are anomaly-free under $PGL\left(\mathbb{Q}_p\right)$. Setting the geodesic distance of $d_{reg}(C,x) = 0$ is the same as factoring $p^{d(C,x) \Delta}$ out from the non-regularized bulk-to-boundary propagator \eqref{bulkbulkrenormalized} with one point at the asymptotic boundary. $p^{d(C,x)\Delta}\rightarrow \infty$ plays the same role as $|\delta_x|^{-\Delta}$.
Therefore, we freely choose one regularization approach and use the method of images to find the bulk-to-boundary propagator. The bulk-to-boundary propagator is given as \cite{Caltech}:
\begin{equation}
    G_{b\partial}(b,x) = p^{-\Delta d_{reg}(b,x)}+ \frac{2p^{-\Delta h}}{p^{\Delta l}-1}.
\end{equation}
For the $n=0$ case, we combine the two propagators to obtain the averaged three-point coefficient 

\begin{equation}
    \label{zero}
    \begin{aligned}
    \overline{\langle E|\mathcal{O}| E\rangle}_{n=0} &\approx C_{\mathcal{O}\chi\chi}\sum_{(0,h)}d(0,h)\left(p^{-\Delta_\mathcal{O}d_{reg}(b,x)}+ \frac{2p^{-\Delta_\mathcal{O}h}}{p^{\Delta_\mathcal{O}l}-1}\right)\frac{2p^{-2\Delta_\chi h}}{p^{\Delta_\chi l}-1},
    \end{aligned}
\end{equation}
where $d(0,h)$ denotes the degeneracy of vertices with the coordinate $(0,h)$. Notice that there is a unique path from the horizon vertex 0 to the boundary point $x$ as well as a unique intersection point between the path from the bulk point $b$ to the boundary point $x$ and the path from vertex 0 to $x$. In order to compute the summation, we introduce one more parameter $i$ to represent the intersection point between the two paths. Additionally, the parameter $i$ will parametrize the bulk point $b$. By using the parameters $(n,h,i)$, we rewrite the summation in terms of a nested geometrical series:
\begin{equation}
    \label{cal1}
    \begin{aligned}
    \overline{\langle E|\mathcal{O}| E\rangle}_{n=0} \approx &C_{\mathcal{O}\chi\chi}\sum_{i=0}^{\infty}\left(p^{\Delta_ \mathcal{O}i}\frac{2p^{-2\Delta_\chi i}}{p^{\Delta_\chi l}-1}+\sum_{h=i+1}^{\infty}(p-2)p^{h-i-1}p^{\Delta_{\mathcal{O}}(2i-h)}\frac{2p^{-2\Delta_\chi h}}{p^{\Delta_\chi l}-1}\right)\\
    &+ C_{\mathcal{O}\chi\chi}\frac{2}{p^{\Delta_{\mathcal{O}}l}-1}\frac{2\left(1+\frac{p-1}{p(p^{\Delta_\mathcal{O}+2\Delta_\chi -1}-1)}\right)}{p^{\Delta_\chi l}-1}\\
    =& C_{\mathcal{O}\chi\chi}\left[\frac{2\left(1+\frac{p-2}{p(p^{\Delta_\mathcal{O}+2\Delta_\chi -1}-1)}\right)}{(p^{\Delta_\chi l}-1)(1-p^{\Delta_{\mathcal{O}}-2\Delta_\chi})} +\frac{4\left(1+\frac{p-1}{p(p^{\Delta_\mathcal{O}+2\Delta_\chi -1}-1)}\right)}{\left(p^{\Delta_{\mathcal{O}}l}-1\right)\left(p^{\Delta_\chi l}-1\right)}\right].
    \end{aligned}
\end{equation}

In order to make the geometrical series converge for the above summations, we find inequalities between the scaling dimensions of operator $\mathcal{O}$ and $\chi$:
\begin{equation}
    \label{eq:dimension condition}
    \Delta_{\mathcal{O}}+2\Delta_{\chi} > 1, \quad \Delta_{\mathcal{O}}<2\Delta_{\chi}.
\end{equation}
The first inequality is automatically satisfied, as mentioned in Section \ref{sec:3}, we use the convention in \cite{Gubser1} that $\Delta = \Delta_{+} > 1/2$. The second inequality adds an extra constraint on the dimension of the operator $\mathcal{O}$. When $\Delta_\mathcal{O}$ is small enough, our calculation is well defined until $\Delta_{\mathcal{O}}$ saturates the inequality (\ref{eq:dimension condition}). Further regularization is required for this. However, the second inequality is only related to coefficients independent of the length of the horizon $l$. Therefore, it will not affect asymptotic behavior for large $l$.
\subsubsection{\texorpdfstring{$n\neq 0$}{} case}
\begin{figure}[ht]
    \centering
    \includegraphics[width = 8cm]{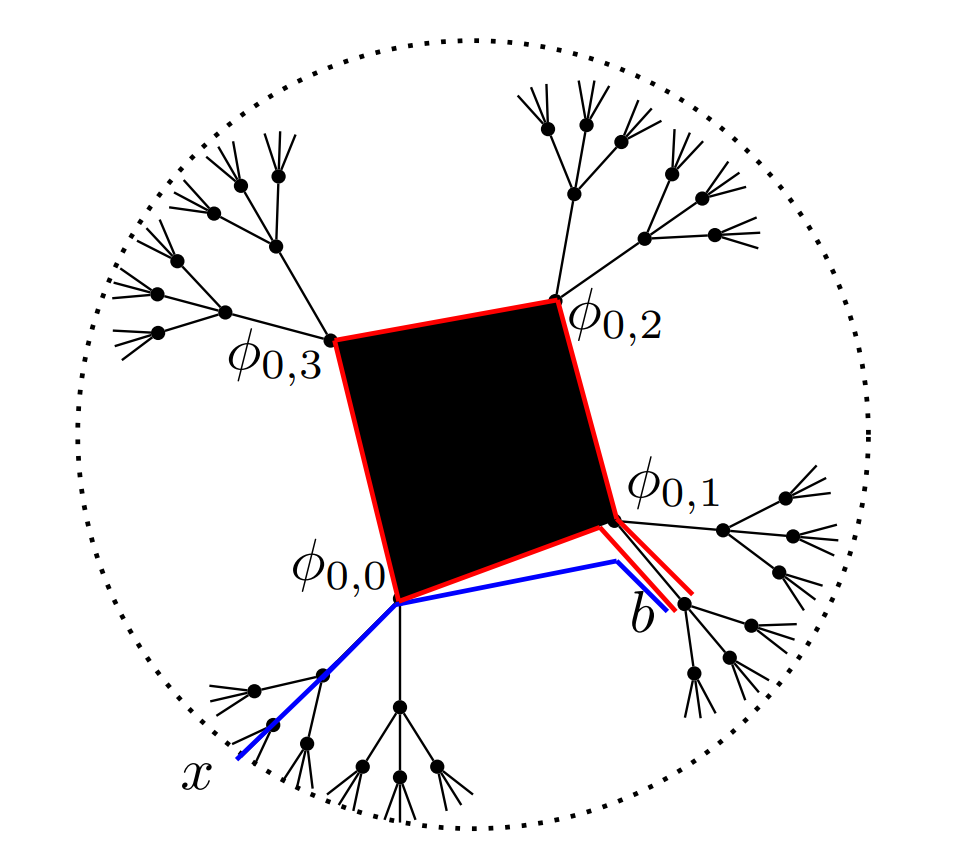}
    \caption{Witten diagram in the $p$-adic BTZ black hole ($p = 3,l=4, n\neq0$). Red line: the bulk-to-bulk propagator. Blue line: the bulk-to-boundary propagator.}
    \label{fig:nn0}
\end{figure}

This case is simpler than $n=0$. The Witten diagram is now visualized as Figure \ref{fig:nn0}.
%\newpage
The bulk-to-bulk propagator is the same as (\ref{bulkbulk}), while the bulk-to-boundary propagator is slightly different \cite{Caltech}. We evaluate the summations (\ref{wittendiagram}) as follows:
\begin{equation}
    \label{cal2}
    \begin{aligned}
    \overline{\langle E|\mathcal{O}| E\rangle}_{n\ne 0}&\approx C_{\mathcal{O}\chi\chi}\sum_{n=1}^{l-1}\sum_{(n,h)}d(n,h)\frac{p^{\Delta_{\mathcal{O}}(l-n)}+p^{\Delta_{\mathcal{O}}n}}{p^{\Delta_{\mathcal{O}}l}-1}p^{-\Delta_{\mathcal{O}}h}\frac{2p^{-2\Delta_\chi h}}{p^{\Delta_\chi l}-1}\\
    &= C_{\mathcal{O}\chi\chi}\sum_{n=1}^{l-1} \frac{p^{\Delta_{\mathcal{O}}(l-n)}+p^{\Delta_{\mathcal{O}}n}}{p^{\Delta_{\mathcal{O}}l}-1}\frac{2\left(1+\frac{p-1}{p\left(p^{\Delta_\mathcal{O}+2\Delta_\chi -1}-1\right)}\right)}{p^{\Delta_\chi l}-1}\\
    &= 4 C_{\mathcal{O}\chi\chi} \frac{p^{\Delta_{\mathcal{O}}l}-p^{\Delta_\mathcal{O}}}{(p^{\Delta_{\mathcal{O}}}-1)(p^{\Delta_{\mathcal{O}}l}-1)}\frac{1+\frac{p-1}{p(p^{\Delta_\mathcal{O}+2\Delta_\chi -1}-1)}}{p^{\Delta_\chi l}-1}
    \end{aligned}
\end{equation}
In this case, we have no issues for divergences in the geometrical series. The only requirement $\Delta_{\mathcal{O}}+2\Delta_{\chi}>1$ has already been shown to be satisfied in previous subsection.

After having the contributions from both $n=0$ and $n\ne0$ cases, we then get the full expression for the averaged three-point coefficient:
\begin{equation}
    \label{eq:wittend}
    \begin{aligned}
    \overline{\langle E|\mathcal{O}| E\rangle} &= \overline{\langle E|\mathcal{O}| E\rangle}_{n=0} + \overline{\langle E|\mathcal{O}| E\rangle}_{n\ne 0}\\
    &= 2C_{\mathcal{O}\chi\chi}\left[\frac{1+\frac{p-2}{p\left(p^{\Delta_\mathcal{O}+2\Delta_\chi -1}-1\right)}}{\left(p^{\Delta_\chi l}-1)(1-p^{\Delta_{\mathcal{O}}-2\Delta_\chi}\right)}+2\frac{1+\frac{p-1}{p\left(p^{\Delta_\mathcal{O}+2\Delta_\chi -1}-1\right)}}{\left(p^{\Delta_\chi l}-1\right)(p^{\Delta_{\mathcal{O}}}-1)}\right]\\
    &= C^\prime_{O\chi\chi}\frac{1}{p^{\Delta_{\chi}l}-1} \xrightarrow{l\rightarrow\infty} \boxed{C^\prime_{\mathcal{O}\chi\chi}p^{-\Delta_{\chi}l}} 
    \end{aligned}
\end{equation}
The coefficient $C^\prime_{\mathcal{O}\chi\chi}$ is viewed as the three-point coefficient $\langle \chi|\mathcal{O}|\chi\rangle$ and absorbs all factors independent of the horizon length $l$. In the last line, we show that as $l\rightarrow \infty$, the averaged three point coefficient $\overline{\langle E|\mathcal{O}|E\rangle}$ has an asymptotic behavior with an exponential dependence on horizon length $l$. 

\subsection{Physical implications}
By comparing (\ref{eq: KrausMaloneywitten}) with our average three-point coefficient (\ref{eq:wittend}), we find that $l$ is a $p$-adic counterpart of $2\pi r_{+}$ which is the outer horizon area of a normal BTZ black hole. If we rewrite $p^{-\Delta_{\chi}l }$ as $e^{-\ln{p}\Delta_{\chi}l}$, it will become reminiscent of $e^{-2\pi\Delta_\chi r_{+}}$ in (\ref{eq: KrausMaloneywitten}). However, in the $p$-adic case, we miss a counterpart to $r_{+}^{\Delta_{\mathcal{O}}}$. This term can be realized as the dominant normalization factor $r_{+}^{\Delta}$ in the bulk-to-boundary propagator of a normal Euclidean BTZ black hole \cite{KrausMaloney}. Physically, it can be thought as the horizon radius being probed by the particle $\mathcal{O}$ entering the bulk from the boundary. In a continuum spacetime, the horizon radius is well defined by a Riemannian metric. In the $p$-adic BTZ graph, the black hole is represented by a polygon, which has no radius measured by the graph's metric. Therefore, when the particle $\phi$ is emanated into the $p$-adic BTZ background, it cannot measure the radius of horizon as well as unable to create a term including the horizon radius and its scaling dimension $\Delta_{\mathcal{O}}$. 

In Section \ref{sec:3}, we provide calculations on the $p$-adic CFT partition function and the density of states. However, our knowledge is primitive on the modular transformations for $p$-adic genus-1 Tate curves. If we understand the modular transformation, we can obtain the averaged three point coefficient entirely from the CFT side. Our averaged three-point coefficient displays an unconventional feature compared to the Euclidean BTZ case to then indicate that the $p$-adic modular transformation is non-trivial. We will explore this aspect further in future work.

On the other hand, our geometries only capture AdS length scale effects and miss contributions coming from ``small loops'' which can be non-trivial, as stressed in \cite{pAdS}. It would be nice to see if the bulk calculation can be reproduced from the $p$-adic CFT side.

\section{\texorpdfstring{$p$}{}-adic representations }
\label{sec:5}
The proposed $p$-adic AdS/CFT correspondence provides tools to understand some features of the boundary $p$-adic CFT. However, for a general (not necessarily holographic) CFT, the bulk/boundary duality cannot allow us to study the theory comprehensively. In order to fully solve a general $p$-adic CFT, a Hilbert space interpretation is necessary. For example, independent of the bulk calculations in Section \ref{sec:4}, if one wants to compute the one-point function of a primary operator $\mathcal{O}$ of $p$-adic CFT, analogous to $\langle \mathcal{O}\rangle_{\tau}=\operatorname{Tr}_{\mathcal{H}} \mathcal{O} q^{L_{0}-\frac{c}{24}}\overline{q}^{\overline{L}_0 - \frac{c}{24}}$ with $q \equiv e^{2 \pi i \tau}$ in an ordinary 2d CFT, one would hope to have $p$-adic exponentials and analogues of Virasoro generators $L_0$ and $L_{\pm 1}$ as well as Verma modules.

In a normal quantum field theory, its Hilbert space could be constructed based on representations of Lie algebra $\mathfrak{g}$ associated to the global or internal symmetry group $G$. In a $p$-adic CFT, the global symmetry group is $PGL\left(2,\mathbb{Q}_p\right)$, so analogous to ordinary CFTs, we should study the Lie algebra representations of this group. Typically, a $p$-adic CFT is a quantum field theory with complex-valued (or real-valued) fields over $\mathbb{Q}_p$, which restricts our interests to a vector space $V$ over $\mathbb{C}$ as the representation space. In \cite{melzer}, Melzer showed the nonexistence of local derivatives over $\mathbb{Q}_{p}$. Meanwhile, in the usual context of the Lie algebra, we can always define the exponential map $\exp: \mathfrak{g}\rightarrow G$, while in the $p$-adic case, the exponential function of $p$-adic numbers does not converge nicely \cite{Koblitz}. Moreover, it is a totally disconnected group, its corresponding would-be Lie algebra ``$\mathfrak{pgl}\left(2,\mathbb{Q}_p\right)$'' does not exist. The Virasoro-like local conformal algebra never shows up.

Although we cannot find any suitable complex representation of Lie algebra, we still hope to directly study representations of the global conformal group $PGL\left(2,\mathbb{Q}_p\right)$. Actually, several recent papers indeed explore the power of group representations in quantizing a theory, such as Jackiw-Teitelboim gravity \cite{JT} and spinors on AdS$_2$ \cite{Kitaev}, in that their Hilbert spaces can be partially\footnote{Some Lie algebra data such as quadratic Casimir are still required.} defined by group representations of $SL(2,\mathbb{R})\times U(1)/\mathbb{Z}$ or $\widetilde{SL(2,\mathbb{R})}$. There are numerous types of $PGL\left(2,\mathbb{Q}_p\right)$ representations, so we add some reasonable assumptions to narrow down our search list. Since all $p$-adic CFTs are unitary \cite{melzer}, we expect a suitable representation to also be \textit{unitary}. Notice that any unitary irreducible representations (irreps) of $PGL\left(2,\mathbb{Q}_p\right)$ naturally induces a $GL\left(2,\mathbb{Q}_p\right)$ unitary irreps, so that we could study unitary irreps of $GL\left(2,\mathbb{Q}_p\right)$ and canonically restrict them onto the subgroup $PGL\left(2,\mathbb{Q}_p\right)$. Another advantage to study $GL\left(2,\mathbb{Q}_p\right)$ comes from the classification theorem on all of its unitary irreps. In the rest of this section, we will analyze this theorem and evaluate the suitability of all unitary irreps as physical Hilbert spaces over $\mathbb{C}$ of $p$-adic CFTs. Rather than being mathematically rigorous, we provide sufficient amount of evidence following \cite{L-function}. 

\subsection{Troubles with Lie algebras}
\label{sec:trouble}
 The usual Iwasawa decomposition\footnote{For real semisimple Lie groups, it is defined via their Lie algebras.} still holds for TDLC (totally disconnected locally compact) groups (as introduced in \footreff{foot:TDLC}) of our interests, such as $SL\left(2,\mathbb{Q}_p\right)$ or $PGL\left(2,\mathbb{Q}_p\right)$. Any element of $SL\left(2,\mathbb{Q}_p\right)$, the commutator subgroup of $GL\left(2,\mathbb{Q}_p\right)$, as presented in \cite{Caltech}, can be decomposed into a product of special conformal transformation, rotation, dilatation, and translation as shown respectively:
\begin{equation}
    \left(\begin{array}{cc}{p^{m} a} & {b} \\ {c} & {p^{-m} a^{-1}(1+b c)}\end{array}\right)=\left(\begin{array}{cc}{1} & {0} \\ {c p^{-m} a^{-1}} & {1}\end{array}\right)\left(\begin{array}{cc}{a} & {0} \\ {0} & {a^{-1}}\end{array}\right)\left(\begin{array}{cc}{p^{m}} & {0} \\ {0} & {p^{-m}}\end{array}\right)\left(\begin{array}{cc}{1} & {b p^{-m} a^{-1}} \\ {0} & {1}\end{array}\right),
\end{equation}
where $a,b,c \in \mathbb{Q}_p$ and $|a|_p = 1$. The decomposition of $PGL\left(2,\mathbb{Q}_p\right)$ is similar, but up to a $\pm$ sign on the total determinant.\footnote{Each sign sector is similar to a connected component of the usual Lorentz group $SL(2,\mathbb{C})$. For the Iwasawa decomposition of $GL\left(2,\mathbb{Q}_p\right)$, see Proposition 4.2.1 in \cite{L-function}.}

One might believe that the exponential map from Lie algebras to the usual matrix group $GL(n,\mathbb{C})$ works for $p$-adic groups as well, but this is unfortunately incorrect. Indeed, one could define a tangent space and Lie algebra functor near the identity of $SL\left(2,\mathbb{Q}_p\right)$ \cite{group}, but the total disconnectedness of the group poses a serious problem. For $z \in \mathbb{Q}_p$, the $p$-adic exponential is defined as
\begin{equation}
    \exp (z)\equiv\sum_{n=0}^{\infty} \frac{z^{n}}{n !},
\end{equation}
which diverges at the identity since the radius of convergence is $|z|_{p}<p^{-1 /(p-1)}$.

Another fundamental reason is as follows. Having a tangent space $T_{e}$ at the identity $e$ of the group analytical manifold $PGL\left(2,\mathbb{Q}_p\right)$, it is natural to introduce a one-parameter subgroup $\phi:\mathbb{F} \rightarrow PGL\left(2,\mathbb{Q}_p\right)$, where $\mathbb{F}$ is a number field, which is $\mathbb{R}$ for the usual connected Lie groups. $\phi$ also defines vector fields on the group manifold. Moreover, one can build an exponential map to recover local features of the group via Lie algebra. Thus, 
\begin{equation}
    \exp: \mathbb{F} \rightarrow PGL\left(2,\mathbb{Q}_p\right), ~ t\mapsto e^{tL},
\end{equation}
with the Lie algebra element $L\in T_{e}$. Consequently, we must select the correct number field $\mathbb{F}$ for the parameter $t$. $\mathbb{R}$ is ruled out due to the disconnectivity of $p$-adic groups. The only remaining candidate is $\mathbb{Q}_p$. However, another issue arises when we consider the representation of $PGL\left(2,\mathbb{Q}_p\right)$. With the representation space $V$ over $\mathbb{C}$, we expect for any $g\in PGL\left(2,\mathbb{Q}_p\right)$, its image $\pi(g)\in GL(V)$ whose entries are all $\mathbb{C}$-valued. From the exponential map, we see that the image can always be written as
\begin{equation}
    \pi(g) = e^{tM},
\end{equation}
where $M = \pi(L)$ is the image of the Lie algebra element $L$.\footnote{The Lie algebra elements are complex-valued matrices.} However, $t$ and entries of $M$ are in different number fields with different norms, so the multiplication $tM$ is forbidden, and the Lie algebra representation over $\mathbb{C}$ cannot exist. Since there is no well-defined Lie algebra or ``infinitesimal generators'' for the dilatation operator $L_0$, it is a little bit dubious to discuss a ``state-operator correspondence'' used in \cite{Ling-Yan} and hence radial quantization.

However, we should also mention the possibility to construct a Lie algebra representation over $\mathbb{Q}_p$ \cite{Verma,BGG}. In these cases, we need to consider Hilbert spaces over $\mathbb{Q}_p$ though, which is inconsistent with Melzer's axioms for $p$-adic CFTs.

\subsection{Admissible representations of \texorpdfstring{$GL\left(2,\mathbb{Q}_p\right)$}{} in general}

Due to the troubles on the existence of $p$-adic Lie algebra, we turn our attention to group representations. The unitarity of $p$-adic CFTs directs us to unitary representations, which are subspaces of the physical Hilbert spaces as usual.

We start from the representation vector space $V$ over $\mathbb{C}$. Let $GL(V)$ be the space of all automorphisms of $V$, and $\pi$ be the following homomorphism
\begin{equation}
 \pi : GL\left(2,\mathbb{Q}_p\right) \rightarrow GL(V).
\end{equation}
Given an inner product\footnote{Formally speaking, this is a positive-definite Hermitian form, and is equivalent to the usual pairing between bras and kets.} $(\cdot,\cdot)$ on $V$, a \textit{unitary} representation $(\pi,V)$ of $G$ satisfies
\begin{equation}
\label{eq:unitary}
    (\pi(g) \cdot v, \pi(g) \cdot w)=(v, w), \quad\forall g \in G,\, v, w \in V.
\end{equation}
Clearly, this definition is \textit{relative} to the inner product prescribed on $V$. If $V$ is not equipped with an inner product that makes $(\pi,V)$ unitary, one can ask if $(\pi,V)$ can be made unitary by choosing an appropriate inner product \cite{L-function}. To this end, a representation $(\pi,V)$ is defined as \textit{unitarizable} if there exists\footnote{Existence of inner products is the first thing to look for in group representations. For example, for $SL(2,\mathbb{R})$ in JT gravity, among four types of its unitary irreps, trivial and complementary series representations are not considered \cite{JT} due to the lack of inner product. All of its finite-dimensional representations are non-unitary as well \cite{Kitaev}.} an inner product $(\cdot,\cdot)$ such that (\ref{eq:unitary}) holds. Moreover, it is straightforward to turn a unitary representation $V$ into a complete metric space \cite{Hilbert,Hilbert2}, and therefore a Hilbert space; in fact, the space of unitary admissible representations of $GL\left(2,\mathbb{Q}_p\right)$ is a proper subspace of the space of $\mathbb{C}$-Hilbert representations of $GL\left(2,\mathbb{Q}_p\right)$. Notice that inner products here do not rely on the dual (or \textit{contragredient}) representation of $V$.

We further assume that we are dealing with irreps. According to the admissibility theorem\footnote{The original Harish-Chandra's admissibility theorem \cite{Harish-Chandra1,Harish-Chandra2,Harish-Chandra3,Harish-Chandra4} only works for real reductive Lie groups.}, all unitary irreps of a $p$-adic reductive group such as $GL(2,\mathbb{Q}_p)$ \cite{Whittaker} are admissible, so we only consider admissible ones. This is also empirically reasonable, because at least for real and complex Lie groups, their irreps naturally appearing in PDEs, geometry, number theory and physics are all admissible \cite{natural}. The admissibility theorem was originally proved in \cite{Berstein1} and later illustrated in \cite{Bernstein}\footnote{In this set of lecture notes, all adjectives ``irreducible'' should be interpreted in the category of unitary representations.} (Section II.2.2). These were recently improved upon to work for more general TDLC groups \cite{First} and \cite{Admissibility1} (Corollary 6.30). For a rigorous definition of an \textit{admissible} representation, we refer the reader to these textbooks \cite{L-function,Admissibility1, Bernstein}. Furthermore, a smooth irrep is admissible \cite{L-function} (Theorem 6.1.11). 

\begin{comment}
Now to be complete, we present the definition of an admissible representation. An \textit{admissible} representation $(\pi, V)$ of $G$ requires that the subspace of $V$ fixed by any compact open subgroup of $G$ is finite-dimensional \cite{L-function,Admissibility1, Bernstein}. It also has to be \textit{smooth}, meaning that for $v \in V$, the function
\begin{equation}
    \left(\begin{array}{ll}{a} & {b} \\ {c} & {d}\end{array}\right) \mapsto \pi\left(\left(\begin{array}{ll}{a} & {b} \\ {c} & {d}\end{array}\right)\right) \cdot v, \quad\forall \left(\begin{array}{ll}{a} & {b} \\ {c} & {d}\end{array}\right) \in G L\left(2, \mathbb{Q}_{p}\right)
\end{equation}
is smooth, i.e., locally constant\footnote{It is absent on usual Lie groups, such as $SU(2)$.} \cite{L-function, autreps, UCLA}. Furthermore, a smooth irrep is admissible \cite{L-function} (Theorem 6.1.11). Dual representations of admissible representations are all admissible \cite{L-function}.

Finally we summarize the relations between various $GL\left(2,\mathbb{Q}_p\right)$ representations in Figure \ref{fig:Venn}. Automorphic representations are not considered at all, because they are adelic over all prime numbers.

\begin{figure}[h!]
    \centering
    \includegraphics[width = 9cm]{Venn.PNG}
    \caption{Relations between different types of representations for $GL\left(2,\mathbb{Q}_p\right)$. \haoyu{May need color code?}}
    \label{fig:Venn}
\end{figure}
\end{comment}
\subsection{Finite-dimensional admissible representations}
We start our discussion on finite-dimensional admissible irreps. These representations appear reasonable at first sight because they are consistent with the absence of descendants in $p$-adic CFTs. This is also reasonable especially when there are only a finite number of primaries. However, all finite-dimensional smooth irreps of $GL\left(2,\mathbb{Q}_p\right)$ are trivial in the sense that they are one-dimensional complex vector spaces such that the images of $GL\left(2,\mathbb{Q}_p\right)$ act as scalar multiplication.
\begin{comment}
as stated below \cite{L-function}.

\textbf{Theorem} \emph{Let $(\pi, V)$ be a finite-dimensional \textbf{smooth irrep} of $GL\left(2,\mathbb{Q}_p\right)$, then $V \cong \mathbb{C}$ and $\exists$ a multiplicative character $\omega$: $\mathbb{Q}^\times_p \rightarrow \mathbb{C}^\times$ such that $\pi(g) \cdot v = \omega (\text{det}\,g) \cdot v ~ \forall ~ g \in GL\left(2,\mathbb{Q}_p\right), v \in V$, where det is the usual determinant.}
\end{comment}

For the group $SL\left(2,\mathbb{Q}_p\right)$, its linear character is 1. On the other hand, $PGL\left(2,\mathbb{Q}_p\right)$ consists of group elements of $GL\left(2,\mathbb{Q}_p\right)$ identified up to a scalar factor so that the linear character $\omega$ must be constant on the determinant in order to be consistent with this identification. Since $\omega$ is trivial, the dilatation transformation cannot be realized in this finite-dimensional admissible representation. Hence it is \emph{not} a desirable physical Hilbert space. However, it would be interesting to see if an \textit{ensemble} of primaries can be viewed as a tensor product of one-dimensional representations.

One of the simplest examples is presented in Section 4.1 of \cite{Caltech}, the free boson on the boundary is viewed as a scalar representation of $PGL(2,\mathbb{Q}_p)$, and conformal dimensions of $\phi$ and Vladimirov derivative of $\phi$ are 0 and 1. However, we hope for more. One hint may come from the recent work on Green's functions of Vladimirov in the context of $p$-adic holography \cite{huang2020green}.

\subsection{Infinite-dimensional admissible representations}
According to the Langlands-like classification theorem \cite{L-function}, there are three classes of infinite-dimensional admissible representations for $GL\left(2,\mathbb{Q}_p\right)$: supercuspidal, principal series, and special.\footnote{All of them enjoy so-called Kirillov models and Whittaker models, which we will not explain or pursue for now. For an accessible exposition on Whittaker models, see these notes \cite{Whittaker}.} Certainly, all of them contain non-unitary cases which do not fall into this classification, and those non-unitary cases are not of physical interests, because $p$-adic CFTs satisfying Melzer's axioms are automatically unitary. Nevertheless, we will introduce their unitarity-independent definitions, and save unitarity-specific definitions to future work. In order to present the classification, we need to introduce the following object first.

\textbf{Definition} For an infinite-dimensional representation $(\pi,V)$ and a unipotent subgroup $N=\left\{\begin{pmatrix}
    1 & *\\0 & 1
    \end{pmatrix}\Bigg| *\in\mathbb{Q}_p\right\}$, consider the subspace
\begin{equation}
    V_N=\{\pi(n)v-v|n\in N,v\in V\},
\end{equation}
then the quotient
\begin{equation}
    V^N\equiv V/V_N
\end{equation}
is called the \textit{Jacquet module} of $V$. The classification of infinite-dimensional admissible representations is completely encoded by the dimension of $V^N$, which is at most two \cite{autreps}. When $\text{dim}_{\mathbb{C}} \,V^N=0,1,2$, the representation is supercuspidal, special or principal series, respectively \cite{Jacquet}. Incidentally, $V^N$ also vanishes for finite-dimensional admissible representations.

For usual 2d CFTs, states with different Virasoro levels are orthogonal and obviously span an infinite-dimensional representation of the Virasoro algebra. Then in $p$-adic CFTs, one naively would think that different vectors in the representation space $V$ have different energy levels. However, since we lack the necessary Casimir operators and algebra structure to define physical observables and quanta for the states, the realization of energy levels in a group representation is still mysterious.

Below are nontechnical descriptions of the classification of infinite-dimensional admissible representations.

\subsubsection*{Principal series and special representations}
Principal series representations arise commonly in physics for non-compact semisimple Lie groups, and they are also present for $GL\left(2,\mathbb{Q}_p\right)$.
According to Jacquet--Langlands \cite{JacquetLanglands}, this representation becomes reducible if the characters obey $\chi_1\chi_2^{-1}=|\cdot|^{\pm1}$. If $\chi_1\chi_2^{-1}=|\cdot|^{-1}$, then the vector space $V(\chi_1,\chi_2)$ contains a 1d invariant subspace $W$ such that $V(\chi_1,\chi_2)/W$ is an irrep called \textit{special representation}; if $\chi_1\chi_2^{-1}=|\cdot|$, then $V(\chi_1,\chi_2)$ contains a 1d admissible subspace also called \textit{special representation}.

\subsubsection*{Supercuspidal representations}

If the Jacquet module $V^N$ vanishes, then $(\pi,V)$ is called a \textit{supercuspidal representation.\footnote{The adjective ``super'' stands for the $p$-adic version of ``cuspidal'' in the finite field $\mathbb{F}_q$ case \cite{Woodbury}, which is presented in Appendix \ref{app:induction}. For an equivalent definition in terms of integrals, see Section 6.13 of \cite{L-function}. Equivalently, any irrep of $GL\left(2,\mathbb{Q}_p\right)$ which is not a subrepresentation of any representation induced from the Borel subgroup is \emph{supercuspidal}.} } Although this one-line definition looks innocent, they are in general notoriously difficult to construct, and we present the simplest case via the so-called ``\textit{compact induction}'' in Appendix \ref{app:induction}. We will use quite qualitative phrases in this short subsection.

However, supercuspidal representations are mathematically desirable due to its handful of nice properties. They are \textit{the} ``native'' representations of $GL\left(2,\mathbb{Q}_p\right)$, because other admissible representations can all be constructed from them, by inducing a representation $\rho=(\rho_1,\rho_2)$ of a parabolic subgroup $P=MN$, where $\rho_i$ is a supercuspidal representation of $GL_1\left(\mathbb{Q}_p\right)$. Another feature is that they have nicer inner products than the other two infinite-dimensional representations \cite{autreps}. %, i.e., a character of $\mathbb{Q}_p^{\times}$, and the Levi subgroup $M\simeq GL_1\left(\mathbb{Q}_1\right)\times GL_1\left(\mathbb{Q}_1\right)\simeq \mathbb{Q}_p^{\times}\times \mathbb{Q}_p^{\times}$.

They are also the most well-behaved representations of $GL\left(2,\mathbb{Q}_p\right)$, i.e., that they behave much like representations of a compact group \cite{Taibi}. Finally, in familiar terms for
$SL(2,\mathbb{R})$, supercuspidal and special representations are analogues of $SL(2,\mathbb{R})$ ``discrete series'' for $GL\left(2,\mathbb{Q}_p\right)$.

\subsection{Key signature for physical representations}
In previous subsections, we enumerated all candidate representations for the $p$-adic CFT Hilbert space. Although we made cogent arguments on the nonexistence of conformal algebra and triviality of finite-dimensional admissible representations, there are still three classes of infinite-dimensional irreps remaining. There is no simple reasoning we could present to determine which one of them is the most suitable physical representation, and the difficulty of explicit construction of supercuspidal representations makes the computation over it tough. Fortunately, we find an important signature which could show clues as to which are true physical representations.

In the Virasoro character formula for normal chiral CFT on a torus $\chi(q)=\text{Tr}_{\mathcal{H}}q^{L_0-\frac{c}{24}}$, $q$ is related to the modulus of $T^2$ torus via $q = e^{2\pi i \tau}$. However in Section \ref{sec:BTZ}, we saw the impossibility of defining a $p$-adic modulus $\tau\in \mathbb{Q}_p$. Moreover, the dilatation generator $L_0$ does not exist as discussed in Section \ref{sec:trouble}, so the ordinary Virasoro character apparently makes no sense in $p$-adic CFTs. Nevertheless, $q^{L_0-\frac{c}{24}}$ viewed as a whole can be interpreted as the representation of the dilatation transformation:
\begin{equation}
    \left(\begin{array}{cc}{q^{\frac{1}{2}}} & {0} \\ {0} & {q^{-\frac{1}{2}}}\end{array}\right),
\end{equation}
which is exactly the same as the Schottky parameter in (\ref{eq:qmatrix}). Meanwhile, a genus-1 curve over $\mathbb{Q}_p$ was similarly constructed via $p$-adic Schottky group $q^{\mathbb{Z}}, q \in \mathbb{Q}_p^{\times}$. Intuitively, we could generalize the Virasoro character to $p$-adic CFTs by considering the image of the Schottky group generator under a $GL\left(2,\mathbb{Q}_p\right)$ representation $(\pi,V)$, and using the new character to write down an analogous partition function for genus-1 $p$-adic CFT:
\begin{equation}
\label{eq:sign}
    Z_{p-\text{adic}\operatorname{CFT}} = \operatorname{Tr}_{V} \pi\left[\left(\begin{array}{cc}{q^{\frac{1}{2}}} & {0} \\ {0} & {q^{-\frac{1}{2}}}\end{array}\right)\right],
\end{equation}
where the trace function always exists because $GL\left(2,\mathbb{Q}_p\right)$ is a TDLC group \cite{Admissibility1}. One thing worth looking at is to define a bounded-from-below $V$ in terms of the Jacquet module.

In Section \ref{sec:3} we have explicitly calculated $p$-adic CFT partition functions from bulk path integral. In principle, we could check results there against (\ref{eq:sign}) for all three classes of infinite-dimensional admissible representations. This check would yield a key signature of physical representations $\mathcal{H}$, and may also demystify the connections between $GL\left(2,\mathbb{Q}_p\right)$ representations and Chebyshev polynomials. Another ambitious thought is to apply group representations to possibly classify $p$-adic CFTs, just like ordinary minimal models, etc.

\section{Summary and Outlook}
 \label{sec:6}
We end with a summary of our results and several open questions for future exploration.
\subsection{Discussion}
In this paper, we found the density of states of genus-1 $p$-adic BTZ black holes in the isotropic sector. Avoiding the assumption on the existence of a state-operator correspondence, we provide a new way to calculate the genus-1 $p$-adic BTZ black hole partitions function via linear recurrence in scalar fields on vertices. Regarding both accounts, we have shown several similarities to their continuum analogues, but still realized features from Melzer's axioms for \textit{non-Archimedean} CFTs. 

Our analytical study on density of states in the high-temperature limit suggest that scalars in BTZ background obey a Bekenstein-Hawking-like area law and the results are analogous to the semiclassical genus-one partition function by Maloney and Witten \cite{MaloneyWitten}. However, one subtlety with our results are that they are unstable when $p=3$. Possibly, this might be explained from our semi-classical analysis omitting gravitational contributions. Including gravitational effects for $p$-adic AdS/CFT was proposed by \cite{Edge} via edge length dynamics, however, will be saved to future work, 

Additionally, we calculated the averaged three-point coefficient in a $p$-adic BTZ black hole background and showed similarity with its ordinary counterpart by Kraus and Maloney, but notion of $p$-adic modular transformations remain unknown \cite{KrausMaloney}, so that one is unable to study the thermal $p$-adic AdS. We hope this calculation could initiate future work on $n$-point coefficients of $p$-adic CFTs on higher-genus Mumford curves, such as heavy-heavy-heavy three-point functions on regular genus-2 surfaces investigated in \cite{Cardy4}. In fact, higher genus $p$-adic BTZ black holes were already developed by \cite{Caltech} using higher rank Schottky groups and Mumford curves. 

Finally, we aim to narrow down the list of candidate Hilbert spaces for $p$-adic CFTs and provide hints for quantization. From the bulk point of view, the Hilbert space over $\mathbb{C}$ seems to be a very exotic one, due to Chebyshev polynomials showing up in Section \ref{sec:3}.
 
\subsection{Open questions}
We provide a few open questions that would be interesting to explore in future work on $p$-adic AdS/CFT.

We have only considered the same species of bulk scalar fields but not the possibility of different species. Extending our bulk techniques to an ensemble of different species of bulk scalars $\phi_i$ would not only be interesting (due to the existence of multi-particle states in ordinary AdS$_3$/CFT$_2$ \cite{Xi}), but might also shed light on $p$-adic CFT Hilbert space representations. A naive guess for the boundary partition function with an ensemble of primaries $\chi_i$ dual to $\phi_i$ would be similar to that of ordinary 2d CFTs, with multiplicities $M_{ij}$ of highest-weight states $|i,j\rangle$:
\begin{equation}
    Z =\sum_{i, j} M_{i j} \chi_{i}(\tau) \chi_{j}(\overline{\tau}),
\end{equation}
i.e., summation over primaries. While from the bulk point of view, since different scalars in the action (\ref{eq:action}) decouple from each other, the total partition function should be a simple product of individual partition functions like (\ref{eq:treepart}) for Bruhat-Tits trees, or (\ref{eq:BTZpart}) for BTZ black holes. The absence of descendants in $p$-adic CFT obscures the connection between the summation over primaries on boundary and the product over them in bulk, which are transparently related in ordinary AdS$_3$/CFT$_2$.

As we have mentioned earlier, the $\mathcal{S}$-transformation on genus-1 Tate curve is still missing, so there is no good analog of thermal $p$-adic AdS. We would like to study these potential $p$-adic modular transformation, and even $p$-adic mapping class groups.

Another question is about the role of $GL\left(n,\mathbb{Q}_p\right)$ in ``$p$-adic'' holography or in ``higher-dimensional'' $p$-adic CFTs, the latter being somewhat studied in \cite{Gervais}. For ordinary higher-dimensional CFTs, their fields can organize  into Virasoro representations by \textit{parabolic (generalized) Verma modules}, as stressed in \cite{Yamazaki}; they have also been used in ordinary affine Lie algebras \cite{parabolic}. Although Verma modules are absent in complex representations of $p$-adic groups, they have been constructed as representations on $p$-adic vector spaces instead of Hilbert spaces \cite{Verma}. Then maybe it is worthwhile looking into the former vector spaces.

As to the connection between calculations in Section \ref{sec:3} and $GL(2,\mathbb{Q}_p)$ representations, unexpected coincidence showed up: the determinant of Laplacian on Bruhat-Tits tree (\ref{eq:coincide}) agrees with the volume of the following double coset \cite{L-function} (Theorem 8.10.19 and Chapter 9.2 therein):
\begin{equation}
GL\left(2,\mathbb{Z}_p\right) \cdot 
\begin{pmatrix}
p^N & 0\\0 & 1
\end{pmatrix}
\cdot GL\left(2, \mathbb{Z}_p\right)
\label{eq:orbit}
\end{equation}
with respect to a Haar measure in the context of principal series representations of $GL(2,\mathbb{Q}_p)$. We will present one explanation for this seeming coincidence in using the graph Laplacian
on a Bruhat-Tits tree in Appendix \ref{sec:Laplacian}. %\textcolor{blue}{Volume of $\mathbb{Q}_p$ in terms of $p$-adic integration in 1605.07639?}

There are more ambitious questions. Since our auxiliary cutoff $N$ is necessary in Section \ref{sec:3}, it is natural to ask what will happen to the boundary $p$-adic CFT when one introduce a physical finite cut-off on the Bruhat-Tits tree? Since there is no stress tensor defined in $p$-adic CFT yet \cite{Gubser3}, an analogue of the $T\overline{T}$ deformation in cut-off AdS$_3$/CFT$_2$ \cite{TTbar,Kraus:2018xrn,Guica:2019nzm,Kraus:2021cwf,Ebert:2022cle,Ebert:2022ehb,Kraus:2022mnu} or NAdS$_2$/NCFT$_1$ \cite{Iliesiu:2020zld,Ebert:2022gyn} seems to be difficult.\footnote{See \cite{Qu:2021fgz} for a recent proposal on the effective theory of the Bruhat-Tits tree at a finite boundary.} However, Gubser et al. \cite{Edge} calculated stress tensor-like 2- and 3-point correlators that are reminiscent to AdS$_3$/CFT$_2$, but lacks a notion of spin and the 3-point correlators unexpectedly vanish.

Finally, beyond AdS/CFT, is it possible to formulate a $p$-adic dS/CFT correspondence?\footnote{A recent proposal for a $p$-adic Euclidean dS$_2$ was found by \cite{Qu:2021huo}.} A precursor was given by \cite{Symmetree} in the context of eteranal inflation with dS vacua, but not in the context of string theory.

\acknowledgments
We are grateful to Ori J. Ganor for providing numerous insightful advice and constant guidance. We are thankful for Matthew T. E. Heydeman and Sarthak Parikh providing detailed comments to the preprint, and an anonymous referee for comments on the anisotropic eigenvectors. We thank Ning Bao, Steven S. Gubser, Petr Ho\v{r}ava, Ling-Yan (Janet) Hung, Per Kraus, and Yehao Zhou for helpful discussions and comments. We especially appreciate Richard E. Borcherds for his advice to study infinite-dimensional admissible representations of $GL\left(2,\mathbb{Q}_p\right)$. S.E. would like to thank the Berkeley Center for Theoretical Physics for hospitality during Summer 2019.

\appendix

\section{The full spectrum of the Bruhat-Tits tree Laplacian}
\label{sec:spectrumBTTree}
In this appendix, we discuss the entire spectrum of the graph Laplacian on the Bruhat-Tits tree. Namely, anisotropic eigenvectors which are absent in Section \ref{sec:3} are included here.

\subsection{$\lambda = p+1$ eigenvalues}

In the situation when the eigenvalues $\lambda$ are $p+1$, then at depth $n$ and vertex $a$, one has:
\begin{equation}
\label{eq:r2d}
\left( p +  1 - \lambda \right) \phi_n = \phi_{n-1} + \sum_a \phi_{n+1, a} \implies \phi_{n-1} = - \sum_a \phi_{n+1, a}.
\end{equation}
To count the multiplicity of this eigenvalue, we first consider the case when the depth $N$ is even. From Dirichlet boundary conditions $\phi_N = 0$, the equations of motion for $\phi_{N-1}, \phi_{N-3}, \dots, \phi_1$ imply that $\phi_{N} = \phi_{N-2} = \dots = \phi_0 = 0$. From these boundary conditions, we do not have to distinguish the vertices at the same even depth since $\phi_{N} = \phi_{N-2} = \dots = \phi_0 = 0$. From specifying values of $\phi_{N-1, a}$, then from \eqref{eq:r2d}, all other field values are fixed which makes the degrees of freedom determined purely from the field values at the $N-1$ level. 

We cannot arbitrarily assign values to the fields to the level $N-1$ vertices because there are different subbranches at level $N-1$ connecting to the same level $N-3$ vertex. The summation of these branches must agree with each other. As an example, if we have $\phi_{N-2, a}$ connecting to the same node at $N-3$ then each nodes $\phi_{N-2, a}$ have subnodes $\phi_{N-1,ab}$ with $a, b = 1, \dots, p$. In order to obtain a consistent solution to \eqref{eq:r2d}, we demand that for all $a, a'$ pairs
\begin{equation}
\label{eq:constrainteq}
    \sum_b \phi_{N-1,ab} = \sum_{b'} \phi_{N-1, a'b'}.
\end{equation}
There are $p-1$ independent constraint equations \eqref{eq:constrainteq} for every $N-3$ vertex and another constraint at the center
\begin{equation}
    \sum_a \phi_{1, a} = 0
\end{equation}
so the constraint equations reduce the degrees of freedom on level $N-1$. After subtracting off the number of independent constraint equations, we obtain the multiplicity for the eigenvectors with eigenvalue $\lambda = p+1$
\begin{equation}
    \begin{aligned}
        d^{\operatorname{even}}_{p+1} = D_{N-1} - (p-1) \left( D_{N-3} + D_{N-5} + \dots + D_1\right) - 1 = p^{N-1},
    \end{aligned}
\end{equation}
where the number of vertices at level $n$ is $D_n = (p+1) p^{n-1}$. 

The odd $N$ case analysis is almost exactly the same except that $\phi_0$ can be nonzero and it constrains $\phi_2$ which yields $p$ independent constraint equations. We write 
\begin{equation}
    d^{\operatorname{odd}}_{p+1} = D_{N-1} - (p-1) (D_{N-3}   + D_{N-5}   + \dots + D_2) - p = p^{N-1}.
\end{equation}
\subsection{Anistropic $\lambda \neq p+1$ eigenvalues}
There is another type of anisotropic eigenvectors that do not have the same eigenvalues $p+1$. We determine all these eigenvectors and their corresponding eigenvalues. Consider the Bruhat-Tits tree at depth $N$ so that this type of anisotropic eigenvectors could be divided into $N-2$ classes: $C_0, C_1, C_2, \dots, C_{N-3}$. For each class $C_n$, all the eigenvectors obey the property that all vertices on the tree within a distance of $n$ to the center have zero field value. With this property, we are able to calculate the eigenvalues associated with each individual class. 

To make the above concrete, we consider the $C_{N-3}$ case. If we set the eigenvectors $\phi_0 = \dots = \phi_{N-3} = 0$, then we have the following:
\begin{equation}
\begin{aligned}
\label{eq:eeqns2}
    \sum_a \phi_{N-2, a} &= 0, \\ (p+1 - \lambda) \phi_{N-2, a} &= \sum_b \phi_{N-1, ab}, \\ (p+1 + \lambda) \phi_{N-1, ab} &= \phi_{N-2, a}
\end{aligned}
\end{equation}
and by substituting the last equation into the second equation in \eqref{eq:eeqns2}, we arrive at the eigenvalues
\begin{equation}
    \lambda = p +1\pm \sqrt{p}.
\end{equation}
We can proceed to determine all eigenvalues for any class $C_n$. Generalizing \eqref{eq:eeqns2} to $C_n$, we arrive at the following equations:
\begin{equation}
    \begin{aligned}
    \label{eq:generalrr3}
        \sum_{a_1} \phi_{n+1, a_1} &= 0, \\
        (p+1-\lambda) \phi_{n+1, a_1} &= \sum_{a_2} \phi_{n+2, a_1, a_2}, \\ (p+1 - \lambda) \phi_{n+2, a_1 a_2} &= \phi_{n+1, a_1} + \sum_{a_3} \phi_{n+3, a_1 a_2 a_3} \\& \vdots \\ (p+1 - \lambda) \phi_{N-2, a_1 \dots a_{N-2-n}}  &= \phi_{N-3, a_1 \dots a_{N-3-n}} + \sum_{a_{N-1-n}} \phi_{N-1, a_1 \dots a_{N-1-n}}, \\ (p+1 + \lambda) \phi_{N-1, a_1 \dots a_{N-1-n}} &= \phi_{N-1, a_1\dots a_{N-2-n}}.
    \end{aligned}
\end{equation}
If we consider the second to last equation in \eqref{eq:generalrr3} and multiply by $(p+1 - \lambda)$ and using the last equation to replace $\phi_{N-1}$ by $\phi_{N-2}$, we find
\begin{equation}
    \left( (p+1 - \lambda)^2 - p \right) \phi_{N-2, a_1 \dots a_{N-2-n}} = \left( p+ 1 - \lambda \right) \phi_{N-3, a_1 \dots a_{N-3-n}}.
\end{equation}
We can define a ratio
\begin{equation}
    c_n \equiv \frac{\phi_{n-1, a_1 \dots a_{n-1}}}{\phi_{n, a_1 \dots a_n}}
\end{equation}
which is independent of the $(a_1, \dots, a_n)$ labels. Thus from \eqref{eq:generalrr3}, we obtain a recursion relation for the ratios
\begin{equation}
    c_n  c_{n-1} = c_n(p+1 - \lambda) - p
\end{equation}
and
\begin{equation}
    c_{N-1} = p+1 - \lambda
\end{equation}
so we can compute the $c_n$ recursively in terms of $c_0$. For a specific class $C_n$, the eigenvalues are obtained by solving 
\begin{equation}
    c_{n+2} (p+1 - \lambda) = p.
\end{equation}
Therefore, the recursion relation implies that $c_1 = \dots = c_{n+1} = 0$ and is consistent with the property claimed initially. From the recursion relation, we see that the eigenvalue $\lambda$ obeys a polynomial relation. For $C_{N-3}, C_{N-4}$, the equations are quadratic. Meanwhile, each time when $n$ decreases by two, the degree of equations increases by two. We denote these degree of equations by 
\begin{equation}
    q_n = 2\left\lfloor\frac{N-1-n}{2}\right\rfloor,
\end{equation}
which also counts the number of roots or eigenvalues for one specific $C_n$ class. To know the multiplicity of the eigenvalues, we need to know the number of eigenvectors contained in class $C_n$. 

This number can be calculated recursively. We first calculate the multiplicity of the $C_{N-3}$ class. We have 
\begin{equation}
    (p+1 + \lambda) \phi_{N-1, a_1 \dots a_{N-1-n}} = \phi_{N-2, a_1 \dots a_{N-2-n}}
\end{equation}
all level $N-1$ connecting to the same level $N-2$ vertex will share the same field value. For a fixed eigenvalue and field configuration at level $N-1$, all other field values are determined. Hence the only degrees of freedom come from the field configuration at level $N-1$. Since all vertices at the same branch are the same, the total number of degrees of freedom is $D_{N-2}$ for the number of vertices at level $N-2$. For the $C_{N-3}$ class, we have $\phi_{N-3} = 0$, so we have $D_{N-3}$ constraint equations and the number of eigenvectors are
\begin{equation}
    d_{N-3} = \left( D_{N-2} - D_{N-3} \right) q_{N-3}.
\end{equation}
Similarly, the degeneracy for the $C_{N-4} $ class is
\begin{equation}
    d_{N-4} = \left( D_{N-3} - D_{N-4} \right) q_{N-4},
\end{equation}
and for general $n$
\begin{equation}
    d_n =  \left( D_{n+1} - D_n \right) q_n.
\end{equation}
\subsection{Multiplicity}
We can check the total multiplicity is correct. For odd $N$,
\begin{equation}
    \begin{aligned}
        \sum^{N-3}_{n=0} d_n &=2(p+1)\left(p^{N-3}-p^{N-4}+p^{N-4}-p^{N-5}+2 p^{N-5}-2 p^{N-6}  + \dots+\frac{N-3}{2}(p-1)\right)\\&\quad+2 p \left( \frac{N-1}{2} \right) \\
&=2(p+1)\left(p^2+p^4+\ldots+p^{N-3}\right)-(N-3)(p+1)+(N-1) p \\
&=2(p+1) \left( \frac{p^{N-1}-p^2}{p^2-1} \right) +2 p+2-(N-1) \\
&=2 \left( \frac{p^{N-1}-1}{p-1} \right) -(N-1).
    \end{aligned}
\end{equation}
Adding the multiplicity of the other two types of eigenvectors, we find
\begin{equation}
    \begin{aligned}
        D&= d_{\text{isotropic}} + d_{p+1} + \sum^{N-3}_{n = 0} d_n \\&= 2 \left( \frac{p^{N-1}-1}{p-1} \right) \\&= \frac{p^N + p^{N-1} - 2}{p -1}.
    \end{aligned}
\end{equation}
%which is exactly equal to the total number of vertices of the graph Laplacian matrix's dimension $N-1$.

For even $N$, we arrive at a similar expression
\begin{equation}
    \begin{aligned}
\sum_{n=0}^{N-3} d_n &=2(p+1)\left(p^{N-3}-p^{N-4}+p^{N-4}-p^{N-5}+2 p^{N-5}-2 p^{N-6} + \dots+\frac{N-2}{2}(p-1)\right)\\&\quad+2 p \left(\frac{N-2}{2} \right) \\
&=2(p+1)\left(p+p^3+\ldots+p^{N-3}\right)-(N-2)(p+1)+(N-2) p \\
&=2(p+1) \left( \frac{p^{N-1}-p}{p^2-1} \right)+2-N \\
&=2 \left( \frac{p^{N-1}-1}{p-1} \right) -N .
\end{aligned}
\end{equation}
Adding all multiplicities together, we obtain the expected result
\begin{equation}
    \begin{aligned}
        D&= d_{\text{isotropic}}  + d_{p+1} + \sum^{N-3}_{n=0} d_n \\&= 2\left( \frac{p^{N-1} - 1}{p -1} \right) + p^{N-1} \\&= \frac{p^N + p^{N-1} - 2}{p-1}.
    \end{aligned}
\end{equation}
Therefore, we have found all eigenvalues of the Laplacian matrix on the Bruhat-Tits tree and a classification of different eigenvector configurations. A similar analysis on the full spectrum of graph Laplacian on the $p$-adic BTZ black hole can be performed in a parallel manner, and we will defer it to a later treatment \cite{spectrum}. 

\section{Laplacian matrix of a multigraph}
\label{sec:Laplacian}
Here we use a graph-theoretic method to obtain the determinant of Laplacian operator on the Bruhat-Tits tree\footnote{We thank Yehao Zhou for helpful comments.}, which has already been calculated in Section \ref{sec:Laptree}.

Let us notice that the result \eqref{eq:coincide} can be viewed as the product of all nonzero eigenvalues of a directed multigraph Laplacian $\widetilde{\Box}$. This multigraph $G$ contains:
\begin{itemize}
\item $N+1$ vertices, labeled by $0,\dots,N$;
\item One arrow from the $i^{\text{th}}$ vertex to $(i+1)^{\text{th}}$ vertex, where $i=0,\dots,N-2$;
\item $p$ arrows from the $j^{\text{th}}$ vertex to the $(j-1)^{\text{th}}$ vertex, where $j=N,N-1,\dots,2$;
\item $p+1$ arrows from the vertex $1$ to the vertex $0$.
\end{itemize}

The product of eigenvalues of $\widetilde{\Box}$ equals the determinant of the adjacency matrix of $G$, with the $(N+1)^{\text{th}}$ row and the $(N+1)^{\text{th}}$ column removed, because there is no arrow going from anywhere else to the vertex $N$. By Kirchhoff's theorem, this determinant equals the number of spanning trees starting from the vertex $N$, which is 
\begin{equation}
\underbrace{p\cdot p\cdot\dotsc\cdot p}_{N-1}\cdot(p+1)=p^N+p^{N-1}.
\end{equation}

In fact, $G$ can be obtained by ``compressing'' the truncated Bruhat-Tits tree in Section \ref{sec:Laptree} onto one ray using the rotational symmetry, if we restrict to the isotropic sector. Therefore, a spanning tree starting from the vertex $N$ on $G$ is equivalent to a path originating from the center to the cut-off boundary of the Bruhat-Tits tree, which in turn is equivalent to choosing a point at depth $N$ on the tree.

Finally, all points at depth $N$ on the Bruhat-Tits tree form an orbit of the Iwahori subgroup of $GL(2,\mathbb{Z}_p)$, called the \textit{Iwahori orbit}. Under the Haar measure, the orbit has volume 1, so the volume of the double coset \eqref{eq:orbit} equals the number of elements in the quotient of \eqref{eq:orbit} by the right action of Iwahori subgroup. This quotient is exactly the Iwahori orbit representing elements
\begin{equation}
    \begin{pmatrix}
    p^N & 0\\
    0 & 1
    \end{pmatrix},
\end{equation}
namely, points at depth $N$. As we discussed on the previous page, there are $p^N+p^{N-1}$ of them. 

However, for the BTZ graph, there is no good rotational symmetry which allows for a ``compression,'' so a similar analysis obtaining $\det\widetilde{\Box}$ cannot be done.

It would also be interesting to understand this volume purely in terms of $p$-adic integration using the Haar measure, say, in Appendix A of \cite{Caltech}.

\section{BTZ graphs revisited}
\label{app:messy}
In Section \ref{subsec:BTZ}, if we do not use the periodic linear recurrence on the horizon (\ref{eq:recursion2}), without loss of generality, we start from the initial condition at the $\phi_{0,s}$ vertex:
\begin{equation}
    (p-1)(\phi_{0,s}-\phi_{1,s})+(\phi_{0,s}-\phi_{0,s-1})+(\phi_{0,s}-\phi_{0,s+1})=\lambda\phi_{0,s},
\end{equation}
where $\phi_{1,s}$ denotes the field value on the outward vertex one edge away from the horizon point $(0,s)$, and hence
\begin{equation}
    \phi_{1,s}=\frac{(p+1-\lambda)\phi_{0,s}-(\phi_{0,s-1}+\phi_{0,s+1})}{p-1}.
\end{equation}

Similar to what we have shown in Section \ref{sec:Laptree}, all field values $\phi_{n,s},n>1,$ away from the event horizon only depend on their depths $n$ and hence isotropic in each subtree rooted at one horizon vertex $(0,s)$. There is no change in the linear recurrence (\ref{eq:recursion}) for all $n>2$, and for $n=2$ we have
\begin{equation}
\begin{aligned}
    \phi_{2,s}&=\frac{(p+1-\lambda)\phi_{1,s}-\phi_{0,s}}{p} \\
    &= \frac{[\lambda(\lambda -2p -2) + p(p+1)+2] \phi_{0,s} + (\lambda - p - 1)(\phi_{0,s+1}+\phi_{0,s-1})}{p(p-1)},
    \end{aligned}
\end{equation}
then the coefficients get uncontrollably complicated as the depth $n$ increases.

\section{Review on the ordinary BTZ modular parameter}
\label{app:modular}
In ordinary Euclidean AdS$_3$, for a genus-1 gravitational saddle configuration, the modular parameter is $\tau=\theta+i\beta$, defined on the upper-half plane $\mathbb{H}^2$, where $\theta$ is the angular potential and $\beta$ is the inverse temperature, then the tree-level partition function is \cite{MaldacenaStrominger}
\begin{equation}
Z=e^{\pi k\frac{\text{Im}\tau}{|\tau|^2}},
\end{equation}
where $k$ is the inverse 3D Newton's constant. For a non-rotating black hole, as in our case $\theta=0$, so
\begin{equation}
    Z=e^{\frac{\pi k}{\beta}}=e^{\pi k r_+}.
\end{equation}

If corrected by the one-loop contribution as in \cite{MaloneyWitten}, we have:
\begin{equation}
Z=\mathcal{Z}(\tau)\bar{\mathcal{Z}}(\bar{\tau}),
\end{equation}
where the holomorphic piece is
\begin{equation}
    \mathcal{Z}(\tau)=\frac{q_-^{-(k-1/24)}(1-q_-)}{\eta(-1/\tau)},
\end{equation}
and $q_-\equiv e^{-2\pi i/\tau}$. Since the partition function of 3D pure gravity is 1-loop exact \cite{MaloneyWitten}, the combined result is
\begin{equation}
    Z_{\text{tot}}=\frac{e^{\frac{4\pi\text{Im}\tau}{|\tau|^2}\left(k-\frac{1}{24}\right)}}{\eta\left(-\frac{1}{\tau}\right)\bar{\eta}\left(-\frac{1}{\bar{\tau}}\right)}\left[1+e^{-\frac{4\pi \text{Im}\tau}{|\tau|^2}}-2\cos\left(\frac{2\pi \text{Im}\tau}{|\tau|^2}\right)e^{-\frac{2\pi\text{Im}\tau}{|\tau|^2}}\right]
\end{equation}

We will use the $q$-Pochhammer symbol specified at $q$ itself
\begin{equation}
    (q;q)_{\infty}\equiv \prod_{n=1}^{\infty}\left(1-q^n\right),
\end{equation}
as well as the fact that $q_-\bar{q}_-=e^{-4\pi\frac{\text{Im}\tau}{|\tau|^2}}$ and $\eta(-1/\tau)\equiv q_-^{1/24}(q_-;q_-)_{\infty}$, a useful expression when $q\in\mathbb{R}$. 

Hence for a non-rotating BTZ black hole, $q_-=\bar{q}_-$, and at large $r_+=1/\beta$, we have
\begin{equation}
    \frac{e^{4\pi k\frac{\text{Im}\tau}{|\tau|^2}}}{(q_-;q_-)_{\infty}(\bar{q}_-;\bar{q}_-)_{\infty}}\approx \boxed{e^{4k\pi r_+}}.
\end{equation}

Instead when $r_+$ is very small, we use the asymptotics \cite{Watson}:
\begin{equation}
    (q;q)_{\infty}\approx\sqrt{\frac{2\pi}{t}}e^{\frac{t}{24}-\frac{\pi^2}{6t}}, \text{for}~ q=e^{-t}, t\rightarrow0,
\end{equation}
so that the partition function is approximately $\boxed{r_+e^{(4k-1/6)\pi r_+}}$.

\section{An appetizer to compact induction for $GL_2$}
\label{app:induction}

Compact induction is among the very first constructions of supercuspidal representations. The standard philosophy is to induce an irrep of the group $G$ from a representation of a {\it compact} subgroup $H \subset G$. Avoiding most technicalities, we demonstrate this for the simplest case, the symmetric group $S_3$, adopting the approach from \cite{Gross}. We will not define terms not shown in our main text.

It is known that for a given $p$, there are $p(p-1)/2$ distinct supercuspidal representations for $SL\left(2,\mathbb{Q}_p\right)$ \cite{Knightly} (Theorem 2.5), so the supercuspidal representation for $SL\left(2,\mathbb{Q}_2\right)$ is unique. We start from the {\it cuspidal representation} of $SL\left(2,\mathbb{F}_2\right) \cong S_3$, i.e., the character $\rho$ with mappings:

%Big matrix equation, uncomment later
\begin{equation}
 \underset{\substack{\big\downarrow \\ 1}}{\left(\begin{array}{ll}{1} & {0} \\ {0} & {1}\end{array}\right)}, \underset{\substack{\big\downarrow \\  -1}}{\left(\begin{array}{ll}{0} & {1} \\ {1} & {0}\end{array}\right)}, \underset{\substack{\big\downarrow \\ -1}}{\left(\begin{array}{ll}{1} & {1} \\ {0} & {1}\end{array}\right)}, \underset{\substack{\big\downarrow \\ -1}}{\left(\begin{array}{ll}{1} & {0} \\ {1} & {1}\end{array}\right)}, \underset{\substack{\big\downarrow \\  1}}{\left(\begin{array}{ll}{1} & {1} \\ {1} & {0}\end{array}\right)}, \underset{\substack{\big\downarrow \\  1}}{\left(\begin{array}{ll}{0} & {1} \\ {1} & {1}\end{array}\right)},
\end{equation} 
and preform compact induction on $S_3$ to obtain the supercuspidal representation of $SL(2,\mathbb{Q}_2)$. 

We use the fact that there is a unique {\it tamely ramified extension} $\mathbb{Q}_2\left(\zeta_3, \sqrt[\leftroot{-3}\uproot{3}3]{2}\right)/\mathbb{Q}_2$ whose Galois group is exactly $S_3$, where $\zeta_3$ is a $3^\text{rd}$ root of unity. Then the {\it Langlands parameter}
\begin{equation}
\phi : \operatorname{Gal} \left(\mathbb{Q}_2 \left(\zeta_3, \sqrt[\leftroot{-3}\uproot{3}3]{2}\right)\middle/\mathbb{Q}_2 \right) \rightarrow S_3 \subseteq PGL(2, \mathbb{C})
\end{equation}
corresponds to two irreps of $SL\left(2, \mathbb{Q}_2\right)$ given by compact induction from
\begin{equation}
    K_1 = SL(2, \mathbb{Z}), \quad \text{and} ~ K_2 = \left(\begin{array}{ll}{2} & {0} \\ {0} & {1}\end{array}\right) K_1 \begin{pmatrix}1/2 & 0 \\ 0 & 1\end{pmatrix} = \begin{pmatrix}* & 2* \\ */2 & *\end{pmatrix}
\end{equation}
of the characters $K_i \rightarrow S_3 \stackrel{\text { sgn }}{\longrightarrow}\{\pm 1\}$.

More generally and abstractly, compact induction can be performed on $\mathbb{Z}_p/p\mathbb{Z}_p\sim \mathbb{Z}/\mathbb{Z}_p\sim \mathbb{F}_p$ as well, and supercuspidal representations obtained are called {\it depth-zero} \cite{autreps}. With this, one can actually enumerate all supercuspidal representations of $GL\left(2,\mathbb{Q}_p\right)$ \cite{Youcis}.

\bibliographystyle{JHEP}
\bibliography{BTZ}

\providecommand{\href}[2]{#2}\begingroup\raggedright\begin{thebibliography}{10}

\bibitem{$p$-adicVenezianoAmplitude}
P.~G.~O. Freund and E.~Witten, {\it {Adelic String Amplitudes}},  {\em Phys.
  Lett. B} {\bf 199} (1987) 191.

\bibitem{BFOW}
L.~Brekke, P.~G.~O. Freund, M.~Olson, and E.~Witten, {\it {Nonarchimedean
  String Dynamics}},  {\em Nucl. Phys. B} {\bf 302} (1988) 365--402.

\bibitem{BogdanScattering}
B.~Stoica, {\it {Closed form expression for the 5-point Veneziano amplitude in
  terms of 4-point amplitudes}},  \href{http://arxiv.org/abs/2111.02423}{{\tt
  arXiv:2111.02423}}.

\bibitem{Jepsen:2022pkn}
C.~B. Jepsen, {\it {Adelic Amplitudes and Intricacies of Infinite Products}},
  \href{http://arxiv.org/abs/2211.01611}{{\tt arXiv:2211.01611}}.

\bibitem{Zabrodin}
A.~V. Zabrodin, {\it {Nonarchimedean Strings and Bruhat-tits Trees}},  {\em
  Commun. Math. Phys.} {\bf 123} (1989) 463.

\bibitem{Gubser1}
S.~S. Gubser, J.~Knaute, S.~Parikh, A.~Samberg, and P.~Witaszczyk, {\it
  {$p$-adic AdS/CFT}},  {\em Commun. Math. Phys.} {\bf 352} (2017), no.~3
  1019--1059, [\href{http://arxiv.org/abs/1605.01061}{{\tt arXiv:1605.01061}}].

\bibitem{Caltech}
M.~Heydeman, M.~Marcolli, I.~Saberi, and B.~Stoica, {\it {Tensor networks,
  $p$-adic fields, and algebraic curves: arithmetic and the AdS$_3$/CFT$_2$
  correspondence}},  {\em Adv. Theor. Math. Phys.} {\bf 22} (2018) 93--176,
  [\href{http://arxiv.org/abs/1605.07639}{{\tt arXiv:1605.07639}}].

\bibitem{Maldacena}
J.~M. Maldacena, {\it {The Large N limit of superconformal field theories and
  supergravity}},  {\em Adv. Theor. Math. Phys.} {\bf 2} (1998) 231--252,
  [\href{http://arxiv.org/abs/hep-th/9711200}{{\tt hep-th/9711200}}].

\bibitem{L-function}
D.~Goldfeld and J.~Hundley, {\em Automorphic Representations and L-Functions
  for the General Linear Group}, vol.~1 of {\em Cambridge Studies in Advanced
  Mathematics}.
\newblock Cambridge University Press, 2011.

\bibitem{MMPS}
M.~Heydeman, M.~Marcolli, S.~Parikh, and I.~Saberi, {\it {Nonarchimedean
  holographic entropy from networks of perfect tensors}},  {\em Adv. Theor.
  Math. Phys.} {\bf 25} (2021), no.~3 591--721,
  [\href{http://arxiv.org/abs/1812.04057}{{\tt arXiv:1812.04057}}].

\bibitem{Ling-Yan}
L.-Y. Hung, W.~Li, and C.~M. Melby-Thompson, {\it {$p$-adic CFT is a
  holographic tensor network}},  {\em JHEP} {\bf 04} (2019) 170,
  [\href{http://arxiv.org/abs/1902.01411}{{\tt arXiv:1902.01411}}].

\bibitem{GubserSpin}
S.~S. Gubser, C.~Jepsen, and B.~Trundy, {\it {Spin in $p$-adic AdS/CFT}},  {\em
  J. Phys. A} {\bf 52} (2019), no.~14 144004,
  [\href{http://arxiv.org/abs/1811.02538}{{\tt arXiv:1811.02538}}].

\bibitem{SYK}
S.~S. Gubser, M.~Heydeman, C.~Jepsen, S.~Parikh, I.~Saberi, B.~Stoica, and
  B.~Trundy, {\it {Melonic theories over diverse number systems}},  {\em Phys.
  Rev. D} {\bf 98} (2018), no.~12 126007,
  [\href{http://arxiv.org/abs/1707.01087}{{\tt arXiv:1707.01087}}].

\bibitem{Melonic}
S.~S. Gubser, C.~Jepsen, Z.~Ji, and B.~Trundy, {\it {Higher melonic theories}},
   {\em JHEP} {\bf 09} (2018) 049, [\href{http://arxiv.org/abs/1806.04800}{{\tt
  arXiv:1806.04800}}].

\bibitem{box2}
A.~Huang, D.~Mao, and B.~Stoica, {\it {From $p$-adic to Archimedean Physics:
  Renormalization Group Flow and Berkovich Spaces}},
  \href{http://arxiv.org/abs/2001.01725}{{\tt arXiv:2001.01725}}.

\bibitem{Hung123}
L.~Chen, X.~Liu, and L.-Y. Hung, {\it {Bending the Bruhat-Tits tree. Part I.
  Tensor network and emergent Einstein equations}},  {\em JHEP} {\bf 06} (2021)
  094, [\href{http://arxiv.org/abs/2102.12023}{{\tt arXiv:2102.12023}}].

\bibitem{Hung3regw43}
L.~Chen, X.~Liu, and L.-Y. Hung, {\it {Bending the Bruhat-Tits tree. Part II.
  The p-adic BTZ black hole and local diffeomorphism on the Bruhat-Tits tree}},
   {\em JHEP} {\bf 09} (2021) 097, [\href{http://arxiv.org/abs/2102.12024}{{\tt
  arXiv:2102.12024}}].

\bibitem{BogdanBranes}
A.~Huang, B.~Stoica, and X.~Zhong, {\it {Massless $p2$-brane modes and the
  critical line}},  \href{http://arxiv.org/abs/2110.15378}{{\tt
  arXiv:2110.15378}}.

\bibitem{melzer}
E.~Melzer, {\it {Nonarchimedean Conformal Field Theories}},  {\em Int. J. Mod.
  Phys. A} {\bf 4} (1989) 4877.

\bibitem{Symmetree}
D.~Harlow, S.~H. Shenker, D.~Stanford, and L.~Susskind, {\it {Tree-like
  structure of eternal inflation: A solvable model}},  {\em Phys. Rev. D} {\bf
  85} (2012) 063516, [\href{http://arxiv.org/abs/1110.0496}{{\tt
  arXiv:1110.0496}}].

\bibitem{Gubser3}
S.~S. Gubser and S.~Parikh, {\it {Geodesic bulk diagrams on the
  Bruhat\textendash{}Tits tree}},  {\em Phys. Rev. D} {\bf 96} (2017), no.~6
  066024, [\href{http://arxiv.org/abs/1704.01149}{{\tt arXiv:1704.01149}}].

\bibitem{MaloneyWitten}
A.~Maloney and E.~Witten, {\it {Quantum Gravity Partition Functions in Three
  Dimensions}},  {\em JHEP} {\bf 02} (2010) 029,
  [\href{http://arxiv.org/abs/0712.0155}{{\tt arXiv:0712.0155}}].

\bibitem{KrausMaloney}
P.~Kraus and A.~Maloney, {\it {A cardy formula for three-point coefficients or
  how the black hole got its spots}},  {\em JHEP} {\bf 05} (2017) 160,
  [\href{http://arxiv.org/abs/1608.03284}{{\tt arXiv:1608.03284}}].

\bibitem{Milne}
J.~S. Milne, {\it Algebraic number theory (v3.08)},  2020.
\newblock Available at
  \href{https://www.jmilne.org/math/CourseNotes/ANT.pdf}{\texttt{https://www.jmilne.org/math/CourseNotes/ANT.pdf}}.

\bibitem{Koblitz}
N.~Koblitz, {\em $p$-adic Numbers, $p$-adic Analysis, and Zeta-Functions},
  vol.~58 of {\em Graduate Texts in Mathematics}.
\newblock Springer-Verlag, 1977.

\bibitem{UCLAOstro}
G.~Gim, {\it Ostrowski's theorem},  2012.
\newblock Available at
  \href{https://www.math.ucla.edu/~ggim/F12-205A.pdf}{\texttt{https://www.math.ucla.edu/~ggim/F12-205A.pdf}}.

\bibitem{padic}
L.~Brekke and P.~G.~O. Freund, {\it {p-adic numbers in physics}},  {\em Phys.
  Rept.} {\bf 233} (1993) 1--66.

\bibitem{pAdS}
F.~Qu and Y.-h. Gao, {\it {Scalar fields on $p$AdS}},  {\em Phys. Lett. B} {\bf
  786} (2018) 165--170, [\href{http://arxiv.org/abs/1806.07035}{{\tt
  arXiv:1806.07035}}].

\bibitem{Gurarie}
V.~Gurarie, {\it {c theorem for disordered systems}},  {\em Nucl. Phys. B} {\bf
  546} (1999) 765, [\href{http://arxiv.org/abs/cond-mat/9808063}{{\tt
  cond-mat/9808063}}].

\bibitem{Ludwig}
V.~Gurarie and A.~W.~W. Ludwig, {\it {Conformal field theory at central charge
  c=0 and two-dimensional critical systems with quenched disorder}},  in {\em
  {From Fields to Strings: Circumnavigating Theoretical Physics: A Conference
  in Tribute to Ian Kogan}}, pp.~1384--1440, 9, 2004.
\newblock \href{http://arxiv.org/abs/hep-th/0409105}{{\tt hep-th/0409105}}.

\bibitem{BTZ}
M.~Banados, C.~Teitelboim, and J.~Zanelli, {\it {The Black hole in
  three-dimensional space-time}},  {\em Phys. Rev. Lett.} {\bf 69} (1992)
  1849--1851, [\href{http://arxiv.org/abs/hep-th/9204099}{{\tt
  hep-th/9204099}}].

\bibitem{KOS}
P.~Kraus, H.~Ooguri, and S.~Shenker, {\it {Inside the horizon with AdS / CFT}},
   {\em Phys. Rev. D} {\bf 67} (2003) 124022,
  [\href{http://arxiv.org/abs/hep-th/0212277}{{\tt hep-th/0212277}}].

\bibitem{Edge}
S.~S. Gubser, M.~Heydeman, C.~Jepsen, M.~Marcolli, S.~Parikh, I.~Saberi,
  B.~Stoica, and B.~Trundy, {\it {Edge length dynamics on graphs with
  applications to $p$-adic AdS/CFT}},  {\em JHEP} {\bf 06} (2017) 157,
  [\href{http://arxiv.org/abs/1612.09580}{{\tt arXiv:1612.09580}}].

\bibitem{bai2020sum}
S.~{Bai}, A.~{Huang}, L.~{Lu}, and S.-T. {Yau}, {\it {On the Sum of
  Ricci-Curvatures for Weighted Graphs}},
  \href{http://arxiv.org/abs/2001.01776}{{\tt arXiv:2001.01776}}.

\bibitem{huang2020bounds}
A.~Huang, B.~Stoica, X.~Xia, and X.~Zhong, {\it {Bounds on the Ricci curvature
  and solutions to the Einstein equations for weighted graphs}},
  \href{http://arxiv.org/abs/2006.06716}{{\tt arXiv:2006.06716}}.

\bibitem{BTZasQuotient}
M.~Banados, M.~Henneaux, C.~Teitelboim, and J.~Zanelli, {\it {Geometry of the
  (2+1) black hole}},  {\em Phys. Rev. D} {\bf 48} (1993) 1506--1525,
  [\href{http://arxiv.org/abs/gr-qc/9302012}{{\tt gr-qc/9302012}}]. [Erratum:
  Phys.Rev.D 88, 069902 (2013)].

\bibitem{GKP}
S.~S. Gubser, I.~R. Klebanov, and A.~M. Polyakov, {\it {Gauge theory
  correlators from noncritical string theory}},  {\em Phys. Lett. B} {\bf 428}
  (1998) 105--114, [\href{http://arxiv.org/abs/hep-th/9802109}{{\tt
  hep-th/9802109}}].

\bibitem{WittenHolography}
E.~Witten, {\it {Anti-de Sitter space and holography}},  {\em Adv. Theor. Math.
  Phys.} {\bf 2} (1998) 253--291,
  [\href{http://arxiv.org/abs/hep-th/9802150}{{\tt hep-th/9802150}}].

\bibitem{spectrum}
S.~Ebert, H.-Y. Sun, and M.~Zhang, {\it {Full partition function of $p$-adic
  CFTs}},  {\em in preparation} (2022).

\bibitem{Afkhami-Jeddi:2019zci}
N.~Afkhami-Jeddi, T.~Hartman, and A.~Tajdini, {\it {Fast Conformal Bootstrap
  and Constraints on 3d Gravity}},  {\em JHEP} {\bf 05} (2019) 087,
  [\href{http://arxiv.org/abs/1903.06272}{{\tt arXiv:1903.06272}}].

\bibitem{Hartman:2019pcd}
T.~Hartman, D.~Maz\'a\v{c}, and L.~Rastelli, {\it {Sphere Packing and Quantum
  Gravity}},  {\em JHEP} {\bf 12} (2019) 048,
  [\href{http://arxiv.org/abs/1905.01319}{{\tt arXiv:1905.01319}}].

\bibitem{Benjamin:2021ygh}
N.~Benjamin, S.~Collier, A.~L. Fitzpatrick, A.~Maloney, and E.~Perlmutter, {\it
  {Harmonic analysis of 2d CFT partition functions}},  {\em JHEP} {\bf 09}
  (2021) 174, [\href{http://arxiv.org/abs/2107.10744}{{\tt arXiv:2107.10744}}].

\bibitem{Lubotzky}
A.~Lubotzky, {\em Discrete Groups, Expanding Graphs and Invariant Measures},
  vol.~125 of {\em Progress in Mathematics}.
\newblock Springer Basel AG, 1994.

\bibitem{Xi}
S.~Giombi, A.~Maloney, and X.~Yin, {\it {One-loop Partition Functions of 3D
  Gravity}},  {\em JHEP} {\bf 08} (2008) 007,
  [\href{http://arxiv.org/abs/0804.1773}{{\tt arXiv:0804.1773}}].

\bibitem{Krasnov}
K.~Krasnov, {\it {Holography and Riemann surfaces}},  {\em Adv. Theor. Math.
  Phys.} {\bf 4} (2000) 929--979,
  [\href{http://arxiv.org/abs/hep-th/0005106}{{\tt hep-th/0005106}}].

\bibitem{Zwillinger}
D.~Zwillinger, {\it Standard mathematical tables and formulae},  CRC Press,
  30th Edition, (1995).

\bibitem{Carlip}
S.~Carlip, {\it {Logarithmic corrections to black hole entropy from the Cardy
  formula}},  {\em Class. Quant. Grav.} {\bf 17} (2000) 4175--4186,
  [\href{http://arxiv.org/abs/gr-qc/0005017}{{\tt gr-qc/0005017}}].

\bibitem{Log}
R.~K. Kaul and P.~Majumdar, {\it {Logarithmic correction to the
  Bekenstein-Hawking entropy}},  {\em Phys. Rev. Lett.} {\bf 84} (2000)
  5255--5257, [\href{http://arxiv.org/abs/gr-qc/0002040}{{\tt gr-qc/0002040}}].

\bibitem{Sorkin}
R.~D. Sorkin, {\it {1983 paper on entanglement entropy: ``On the Entropy of the
  Vacuum outside a Horizon''}},  in {\em {10th International Conference on
  General Relativity and Gravitation}}, vol.~2, pp.~734--736, 1984.
\newblock \href{http://arxiv.org/abs/1402.3589}{{\tt arXiv:1402.3589}}.

\bibitem{Cardy1}
J.~L. Cardy, {\it {Operator Content of Two-Dimensional Conformally Invariant
  Theories}},  {\em Nucl. Phys. B} {\bf 270} (1986) 186--204.

\bibitem{Cardy2}
J.~L. Cardy, {\it {Effect of Boundary Conditions on the Operator Content of
  Two-Dimensional Conformally Invariant Theories}},  {\em Nucl. Phys. B} {\bf
  275} (1986) 200--218.

\bibitem{Cardy3}
H.~W.~J. Bloete, J.~L. Cardy, and M.~P. Nightingale, {\it {Conformal
  Invariance, the Central Charge, and Universal Finite Size Amplitudes at
  Criticality}},  {\em Phys. Rev. Lett.} {\bf 56} (1986) 742--745.

\bibitem{BH}
J.~D. Brown and M.~Henneaux, {\it {Central Charges in the Canonical Realization
  of Asymptotic Symmetries: An Example from Three-Dimensional Gravity}},  {\em
  Commun. Math. Phys.} {\bf 104} (1986) 207--226.

\bibitem{Yau}
Y.~Lin, L.~Lu, and S.-T. Yau, {\it {Ricci curvature of graphs}},  {\em Tohoku
  Mathematical Journal} {\bf 63} (2011), no.~4 605 -- 627.

\bibitem{Vazirani1}
{Z. Landau and U. Vazirani and T. Vidick}, {\it {A polynomial-time algorithm
  for the ground state of 1D gapped local Hamiltonians}},
  \href{http://arxiv.org/abs/1307.5143}{{\tt arXiv:1307.5143}}.

\bibitem{Vazirani2}
I.~Arad, Z.~Landau, U.~Vazirani, and T.~Vidick, {\it {Rigorous RG algorithms
  and area laws for low energy eigenstates in 1D}},  {\em Commun. Math. Phys.}
  {\bf 356} (2017) 65--105, [\href{http://arxiv.org/abs/1602.08828}{{\tt
  arXiv:1602.08828}}].

\bibitem{JT}
L.~V. Iliesiu, S.~S. Pufu, H.~Verlinde, and Y.~Wang, {\it {An exact
  quantization of Jackiw-Teitelboim gravity}},  {\em JHEP} {\bf 11} (2019) 091,
  [\href{http://arxiv.org/abs/1905.02726}{{\tt arXiv:1905.02726}}].

\bibitem{Kitaev}
A.~Kitaev, {\it {Notes on $\widetilde{\mathrm{SL}}(2,\mathbb{R})$
  representations}},  \href{http://arxiv.org/abs/1711.08169}{{\tt
  arXiv:1711.08169}}.

\bibitem{group}
H.~Glockner, {\it {Lectures on Lie groups over local fields}},
  \href{http://arxiv.org/abs/0804.2234}{{\tt arXiv:0804.2234}}.

\bibitem{Verma}
S.~Orlik and M.~Strauch, {\it {On the irreducibility of locally analytic
  principal series representations}},  {\em Represent. Theory} {\bf 14} (12,
  2010) 713--746, [\href{http://arxiv.org/abs/0612809}{{\tt 0612809}}].

\bibitem{BGG}
O.~T.~R. Jones, {\it {An analogue of the BGG resolution for locally analytic
  principal series}},  {\em Journal of Number Theory} {\bf 131} (9, 2011)
  1616--1640, [\href{http://arxiv.org/abs/0912.4756}{{\tt arXiv:0912.4756}}].

\bibitem{Hilbert}
A.~J. Silberger, {\em Introduction to Harmonic Analysis on Reductive P-adic
  Groups. (MN-23): Based on lectures by Harish-Chandra at The Institute for
  Advanced Study, 1971-73}.
\newblock Princeton University Press, 1979.

\bibitem{Hilbert2}
J.~D. Rogawski, ``Modular forms, the ramanujan conjecture and the
  jacquet-langlands correspondence.''
\newblock an \href{https://www.math.ucla.edu/~jonr/eprints/lub.pdf}{Appendix}
  to \cite{Lubotzky}.

\bibitem{Harish-Chandra1}
Harish-Chandra, {\it Representations of semisimple lie groups on a banach
  space},  {\em Proceedings of the National Academy of Sciences} {\bf 37}
  (1951), no.~3 170--173,
  [\href{http://arxiv.org/abs/https://www.pnas.org/doi/pdf/10.1073/pnas.37.3.170}{{\tt
  https://www.pnas.org/doi/pdf/10.1073/pnas.37.3.170}}].

\bibitem{Harish-Chandra2}
Harish-Chandra, {\it Representations of semisimple lie groups on a banach
  space},  {\em Proceedings of the National Academy of Sciences} {\bf 37}
  (1951), no.~3 366--369,
  [\href{http://arxiv.org/abs/https://www.pnas.org/doi/pdf/10.1073/pnas.37.3.170}{{\tt
  https://www.pnas.org/doi/pdf/10.1073/pnas.37.3.170}}].

\bibitem{Harish-Chandra3}
Harish-Chandra, {\it Representations of semisimple lie groups on a banach
  space},  {\em Proceedings of the National Academy of Sciences} {\bf 37}
  (1951), no.~3 691--694,
  [\href{http://arxiv.org/abs/https://www.pnas.org/doi/pdf/10.1073/pnas.37.3.170}{{\tt
  https://www.pnas.org/doi/pdf/10.1073/pnas.37.3.170}}].

\bibitem{Harish-Chandra4}
Harish-Chandra, {\it Representations of semisimple lie groups iv},  {\em
  American Journal of Mathematics} {\bf 77} (1955), no.~4 743--777.

\bibitem{Whittaker}
P.~Garrett,
  ``\href{http://www-users.math.umn.edu/~garrett/m/v/toy_GL2.pdf}{Representations
  of $GL(2)$ and $SL(2)$ over finite fields}.''
\newblock in \href{http://www-users.math.umn.edu/~garrett/m/v/}{Vignettes on
  automorphic forms, representations, L-functions, and number theory}.

\bibitem{natural}
M.~Libine, {\it {Introduction to Representations of Real Semisimple Lie
  Groups}},  \href{http://arxiv.org/abs/1212.2578}{{\tt arXiv:1212.2578}}.

\bibitem{Berstein1}
I.~N. Bern\v{s}te\u{i}n, ``All reductive $p$-adic groups are of type \rom{1}.''
\newblock
  \href{http://www.mathnet.ru/php/archive.phtml?wshow=paper&jrnid=faa&paperid=2324&option_lang=eng}{\emph{Funkcional.
  Anal. i Prilo\v zen.} {\bf 8} (1974) 3-6; \emph{Funct. Anal. Appl.} {\bf 8}
  (1974) 91-93}.

\bibitem{Bernstein}
I.~N. Bern\v{s}te\u{i}n, ``Representations of $p$-adic groups.''
\newblock Available at
  \href{http://www.math.harvard.edu/~gaitsgde/Jerusalem_2010/GradStudentSeminar/p-adic.pdf}{\texttt{\url{http://www.math.harvard.edu/~gaitsgde/Jerusalem_2010/GradStudentSeminar/p-adic.pdf}}}.

\bibitem{First}
U.~A. First and T.~Rud, {\it {On Uniform Admissibility of Unitary and Smooth
  Representations}},  {\em Archiv der Mathematik volume} {\bf 112} (2019)
  169--179, [\href{http://arxiv.org/abs/1801.08719}{{\tt arXiv:1801.08719}}].

\bibitem{Admissibility1}
T.~Rud, ``Admissibility of representations of totally disconnected locally
  compact groups.''
\newblock
  \href{http://www.math.ubc.ca/~thomas/TeXthings/Projet_Master_Thomas_Rud_v3.pdf}{Master
  Thesis}, EPFL and UBC, 2016.

\bibitem{huang2020green}
A.~Huang, B.~Stoica, S.-T. Yau, and X.~Zhong, {\it {Green\textquoteright{}s
  functions for Vladimirov derivatives and Tate\textquoteright{}s thesis}},
  {\em Commun. Num. Theor. Phys.} {\bf 15} (2021), no.~2 315--361,
  [\href{http://arxiv.org/abs/2001.01721}{{\tt arXiv:2001.01721}}].

\bibitem{autreps}
K.~Martin, ``Automorphic representations, $\operatorname{Fall}$ 2011 notes.''
\newblock Available at
  \href{http://www2.math.ou.edu/~kmartin/autreps/autreps.pdf}{\texttt{http://www2.math.ou.edu/~kmartin/autreps/autreps.pdf}}.

\bibitem{Jacquet}
H.~Jacquet, ``Repr\'esentations des groupes lin\'eaires $p$-adiques.''
\newblock in \emph{Theory of group representations and Fourier analysis
  (proceedings of a conference at Monecatini)}, C.I.M.E. Edizioni Cremonese,
  Rome, 1971.

\bibitem{JacquetLanglands}
H.~Jacquet and R.~P. Langlands, {\em Automorphic Forms on GL (2)}, vol.~114 of
  {\em Lecture Notes in Mathematics}.
\newblock Springer-Verlag, 1970.

\bibitem{Woodbury}
M.~C. Woodbury, ``Representation theory and number theory.''
\newblock Available at
  \href{http://www.mi.uni-koeln.de/~woodbury/research/Grossnotes.pdf}{\texttt{http://www.mi.uni-koeln.de/~woodbury/research/Grossnotes.pdf}}.

\bibitem{Taibi}
O.~Ta\"ibi, ``The $\operatorname{J}$acquet-$\operatorname{L}$anglands
  correspondence for $\operatorname{GL}_2\left(\mathbb{Q}_p\right)$.''
\newblock \href{https://otaibi.perso.math.cnrs.fr/notesJL.pdf}{\emph{Notes for
  the M2 course}}.

\bibitem{Cardy4}
J.~Cardy, A.~Maloney, and H.~Maxfield, {\it {A new handle on three-point
  coefficients: OPE asymptotics from genus two modular invariance}},  {\em
  JHEP} {\bf 10} (2017) 136, [\href{http://arxiv.org/abs/1705.05855}{{\tt
  arXiv:1705.05855}}].

\bibitem{Gervais}
J.-L. Gervais, {\it {$p$-adic Analyticity and Virasoro Algebras for Conformal
  Theories in More Than Two-dimensions}},  {\em Phys. Lett. B} {\bf 201} (1988)
  306--310.

\bibitem{Yamazaki}
M.~Yamazaki, {\it {Comments on Determinant Formulas for General CFTs}},  {\em
  JHEP} {\bf 10} (2016) 035, [\href{http://arxiv.org/abs/1601.04072}{{\tt
  arXiv:1601.04072}}].

\bibitem{parabolic}
C.~Schweigert, ``Introduction to conformal field theory.''
\newblock
  \href{https://www.math.uni-hamburg.de/home/schweigert/ws13/cft.html}{65-405,
  Winter 2013/14}.

\bibitem{TTbar}
L.~McGough, M.~Mezei, and H.~Verlinde, {\it {Moving the CFT into the bulk with
  $ T\overline{T} $}},  {\em JHEP} {\bf 04} (2018) 010,
  [\href{http://arxiv.org/abs/1611.03470}{{\tt arXiv:1611.03470}}].

\bibitem{Kraus:2018xrn}
P.~Kraus, J.~Liu, and D.~Marolf, {\it {Cutoff AdS$_{3}$ versus the $
  T\overline{T} $ deformation}},  {\em JHEP} {\bf 07} (2018) 027,
  [\href{http://arxiv.org/abs/1801.02714}{{\tt arXiv:1801.02714}}].

\bibitem{Guica:2019nzm}
M.~Guica and R.~Monten, {\it {$T\bar T$ and the mirage of a bulk cutoff}},
  {\em SciPost Phys.} {\bf 10} (2021), no.~2 024,
  [\href{http://arxiv.org/abs/1906.11251}{{\tt arXiv:1906.11251}}].

\bibitem{Kraus:2021cwf}
P.~Kraus, R.~Monten, and R.~M. Myers, {\it {3D Gravity in a Box}},  {\em
  SciPost Phys.} {\bf 11} (2021) 070,
  [\href{http://arxiv.org/abs/2103.13398}{{\tt arXiv:2103.13398}}].

\bibitem{Ebert:2022cle}
S.~Ebert, E.~Hijano, P.~Kraus, R.~Monten, and R.~M. Myers, {\it {Field Theory
  of Interacting Boundary Gravitons}},
  \href{http://arxiv.org/abs/2201.01780}{{\tt arXiv:2201.01780}}.

\bibitem{Ebert:2022ehb}
S.~Ebert, C.~Ferko, H.-Y. Sun, and Z.~Sun, {\it {$T\bar{T}$ in JT Gravity and
  BF Gauge Theory}},  \href{http://arxiv.org/abs/2205.07817}{{\tt
  arXiv:2205.07817}}.

\bibitem{Kraus:2022mnu}
P.~Kraus, R.~Monten, and K.~Roumpedakis, {\it {Refining the Cutoff 3d Gravity /
  $T\bar{T}$ Correspondence}},  \href{http://arxiv.org/abs/2206.00674}{{\tt
  arXiv:2206.00674}}.

\bibitem{Iliesiu:2020zld}
L.~V. Iliesiu, J.~Kruthoff, G.~J. Turiaci, and H.~Verlinde, {\it {JT gravity at
  finite cutoff}},  {\em SciPost Phys.} {\bf 9} (2020) 023,
  [\href{http://arxiv.org/abs/2004.07242}{{\tt arXiv:2004.07242}}].

\bibitem{Ebert:2022gyn}
S.~Ebert, H.-Y. Sun, and Z.~Sun, {\it {$ T\overline{T} $-deformed free energy
  of the Airy model}},  {\em JHEP} {\bf 08} (2022) 026,
  [\href{http://arxiv.org/abs/2202.03454}{{\tt arXiv:2202.03454}}].

\bibitem{Qu:2021fgz}
F.~Qu, {\it {Effective field theory on a finite boundary of the Bruhat-Tits
  tree}},  {\em Phys. Rev. D} {\bf 103} (2021), no.~8 086015,
  [\href{http://arxiv.org/abs/2103.02882}{{\tt arXiv:2103.02882}}].

\bibitem{Qu:2021huo}
F.~Qu, {\it {Euclidean (A)dS spaces over $p$-adic numbers}},
  \href{http://arxiv.org/abs/2105.10183}{{\tt arXiv:2105.10183}}.

\bibitem{MaldacenaStrominger}
J.~M. Maldacena and A.~Strominger, {\it {AdS(3) black holes and a stringy
  exclusion principle}},  {\em JHEP} {\bf 12} (1998) 005,
  [\href{http://arxiv.org/abs/hep-th/9804085}{{\tt hep-th/9804085}}].

\bibitem{Watson}
G.~N. Watson, ``The final problem: an account of the mock theta functions.''
\newblock
  \href{https://londmathsoc.onlinelibrary.wiley.com/doi/abs/10.1112/jlms/s1-11.1.55}{\emph{J.
  London Math. Soc.} {\bf 11} (1936) 55-80}.

\bibitem{Gross}
C.~Li, ``Lecture notes on representation theoy and number theory.''
\newblock by Benedict Gross at Columbia (Fall 2011). Available at
  \href{http://www.math.columbia.edu/~chaoli/docs/EilenbergLectures.html}{\texttt{http://www.math.columbia.edu/~chaoli/docs/EilenbergLectures.html}}.

\bibitem{Knightly}
A.~Knightly and C.~Ragsdale, ``Matrix coefficients of depth-zero supercuspidal
  representations of $\operatorname{GL}(2)$.''
\newblock \href{https://msp.org/involve/2014/7-5/p08.xhtml}{\emph{Involve} {\bf
  7} (2014) 669-690}.

\bibitem{Youcis}
A.~Youcis, ``Notes for $p$-adic representations.''
\newblock Available at
  \href{https://ayoucis.files.wordpress.com/2016/10/notes-p-adic.pdf}{\texttt{https://ayoucis.files.wordpress.com/2016/10/notes-p-adic.pdf}}.

\end{thebibliography}\endgroup
\end{document}